\documentclass[12pt]{article}

\usepackage{array}
\usepackage{epsfig}
\usepackage{amssymb}
\usepackage{graphics,graphpap}

\renewcommand{\thefootnote}{\fnsymbol{footnote}}

\newcommand\ud{\uparrow\downarrow}
\newcommand\du{\downarrow\uparrow}

\newcommand{\ub}{\bar{u}}
\newcommand{\quark}{\langle \bar q q\rangle}
\newcommand{\mixed}{\langle \bar q \sigma gG q\rangle}
\newcommand{\squark}{\langle \bar s s\rangle}

\newcommand{\smixed}{\langle \bar s \sigma gG s\rangle}

\newcommand{\gluon}{\left\langle \frac{\alpha_s}{\pi}\,G^2\right\rangle}
\newcommand{\bra}{\langle}
\newcommand{\ket}{\rangle}
\newcommand{\ds}{\displaystyle}
\newcommand{\wt}{\widetilde}

\def\s#1{\setbox0=\hbox{$#1$}%
  \rlap{\ifdim\wd0>.7em\kern.22\wd0\else\kern.1\wd0\fi /}#1}


\begin{document}

\begin{titlepage}
\begin{flushright}
\begin{tabular}{l}
IPPP/07/33\\
DCPT/07/66
\end{tabular}
\end{flushright}
\vskip1.5cm
\begin{center}
{\Large \bf \boldmath
Twist-4 Distribution Amplitudes\\[7pt] of the K$^*$ and $\phi$ Mesons in QCD}
\vskip1.3cm 
{\sc
Patricia Ball\footnote{Patricia.Ball@durham.ac.uk}$^{,1}$,
V.M.~Braun\footnote{Vladimir.Braun@physik.uni-regensburg.de}$^{,2}$
and
A. Lenz\footnote{Alexander.Lenz@physik.uni-regensburg.de}$^{,2}$}
  \vskip0.5cm
        $^1$ {\em IPPP, Department of Physics,
University of Durham, Durham DH1 3LE, UK} \\
\vskip0.4cm
$^2$ {\em Institut f\"ur Theoretische Physik, \\ Universit\"at Regensburg,
D--93040 Regensburg, Germany}

\vskip2cm


\vskip3cm

{\large\bf Abstract\\[10pt]} \parbox[t]{\textwidth}{
We present a systematic study of twist-4 light-cone distribution 
amplitudes of the $K^*$ and $\phi$ meson in QCD. 
The structure of SU(3)-breaking corrections is studied in
detail. Non-perturbative input parameters are estimated from QCD sum
rules and a renormalon based model. 
As a by-product, we give a complete reanalysis
of the parameters of the
twist-4 $\rho$-meson distribution amplitudes.
}

\vfill

\end{center}
\end{titlepage}

\setcounter{footnote}{0}
\renewcommand{\thefootnote}{\arabic{footnote}}
\renewcommand{\theequation}{\arabic{section}.\arabic{equation}}

\newpage

\section{Introduction}\label{sec:1}
\setcounter{equation}{0}

The notion of distribution amplitudes (DAs) \cite{exclusive} refers to 
matrix elements of nonlocal light-ray operators sandwiched between the 
hadron state and the vacuum. The physical interpretation of DAs is 
transparent in the infinite momentum frame \cite{BLreport},
in which case DAs correspond  to momentum-fraction distributions of
partons in a hadron at small transverse separation. Equivalently, 
DAs can be related to the transverse momentum integrals of the
hadron's Bethe-Salpeter wave functions that appear e.g. using the
formalism of the light-cone quantization~\cite{LCQ}. Schematically,
$$
\phi(x) \sim \int^{|k_\perp|<\mu} d^2k_\perp \phi_{BS}(x,k_\perp)\,.
$$

The natural application of DAs in phenomenology are exclusive hard processes
with large momentum transfer which can be calculated using
factorisation methods. Apart from meson/baryon electromagnetic form factors, 
this includes a large class of
phenomenologically very interesting $B$ meson decays, such as
weak decay form factors \cite{LCSR}, 
the non-leptonic decays $B\to M_1 M_2$, where $M_i$ is a light meson 
\cite{BBNS},
and also rare radiative and
semileptonic decays, $B\to M\gamma$ \cite{BB00} and $B\to
M\ell^+\ell^-$ \cite{BFS}, which
involve flavour-changing neutral currents and
are heavily suppressed in the Standard Model, but sensitive to
new-physics effects. For these processes, the shape of the DAs is very important: in 
$B\to\rho\gamma$ vs.\ $B\to
K^*\gamma$, for example, the size of SU(3) breaking in the relevant DAs is presently the
dominant source of theoretical uncertainty \cite{BJZ}. All these decays will
be studied in detail at the forthcoming LHC, which makes the detailed
investigation of the various leading and higher-twist DAs both timely and relevant. 

The crucial point and main technical difficulty in the construction of higher-twist DAs is the 
necessity to satisfy the exact equations of motion (EOM), which yield
relations between physical effects 
of different origin: for example, using EOM, the contributions of
orbital angular momentum 
in the  valence component of the wave function can be expressed (for mesons)
in terms of contributions of higher 
Fock states. An appropriate framework for implementing these
constraints was developed in Ref.~\cite{BF90}: it is based on the derivation 
of EOM relations for non-local light-ray operators~\cite{string},
which are solved order by order in 
the conformal expansion; see Ref.~\cite{BKM03} for a review and
further references. In this way it is possible to construct 
self-consistent approximations for the DAs, which involve
a minimum number of hadronic parameters. Another approach, based on 
the study of renormalons, 
was suggested for twist 4 in Refs.~\cite{Andersen99,BGG04}: this
technique is appealing as it allows one to 
obtain an estimate of  high-order contributions to the conformal
expansion which are usually omitted.
In Ref.~\cite{BBL06} we have generalized this approach to include 
SU(3)-breaking 
corrections and have shown how to combine renormalon-based estimates of 
``genuine'' twist-4 effects with meson mass corrections.

In a series of previous papers, Refs.~\cite{BBKT,BB98}, we have
developed the corresponding formalism for vector mesons. 
In Ref.~\cite{BJ07} the analysis of twist-3 DAs of light vector
mesons, $\rho$, $K^*$ and $\phi$,
was completed,  including all SU(3) and G-parity-breaking effects. 
In this paper, we put together the last missing pieces, completing 
the study of the corresponding DAs of twist
4, with the main emphasis on the calculation of SU(3)-breaking effects in the
relevant hadronic matrix elements. These corrections come from
different sources:
\begin{itemize}
\item SU(3) breaking of hadronic parameters: these effects are
  known for twist-2 and -3 parameters, see
  Refs.~\cite{BJ07,BB03}, 
  but have not been studied for
  twist-4 parameters before;
\item G-parity-breaking parameters: these are of parametric order 
$m_s-m_q$ and vanish in the
  limit of equal quark mass, i.e.\ for $\rho$ and $\phi$. For twist-2
  DAs, they  have been calculated, 
  to lowest order in the conformal expansion in
  Refs.~\cite{BBKT,BB03,BL04,BZ05,BZ06a}, and for
  twist-3 DAs in Ref.~\cite{BJ07}; they are unknown for twist-4 DAs;
\item explicit quark-mass corrections in $m_s\pm m_q$ to DAs and
  evolution equations: these affect only higher-twist DAs and are
  induced by the QCD EOM which relate twist-4
  DAs to each other and to twist-2 and -3 DAs. The mass
  corrections to vector meson DAs
  have been calculated to twist-4 accuracy in Ref.~\cite{BB98}. 
Quark-mass corrections to the evolution of DAs under a change of 
the renormalisation scale have so far only been calculated for
  twist-3 DAs \cite{BJ07}.  
\end{itemize}
We shall study all these effects in this paper.
The corresponding analysis of light pseudoscalar meson DAs, $\pi$ and
$K$, can be found in
Refs.~\cite{BF90,BBL06,PB98}.

In addition, in this work we present a new analysis of the parameters
of the twist-4 DAs of the $\rho$ meson. This update is long overdue: 
the ``standard'' values for these parameters can be traced back to a
nearly 20-years-old work, Ref.~\cite{BBK89} (see also \cite{BB98}),
and are in fact crude estimates 
obtained by dividing the leading QCD contribution to the relevant 
correlation functions by the typical hadronic scale. In this work we 
present, for the first time, a complete treatment of the 
twist-4 matrix elements within the QCD sum rule method. In particular, 
we resolve a puzzling discrepancy between the estimate of a certain 
next-to-leading order (NLO) parameter, in conformal spin,
in Ref.~\cite{BB98} and in the 
renormalon model \cite{BGG04}.  

The presentation is organised as follows: in Sec.~\ref{sec:2}  we introduce
notations and shortly review twist-2 and -3 DAs. In
Sec.~\ref{sec:3} we introduce the complete set of  chiral-even and in
Sec.~\ref{sec:4} chiral-odd twist-4 DAs. The conformal expansions of all DAs
are worked out to NLO accuracy in conformal spin and reduced to 
a minimum number
of non-perturbative hadronic parameters by solving the EOM constraints. 
In Sec.~\ref{sec:5}, we present models for these
DAs, based on the calculation of the hadronic parameters from QCD sum
rules. We summarise and conclude in
Sec.~\ref{sec:6}. Details of the QCD sum rule calculations are
given in the appendices.

\section{General Framework}\label{sec:2}
\setcounter{equation}{0}


\subsection{Kinematics and Notations}
Light-cone meson DAs are defined in terms of matrix elements of
non-local light-ray operators extended along a certain light-like 
direction $z_\mu$, $z^2=0$, and sandwiched between the vacuum and the 
meson state.
Following Ref.~\cite{BBL06}, we adopt the generic notations
\begin{equation}
\phi^\lambda_{t;M}(u),\ \psi^\lambda_{t;M}(u),\ \ldots
\end{equation}
and
\begin{equation}
\Phi^\lambda_{t;M}({\underline{\alpha}}),\ 
\Psi^\lambda_{t;M}({\underline{\alpha}}),\
\ldots
\end{equation}
for two- and  three-particle DAs, respectively. The
superscript $\lambda$ denotes the polarisation of the vector meson:
$\lambda=\parallel(\perp)$ for longitudinal (transverse)
polarisation. The first
subscript $t=2,3,4$ stands for the twist; the second,
$M=\rho,K^*,\ldots$, specifies the meson. For definiteness, we will write
most expressions for $K^*$ mesons, i.e.\  $s\bar q$ bound states
with $q =u,d$. Whenever relevant, we will include quark-mass
corrections in the form $m_s\pm m_q$, which allows one to obtain the
results for $\phi$ mesons by $m_q\to m_s$. 
The variable $u$ in the definition of two-particle DAs
always refers to the momentum fraction carried by the quark, $u =
u_s$, whereas
$\bar u \equiv 1-u = u_{\bar q}$ is the antiquark momentum fraction.
The set of variables in the three-particle DAs,
$\underline{\alpha} = \{\alpha_1,\alpha_2,\alpha_3\} = \{\alpha_{s},
\alpha_{\bar q},\alpha_g\}$, corresponds to the momentum fractions 
carried by the quark, antiquark and gluon, respectively.

To facilitate the light-cone expansion, it is
convenient to use light-like vectors $p_\mu$ and $z_\mu$ instead of
the meson's four-momentum $P_\mu$ and the coordinate $x_\mu$:
\begin{eqnarray}
  z_\mu &=& 
x_\mu-P_\mu\,\frac{1}{m_{K^*}^2}\left[xP -\sqrt{(xP)^2-x^2m^2_{K^*}}\,\right]
= x_\mu\left[1-\frac{x^2m_{{K^*}}^2}{4(zp)^2}\right]
-\frac{1}{2}p_\mu\,\frac{x^2}{zp}
    + \mbox{\cal O}(x^4)\,,
\nonumber\\
p_\mu &=& P_\mu-\frac{1}{2}\,z_\mu\, \frac{m^2_{K^*}}{pz}\,.
\label{lc-variables}
\end{eqnarray}
The meson's polarization vector 
$e^{(\lambda)}$ can be decomposed into projections 
onto the two light-like vectors and the orthogonal plane as follows:
\begin{eqnarray}
 e^{(\lambda)}_\mu &=& \frac{e^{(\lambda)}z}{pz}\, p_\mu +
                     \frac{e^{(\lambda)} p}{pz}\, z_\mu +
                     e^{(\lambda)}_{\perp\mu}
=  \frac{e^{(\lambda)} z}{pz}\left( p_\mu -\frac{m^2_{K^*}}{2pz}\, z_\mu
                                         \right)+e^{(\lambda)}_{\perp\mu}\,. 
\label{polv}
\end{eqnarray}
We also need the projector $g_{\mu\nu}^\perp$ 
onto the directions orthogonal to $p$ and $z$,
\begin{equation}
      g^\perp_{\mu\nu} = g_{\mu\nu} -\frac{1}{pz}(p_\mu z_\nu+ p_\nu z_\mu)\,, 
\end{equation}
and will often use the notations
\begin{equation}
    a_z\equiv a_\mu z^\mu, \qquad b_p\equiv b_\mu p^\mu
\end{equation}
for  arbitrary four-vectors $a_\mu$ and $b_\mu$.

The dual gluon field strength
tensor is defined as $\widetilde{G}_{\mu\nu} =
\frac{1}{2}\epsilon_{\mu\nu \rho\sigma} G^{\rho\sigma}$. Our
convention for the covariant derivative is $D_\mu = \partial_\mu - i g
A_\mu$. Sometimes, a different convention for the sign of $g$ is used in the
literature: $D_\mu = \partial_\mu + i g
A_\mu$. The sign of $g$ is relevant for the parameters of
three-particle DAs.

\subsection{Conformal Expansion}

A convenient tool to study DAs is provided by conformal expansion, see
Ref.~\cite{BKM03} for a review.
The underlying idea is similar to partial-wave decomposition in 
quantum mechanics and allows one to separate
transverse and longitudinal variables in the Bethe-Salpeter 
wave-function.  The
dependence on transverse coordinates is formulated as scale dependence
of the relevant operators and is governed by
renormalisation-group equations, the dependence on the longitudinal
momentum fractions is described in terms of irreducible
representations of the corresponding symmetry group, the collinear
conformal group SL(2,$\mathbb R$).
The main rationale behind using the conformal expansion in the present context
is that the EOM always relate contributions of the same spin;
a truncation of the conformal expansion to a certain order is,
therefore, consistent with the EOM.

To construct the conformal expansion for an arbitrary multi-particle
distribution, one first has to decompose each constituent field into
components with fixed Lorentz-spin projection onto the
light-cone. Each such component has conformal spin
$$
j=\frac{1}{2}\, (l+s),
$$
where $l$ is the canonical dimension  and $s$ the (Lorentz-) spin
projection. In particular, $l=3/2$ for quarks and $l=2$ for gluons.
A  quark field is decomposed as $\psi_+ \equiv
\Lambda_+\psi$ and $\psi_-=
\Lambda_-\psi$ with spin projection operators $\Lambda_+ = 
\gamma_p\gamma_z/(2pz)$ and 
 $\Lambda_- = \gamma_z\gamma_p/(2pz)$, corresponding to
$s=+1/2$ and $s=-1/2$, respectively. For the gluon
field strength there are three possibilities:
$G_{z\perp}$ corresponds to $s=+1$,
$G_{p\perp}$ to $s=-1$, and both
$G_{\perp\perp}$ and $G_{zp}$ correspond to $s=0$.
Multi-particle states built of fields with definite Lorentz-spin
projection can be expanded in
irreducible  representations of SL(2,$\mathbb R$) 
with increasing conformal spin.
The explicit expression for the DA
of an $m$-particle state with the lowest possible conformal spin
 $j=j_1+\ldots+j_m$, the so-called asymptotic DA, is given by \cite{BF90}
\begin{equation}
\phi_{as}(\alpha_1,\alpha_2,\cdots,\alpha_m) =
\frac{\Gamma(2j_1+\cdots +2j_m)}{\Gamma(2j_1)\cdots \Gamma(2j_m)}\,
\alpha_1^{2j_1-1}\alpha_2^{2j_2-1}\ldots \alpha_m^{2j_m-1}.
\label{asym}
\end{equation}
Multi-particle irreducible representations with higher spin
$j+n,n=1,2,\ldots$,
are given by  polynomials of $m$ variables (with the constraint
$\sum_{k=1}^m \alpha_k=1$ ), which are orthogonal over
 the weight function (\ref{asym}). For the two-particle
 DAs these are Gegenbauer polynomials, a convenient basis of
 orthogonal conformal 
 polynomials for the three-particle DAs is given in App.~A 
of Ref.~\cite{BKM03} (see also \cite{Cbasis}).

The anomalous dimensions of higher conformal amplitudes do,
generally, increase logarithmically with the conformal spin, but a 
complete analysis of twist-4 anomalous dimensions is still lacking.
It follows that all DAs approach their asymptotic form in the
asymptotic limit $\alpha_s\to 0$, i.e.\ at the scale $\mu\to\infty$. 
For practical applications one usually assumes that
the conformal expansion is converging fast enough, so that a
truncation after the few first terms is 
sufficient. The renormalon model of Refs.~\cite{Andersen99,BGG04} 
presents an attempt to test this assumption,
by giving an upper bound for possible higher-spin contributions. 

\subsection{Twist-2 Distributions}

The twist-2 DAs $\phi_{2;K^*}^{\parallel,\perp}$ of $K^*$ mesons are
defined in terms of the following matrix elements of non-local
operators ($\xi=2u-1$) \cite{BBKT}:
\begin{eqnarray}
\lefteqn{\hspace*{-1.3cm}\langle 0 | \bar q(x) \gamma_\mu s(-x) |
 K^*(P,\lambda)\rangle = 
f_{K^*}^\parallel m_{K^*} \left\{ \frac{e^{(\lambda)}x}{Px}\,P_\mu
\int_0^1 du\, e^{i\xi Px} \left[ \phi_{2;K^*}^\parallel(u) +
  \frac{1}{4}\, m_{K^*}^2 x^2
  \phi^\parallel_{4;K^*}(u)\right]\right.}\nonumber\\
&&{} + \left( e^{(\lambda)}_\mu -
P_\mu\,\frac{e^{(\lambda)}x}{Px}\right) \int_0^1 du\,e^{i\xi
  Px}\,\phi_{3;K^*}^\perp(u)\nonumber\\
&&\left. - \frac{1}{2}\,x_\mu
\,\frac{e^{(\lambda)}x}{(Px)^2} \, m_{K^*}^2 \int_0^1 du\,e^{i\xi Px}\,
\left[ \psi_{4;K^*}^\parallel(u) + \phi_{2;K^*}^\parallel(u) - 2
  \phi_{3;K^*}^\perp(u)\right] + \dots\right\},\label{2.9}
\end{eqnarray}
\begin{eqnarray}
\lefteqn{
\langle 0 | \bar q(x) \sigma_{\mu\nu} s(-x) | K^*(P,\lambda)\rangle 
=}\hspace*{1cm}
\nonumber\\
&&{}i f_{K^*}^\perp \left\{ (e_\mu^{(\lambda)} P_\nu - e_\nu^{(\lambda)}
  P_\mu) \int_0^1 du \, e^{i\xi Px} \left[ \phi_{2;K^*}^\perp(u) +
  \frac{1}{4}\, m_{K^*}^2 x^2
  \phi^\perp_{4;K^*}(u)\right]\right.
\nonumber\\
&&{}
+ (P_\mu x_\nu - P_\nu
x_\mu)\,\frac{e^{(\lambda)}x}{(Px)^2}\,m_{K^*}^2 \int_0^1 du \,
e^{i\xi Px} \left[ \phi_{3;K^*}^\parallel(u) -
  \frac{1}{2}\phi_{2;K^*}^\perp(u) -
  \frac{1}{2}\psi_{4;K^*}^\perp(u)\right]
\nonumber\\
&&{}\left.
+ \frac{1}{2}\,(e_\mu^{(\lambda)} x_\nu - e_\nu^{(\lambda)}x_\mu)\,
\frac{m_{K^*}^2}{Px} \int_0^1 du \,
e^{i\xi Px} \left[ \psi_{4;K^*}^\perp(u)-\phi_{2;K^*}^\perp(u)
  \right]+\dots\right\}.\label{2.10}
\end{eqnarray}
The above relations also include twist-3 and -4 two-particle DAs. The
dots stand for further terms in $x^2$ which are of twist 5 or higher.
The normalisation of all these DAs is given by
\begin{equation}
\int_0^1 du\, \phi(u) = 1\,.
\end{equation}
The conformal expansion goes in terms of Gegenbauer polynomials:
\begin{equation}\label{eq:confexp}
\phi_2^{\parallel,\perp}(u,\mu) = 6 u \bar u \left\{ 1 + \sum_{n=1}^\infty
a_n^{\parallel,\perp}(\mu) C_n^{3/2}(2u-1)\right\}.
\end{equation}
In this paper, we include terms up to NLO in
conformal spin, i.e.\ truncate after $n=2$. The dependence of the
Gegenbauer moments $a_n$ on the renormalisation-scale $\mu$ has been
reviewed in Ref.~\cite{BJ07}, together with the numerical values of
$a_n$ and the decay constants $f_V^{\parallel,\perp}$; these values
are given in Sec.~\ref{sec:5}. 

\subsection{Twist-3 Distributions}

To twist-3 accuracy, there is a total of four two-particle DAs and three
three-particle DAs. Two of the former, $\phi^\perp_{3;K^*}$ and
$\phi_{3;K^*}^\parallel$, have already been defined in the previous
subsection. The other two are given by
\begin{eqnarray}
\langle 0|\bar q(z) \gamma_{\mu} \gamma_{5}
s(-z)|K^*(P,\lambda)\rangle
&=& \frac{1}{2}\,f_{K^*}^\parallel
m_{K^*} \epsilon_{\mu}^{\phantom{\mu}\nu \alpha \beta}
e^{(\lambda)}_{\nu} p_{\alpha} z_{\beta}
\int_{0}^{1} \!du\, e^{i \xi p  x} \psi_{3;K^*}^\perp(u)\,,\\
\langle 0|\bar q(z)s(-z)|K^*(P,\lambda)\rangle
 & = & {} -i f_{K^*}^\perp(e^{(\lambda)} z) m_{K^*}^{2}
\int_{0}^{1} \!du\, e^{i \xi p  z} \psi_{3;K^*}^\parallel(u)\label{2.13}
\end{eqnarray}
with the normalisation
\begin{equation}
\int_0^1 du\, \psi_{3;K^*}^{\parallel(\perp)}(u) = 1-
\frac{f_{K^*}^{\parallel(\perp)}}{f_{K^*}^{\perp(\parallel)}}\,
\frac{m_s+m_q}{m_{K^*}}\,.
\end{equation}
The three-particle DAs are given by:
\begin{eqnarray}
\langle 0 | 
\bar q(z) g \widetilde{G}_{\beta z}(vz)\gamma_z\gamma_5 s(-z) |
K^*(P,\lambda)\rangle & = & f_{K^*}^\parallel m_{K^*} (pz)^2
e^{(\lambda)}_{\perp\beta}
\widetilde\Phi_{3;K^*}^\parallel(v,pz)+\dots\,,
\nonumber\\
\langle 0 | 
\bar q(z) g G_{\beta z}(vz)i\gamma_z s(-z) | K^*(P,\lambda)\rangle & = & 
f_{K^*}^\parallel m_{K^*} (pz)^2
e^{(\lambda)}_{\perp\beta} \Phi_{3;K^*}^\parallel(v,pz)+\dots\,,\nonumber\\
\langle 0 | 
\bar q(z) g G_{z\beta}(vz)\sigma_{z\beta} s(-z) |
K^*(P,\lambda)\rangle & = & f_{K^*}^\perp m_{K^*}^2
(e^{(\lambda)}z)(pz) \Phi_{3;K^*}^\perp(v,pz)\,,\label{3.4}
\end{eqnarray}
where the dots denote terms of higher twist and we use the short-hand notation 
\begin{equation}
{\cal F}(v,pz) = \int {\cal
  D}\underline{\alpha}\,e^{-ipz(\alpha_2-\alpha_1+v\alpha_3)}
  {\cal F}(\underline{\alpha}) 
\end{equation}
with ${\cal F}(\underline{\alpha})$ a three-particle DA.
$\underline{\alpha}$ is the set of parton momentum fractions
$\underline{\alpha} = \{\alpha_1,\alpha_2,\alpha_3\}$ and the integration
measure ${\cal D}\underline{\alpha}$ is defined as
\begin{equation}\label{eq:measure}
\int{\cal D}\underline{\alpha} \equiv \int_0^1 d\alpha_1 d\alpha_2
d\alpha_3\, \delta\left(1-\sum \alpha_i\right).
\end{equation}

The above DAs are not independent of each other, and their 
mutual interrelations have been unravelled in
Ref.~\cite{BBKT}, including quark-mass corrections.
Explicit expressions for the conformal expansion are
given in Ref.~\cite{BJ07}, together with the $\mu$-dependence of
hadronic parameters. Numerical values are
given in Sec.~\ref{sec:5}.

\section{Chiral-Even Twist-4 Distributions}\label{sec:3}
\setcounter{equation}{0}

In this section we derive expressions for the chiral-even
two- and three-particle
twist-4 DAs of the $K^*$ to NLO 
in the conformal expansion. The
corresponding expressions for the $\rho$ were obtained in
Ref.~\cite{BB98}. In this paper we  include also G-parity-violating
 and explicit quark-mass corrections. We 
compare the resulting DAs obtained in conformal expansion with
those following from  the renormalon model developed in
Ref.~\cite{BGG04}. Numerical estimates of the hadronic input
parameters and the resulting DAs are discussed in Sec.~\ref{sec:5}.

We start with the three\--par\-ticle distributions. The analysis closely
follows that of Ref.\cite{BB98}.
There are four chiral-even $K^*$ three-particle DAs of twist 4, 
defined as~\cite{BB98}:%
\footnote{
   In  the notation of  Ref.~\cite{BB98},
   $\Phi_{4;K^*}^\parallel = \Phi$, 
   $\Psi_{4;K^*}^\parallel = \Psi$, 
   $\widetilde\Phi_{4;K^*}^\parallel = \widetilde\Phi$, 
   $\widetilde\Psi_{4;K^*}^\parallel = \widetilde\Psi$.} 
\begin{eqnarray}
\lefteqn{\langle 0 | \bar q(z)\gamma_\mu\gamma_5
g\widetilde G_{\alpha\beta}(vz)s(-z)|K^*(P,\lambda)\rangle\ =}
\nonumber\\
& = & p_\mu \left(e^{(\lambda)}_{\perp\alpha} p_\beta - 
e^{(\lambda)}_{\perp\beta}
p_\alpha\right) f_{K^*}^\parallel m_{K^*}
\widetilde\Phi_{3;K^*}^\parallel(v,pz)
+ \left( p_\alpha g^\perp_{\beta\mu} - p_\beta
g^\perp_{\alpha\mu}\right) \frac{e^{(\lambda)}z}{pz}\,f_{K^*}^\parallel
m_{K^*}^3\widetilde\Phi^\parallel_{4;K^*}(v,pz)\nonumber\\
&&{} + p_\mu \left( p_\alpha z_\beta - p_\beta z_\alpha\right) 
\frac{e^{(\lambda)}z}{(pz)^2}\,f_{K^*}^\parallel
m_{K^*}^3\widetilde\Psi^\parallel_{4;K^*}(v,pz)+\dots,\label{4.1}\\
\lefteqn{\langle 0 | \bar q(z)i\gamma_\mu
g G_{\alpha\beta}(vz)s(-z)|K^*(P,\lambda)\rangle\ =}
\hspace*{0.5cm}\nonumber\\
& = & p_\mu \left(e^{(\lambda)}_{\perp\alpha} p_\beta - 
e^{(\lambda)}_{\perp\beta}
p_\alpha\right) f_{K^*}^\parallel m_{K^*}\Phi_{3;K^*}^\parallel(v,pz)
+ \left( p_\alpha g^\perp_{\beta\mu} - p_\beta
g^\perp_{\alpha\mu}\right) \frac{e^{(\lambda)}z}{pz}\,f_{K^*}^\parallel
m_{K^*}^3\Phi^\parallel_{4;K^*}(v,pz)\nonumber\\
&&{} + p_\mu \left( p_\alpha z_\beta - p_\beta z_\alpha\right) 
\frac{e^{(\lambda)}z}{(pz)^2}\,f_{K^*}^\parallel
m_{K^*}^3\Psi^\parallel_{4;K^*}(v,pz)+\dots\,;\label{4.2}
\end{eqnarray}
the dots denote terms of twist 5 and higher.

$\Psi_{4;K^*}^\parallel$ and  $\widetilde
\Psi_{4;K^*}^\parallel$ correspond to the light-cone projection
$\gamma_z G_{z p}$ which picks up the $s=+1/2$ components of both
quark and antiquark field and the $s=0$ component of
 the gluon field. The conformal expansion then reads:
\begin{eqnarray}
\Psi_{4;K^*}^\parallel(\underline{\alpha}) & = & 120
\alpha_1\alpha_2\alpha_3 [ \psi_0^\parallel + 
\psi_1^\parallel(\alpha_1-\alpha_2) + 
\psi_2^\parallel (3\alpha_3-1) + \dots],\nonumber\\
\widetilde\Psi_{4;K^*}^\parallel(\underline{\alpha}) & = & 120
\alpha_1\alpha_2\alpha_3 [ \widetilde\psi_0^\parallel +
  \widetilde\psi_1^\parallel(\alpha_1-\alpha_2) + 
\widetilde\psi_2^\parallel (3\alpha_3-1) +
\dots].
\label{eq:phi}
\end{eqnarray}
G-parity implies that, for the $\rho$ and $\phi$ meson, $\psi_0^\parallel =
\psi_2^\parallel = \widetilde\psi_1^\parallel = 0$, 
whereas  for the $K^*$ meson $\psi_0^\parallel$, $\psi_2^\parallel$ and 
$\widetilde\psi_1^\parallel$
are ${\mathcal O}(m_s-m_q)$.

In turn, the DAs $\Phi_{4;K^*}$ and  $\widetilde \Phi_{4;K^*}$
correspond to the light-cone projection $\gamma_\perp G_{z \perp}$, 
which is a mixture of different quark-spin states with $s_q=+1/2, 
s_{\bar q}=-1/2$ and  $s_q=-1/2, s_{\bar q}=+1/2$, respectively. 
In both cases $s=+1$ for the gluon. We separate the different quark-spin
projections by introducing the auxiliary amplitudes 
${\Phi}^{\uparrow\downarrow}$ and  ${\Phi}^{\downarrow\uparrow}$, defined as
\begin{eqnarray}
\langle 0 | \bar q(z) g\widetilde{G}_{\mu\nu}(vz) \gamma_z
\gamma_\alpha\gamma_5\gamma_p s(-z) | K^*(P,\lambda)\rangle 
& = & f_{K^*}^\parallel m_{K^*}^3
(e^{(\lambda)}z) (p_\mu g^\perp_{\alpha\nu} - p_\nu
g^\perp_{\alpha\mu}) \Phi^{\uparrow\downarrow}(v,pz)\,,\nonumber\\
\langle 0 | \bar q(z) g\widetilde{G}_{\mu\nu}(vz) \gamma_p
\gamma_\alpha\gamma_5\gamma_z s(-z) | K^*(P,\lambda)\rangle 
& = & f_{K^*}^\parallel m_{K^*}^3
(e^{(\lambda)}z) (p_\mu g^\perp_{\alpha\nu} - p_\nu
g^\perp_{\alpha\mu}) \Phi^{\downarrow\uparrow}(v,pz)\,.\nonumber\\[-15pt]
\end{eqnarray}
The distributions  $\Phi_{4;K^*}^\parallel$ and  $\widetilde
\Phi_{4;K^*}^\parallel$ are then given by
\begin{eqnarray}
{\widetilde\Phi}_{4;K^*}(\underline{\alpha}) & = & \frac{1}{2}\Big[ {\Phi}^{
\uparrow\downarrow}(\underline{\alpha}) + 
{\Phi}^{\downarrow\uparrow}(\underline{\alpha})\Big],\quad
{\Phi}_{4;K^*}(\underline{\alpha}) = 
\frac{1}{2}\Big[ {\Phi}^{\uparrow\downarrow}(\underline{\alpha}) - 
{\Phi}^{\downarrow\uparrow}(\underline{\alpha})\Big].
\label{eq:original}
\end{eqnarray}
${\Phi}^{\uparrow\downarrow}$ and ${\Phi}^{\downarrow\uparrow}$
have a regular expansion in terms of conformal  polynomials:
\begin{eqnarray}
{\Phi}^{\uparrow\downarrow}(\underline{\alpha}) & = &
60\alpha_2\alpha_3^2 \left[ \phi_{0}^{\uparrow\downarrow} + 
\phi_{1}^{\uparrow\downarrow}(\alpha_3-3\alpha_1) +
\phi_{2}^{\uparrow\downarrow}\left(\alpha_3-\frac{3}{2}\alpha_2\right)\right],
\nonumber\\
{\Phi}^{\downarrow\uparrow}(\underline{\alpha}) & = &
60\alpha_1\alpha_3^2 \left[ \phi_{0}^{\downarrow\uparrow} +
\phi_{1}^{\downarrow\uparrow}(\alpha_3-3\alpha_2) +
\phi_{2}^{\downarrow\uparrow}\left(\alpha_3-\frac{3}{2}\alpha_1\right)\right].
\end{eqnarray}
For the $\rho$ and $\phi$ meson, G-parity implies
\begin{equation}
{\Phi}^{\uparrow\downarrow}_{4;\rho(\phi)}(\alpha_1,\alpha_2)=
{\Phi}^{\downarrow\uparrow}_{4;\rho(\phi)}(\alpha_2,\alpha_1)\,,
\end{equation}
so that $\phi_i^{\uparrow\downarrow}\equiv \phi_i^{\downarrow\uparrow}$.%
\footnote{This implies, in particular, that only one of the DAs
  $\Phi_{4;\rho(\phi)}$ and $\widetilde\Phi_{4;\rho(\phi)}$
is dynamically independent.}  For $K^*$, we write
\begin{equation}
\phi_i^{\uparrow\downarrow} = \phi^\parallel_i + \theta^\parallel_i,\qquad 
\phi_i^{\downarrow\uparrow} = \phi^\parallel_i - \theta^\parallel_i,
\end{equation}
where the $\theta_i^\parallel$ are the G-parity-violating
corrections. 
Using (\ref{eq:original}), we readily
derive the following expressions:
\begin{eqnarray}
\lefteqn{{\widetilde\Phi}_{4;K^*}^\parallel(\underline{\alpha}) = 
 30 \alpha_3^2\left\{ \phi^\parallel_{0}(1-\alpha_3)
     +\phi^\parallel_{1}\left[\alpha_3(1-\alpha_3)-
6\alpha_1\alpha_2\right]\right.}
\nonumber\\
& & \left.
            {} +\phi^\parallel_{2}\left[\alpha_3(1-\alpha_3)-
\frac{3}{2}(\alpha_1^2
                               +\alpha_2^2)\right]-
		    (\alpha_1-\alpha_2)\left[ \theta^\parallel_0 + \alpha_3
		      \theta^\parallel_1 + \frac{1}{2}\,(5 \alpha_3-3)
		      \theta^\parallel_2\right]\right\},
\nonumber\\
\lefteqn{ {\Phi}_{4;K^*}^\parallel(\underline{\alpha})  = 
 30 \alpha_3^2\left\{ \theta^\parallel_{0}(1-\alpha_3)
          +\theta^\parallel_{1}\left[\alpha_3(1-\alpha_3)-
6\alpha_1\alpha_2\right]
\right.}\nonumber\\
& &
   \left. {}  +\theta^\parallel_{2}\left[\alpha_3(1-\alpha_3)-
\frac{3}{2}(\alpha_1^2
                               +\alpha_2^2)\right]-
		    (\alpha_1-\alpha_2)\left[ \phi^\parallel_0 + \alpha_3
		      \phi^\parallel_1 + \frac{1}{2}\,(5 \alpha_3-3)
		    \phi^\parallel_2\right]\right\}.
\hspace*{20pt}\label{eq:psi}
\end{eqnarray}
In addition, we introduce one more three-particle DA 
$\Xi_{4;K^*}^\parallel(\underline{\alpha})$ \cite{BGG04}:
\begin{equation}
\label{Xi} 
\left\langle 0\left \vert \bar{q}(z)
\gamma_{\alpha} gD_{\mu}G_{\mu\nu}(vz)s(-z)
\right\vert K^*(P,\lambda)\right\rangle
= f_{K^*}^\parallel m_{K^*}^3 p_{\alpha}p_{\nu}
\,\frac{e^{(\lambda)}z}{pz}\,
\Xi_{4;K^*}^\parallel(v,pz)+\dots
\end{equation}
The Lorentz structure $p_{\alpha}p_{\nu}$ is the only one relevant
at twist 4. Because of the EOM
$D^{\alpha}G^A_{\alpha\beta} = -g\sum_q \bar q t^A \gamma_\beta q$,
where the summation goes over all light flavors, 
$\Xi_{4;K^*}^\parallel$
can be viewed as describing  either a quark-antiquark-gluon or a
specific four-quark Fock-state of the $K^*$, with the
quark-antiquark pair in a colour-octet state and at the same
space-time point. 
The conformal expansion of $\Xi_{4;K^*}^\parallel$ starts with the
conformal spin $j=4$ and reads
\begin{equation}
  \Xi_{4;K^*}^\parallel(\underline{\alpha}) = 
840 \alpha_1\alpha_2\alpha_3^3\Big[\xi^\parallel_0 +\ldots\Big],
\label{Xi1}
\end{equation}
where $\xi^\parallel_0$ is dimensionless. The dots stand for terms with  
higher conformal spin $j=5,6,\ldots$, which are beyond our accuracy.
$\Xi^\parallel_4$ 
was not considered in Ref.~\cite{BB98} because  $\xi^\parallel_0$ is
G-odd and the first G-even term only occurs at the next order in the conformal
expansion, for $j=5$. 

Eqs.~(\ref{eq:phi}),
(\ref{eq:psi}) and (\ref{Xi1}) represent the most general
pa\-ra\-metri\-za\-tion of the chiral-even
twist-4 DAs to  NLO in the conformal-spin
expansion and involve 
13 non-per\-tur\-ba\-tive parameters. Not all of them are independent,
though. In the following, we shall establish their mutual relations
and also  express all leading order (LO) and the G-even NLO 
expansion coefficients in terms of matrix
elements of local operators.

Except for $\Xi^\parallel_{4;K^*}$, 
the asymptotic three-particle DAs correspond to contributions of the 
lowest conformal spin 
$j = j_s+j_{\bar q}+ j_{g} = 3$. The parameters 
$\psi_0^\parallel$, $\widetilde\psi_0^\parallel$, $\phi_0^\parallel$
and $\theta_0^\parallel$
multiplying the asymptotic DAs can be
expressed in terms of local matrix elements as 
\begin{eqnarray}
\langle 0 | \bar q g \widetilde{G}_{\alpha\beta}\gamma_\mu\gamma_5 s |
K^*(P,\lambda)\rangle & = & 
f_K^\parallel m_{K^*} \zeta^\parallel_{3K^*} \left\{
e^{(\lambda)}_\alpha \left( P_\beta P_\mu - \frac{1}{3}\,m_{K^*}^2
g_{\beta\mu}\right) -
\left(\alpha\leftrightarrow\beta\right)\right\}\nonumber\\
& & {}+\frac{1}{3}\,f_K^\parallel m_{K^*}^3 \zeta^\parallel_{4K^*}
\left( e^{(\lambda)}_\alpha g_{\beta\mu} - e^{(\lambda)}_\beta
g_{\alpha\mu}\right),\label{3.12}\\
\langle 0 | \bar q g G_{\alpha\beta}i\gamma_\mu s |
K^*(P,\lambda)\rangle & = & 
f_K^\parallel m_{K^*} \kappa^\parallel_{3K^*} \left\{
e^{(\lambda)}_\alpha \left( P_\beta P_\mu - \frac{1}{3}\,m_{K^*}^2
g_{\beta\mu}\right) -
\left(\alpha\leftrightarrow\beta\right)\right\}\nonumber\\
& & {}+\frac{1}{3}\,f_K^\parallel m_{K^*}^3 \kappa^\parallel_{4K^*}
\left( e^{(\lambda)}_\alpha g_{\beta\mu} - e^{(\lambda)}_\beta
g_{\alpha\mu}\right).\label{4.13}
\end{eqnarray}
Here we adopt the generic notation that $\zeta$ are G-conserving and
$\kappa$ G-breaking parameters. $\zeta_3$, $\kappa_3$ are twist-3 and 
$\zeta_4$, $\kappa_4$ twist-4 parameters.

Taking the local limit of  Eqs.~(\ref{4.1}) and (\ref{4.2}), and
comparing with the above definitions, one obtains
\begin{equation}
\begin{array}[b]{l@{\quad}l}
\displaystyle\phi_0^\parallel = -\frac{1}{3}\,\zeta_{3K^*}^\parallel +
\frac{1}{3}\,\zeta_{4K^*}^\parallel\,, &
\displaystyle\theta_0^\parallel = -\frac{1}{3}\,\kappa_{3K^*}^\parallel +
\frac{1}{3}\,\kappa_{4K^*}^\parallel\,,\\[10pt]
\displaystyle\psi_0^\parallel = 
\phantom{-}\frac{2}{3}\,\kappa_{3K^*}^\parallel +
\frac{1}{3}\,\kappa_{4K^*}^\parallel\,, &
\displaystyle\widetilde\psi_0^\parallel = \phantom{-}
\frac{2}{3}\,\zeta_{3K^*}^\parallel +\frac{1}{3}\,\zeta_{4K^*}^\parallel\,.
\end{array}
\label{4.14}\end{equation}
The results for $\phi_0^\parallel$ and $\widetilde\psi_0^\parallel$ agree with
those obtained in Ref.~\cite{BB98}, the others are new. Note that the
``twist-4'' DAs receive contributions from both twist-3 and -4
operators. This is due to the fact that the standard counting of twist
in terms of ``good'' and ``bad'' components introduced in
Ref.~\cite{soper} differs from the definition of twist as ``dimension minus
spin'' of an operator. See also the discussion in Sec.~2.2 of
Ref.~\cite{BBKT}. 

What about the scale-dependence of these parameters?
The relevant local twist-4 operator mixes with operators of lower
twist for $m_s\neq 0$. Neglecting ${\mathcal O}(m_s^2)$ corrections, 
the mixing is given by
\begin{equation}\label{T4mix}
(\bar q \gamma_\alpha\gamma_5 g \widetilde{G}_{\mu\alpha}s)^{\mu^2} = (\bar q
  \gamma_\alpha\gamma_5 g \widetilde{G}_{\mu\alpha}s)^{\mu_0^2}
  \left(1-\frac{8}{9}\,\frac{\alpha_s}{\pi}\,\ln\,\frac{\mu^2}{\mu_0^2}
  \right) +
  \frac{1}{9}\,\frac{\alpha_s}{\pi}\,\ln\,\frac{\mu^2}{\mu_0^2}\, m_s
  \left[\partial_\mu (\bar q i s)\right]^{\mu_0^2}.
\end{equation}
The matrix element of the derivative operator on the right-hand side vanishes
for vector mesons, so that $\zeta_{4K^*}^\parallel$ renormalises
multiplicatively even for $m_s\neq 0$.
Resumming the logarithm, to LO accuracy, one has
\begin{equation}
\zeta_{4K^*}^\parallel(\mu^2) =
 L^{32/(9\beta_0)}\zeta_{4K^*}^\parallel(\mu_0^2)\,,
\end{equation}
with $L = \alpha_s(\mu^2)/\alpha_s(\mu_0^2)$.

The scale dependence of $\kappa_{4K^*}^\parallel$ can most easily be derived
by observing that this parameter is related to
$a_1^{\parallel}$ and quark masses by the QCD 
EOM \cite{BL04}:
\begin{equation}
\kappa_{4K^*}^\parallel = - \frac{3}{20}\,a_1^{\parallel} 
-\frac{f_{K^*}^\perp}{f_{K^*}^\parallel}\,\frac{m_s-m_q}{4m_{K^*}} +
\frac{m_s^2-m_q^2}{2m_{K^*}^2}\,.
\label{eq:BL04}
\end{equation}
Taking into account the known scale dependence
of $a_1^{\parallel}$, $f_{K^*}^\perp$ and $m_{s,q}$, one obtains
\begin{eqnarray}
\kappa_{4K^*}^\parallel(\mu^2) &=& \kappa_{4K^*}^\parallel(\mu_0^2) -
\frac{3}{20}\,\left(L^{32/(9\beta_0)} - 1\right)
a_1^{\parallel}(\mu_0^2) 
\nonumber\\
&&{}-\left(L^{16/(3\beta_0)} - 1\right)\frac{f_{K^*}^\perp(\mu_0^2)}{
f_{K^*}^\parallel}\,\frac{[m_s-m_q](\mu_0^2)}{4m_{K^*}} +
\left(L^{8/\beta_0} -
1\right)\frac{[m_s^2-m_q^2](\mu_0^2)}{2m_{K^*}^2}\,.
\hspace*{10pt}
\end{eqnarray}

To NLO in conformal spin, the discussion becomes more involved. 
As explained in Ref.~\cite{BF90}, for massless quarks the 
corresponding contributions can be expressed in terms of matrix 
elements of the three possible G-parity-even local quark-antiquark-gluon 
operators of twist 4. These three operators are not independent, however,
but related by the QCD EOM. One is left with 
only one new non-perturbative parameter, 
$\widetilde\omega_{4K^*}^\parallel$,%
\footnote{In the notation of Ref.~\cite{BB98}, 
$\widetilde\omega_{4K^*}^\parallel = \zeta_4 \omega_{4}^A$.}
which can be defined as     
\begin{eqnarray}
 \lefteqn{\langle0|
 \bar q \left [iD_\mu, g\tilde G_{\nu\xi}\right]\gamma_\xi \gamma_5 s
-\frac{4}{9}(i\partial_\mu)
\bar q g\tilde G_{\nu\xi}\gamma_\xi \gamma_5 s
|K^*(P,\lambda)\rangle +(\mu\leftrightarrow\nu) =}
\nonumber\\
&&{}\hspace*{6cm}=
2 f_{K^*}^\parallel m_{K^*}^3\, \widetilde\omega_{4K^*}^\parallel
 \left(e^{(\lambda)}_\mu P_\nu + e^{(\lambda)}_\nu P_\mu \right).
\mbox{\hspace{0cm}}\label{eq:w4A}
\end{eqnarray}
The scale dependence of $\widetilde\omega_{4K^*}^\parallel$, for
massless quarks, 
is given by
$$
\widetilde\omega_{4K^*}^\parallel(\mu^2) = L^{10/\beta_0}\, 
\widetilde\omega_{4K^*}^\parallel(\mu_0^2)\,.
$$
For massive quarks, the twist-4 operator mixes with operators of lower
twist. These lower-twist operators have the same Dirac structure as in
(\ref{T4mix}), but additional derivatives acting on the fields. This
means that, in terms of DAs, $\widetilde{\Phi}^\parallel_{4;K^*}$
mixes with $\psi^\parallel_{3;K^*}$, Eq.~(\ref{2.13}), which in turn
mixes with the three-particle twist-3 DA
$\widetilde{\Phi}^\parallel_{3;K^*}$, Eq.~(\ref{3.4}), 
which itself mixes with twist-2 DAs \cite{BJ07}. As the numerical
impact of the admixture of $m_s$ times lower-twist parameters 
is negligible for all cases investigated so far
(twist-3 and -4 pseudoscalar parameters \cite{BBL06} and twist-3
vector parameters \cite{BJ07}), we refrain from working out these relations.

For massive quarks, one has to distinguish between G-parity-conserving and
G-pa\-ri\-ty-brea\-king contributions. G-parity-conserving corrections do not 
involve new operators, and the difference to the massless case is mainly due 
to corrections proportional to the meson mass. 
This case is described in detail in Refs.~\cite{BBKT,BB98}.
Here we just quote the results obtained in
Ref.~\cite{BB98}:
\begin{eqnarray}
 \phi_{1}^\parallel&=& \phantom{-}\frac{1}{12}\,a_2^\parallel 
              -\frac{5}{12}\,\zeta_{3K^*}^\parallel
              +\frac{3}{16}\,\widetilde\omega_{3K^*}^\parallel
              + \frac{1}{8}\,\omega_{3K^*}^\parallel
              + \frac{7}{2}\,\widetilde\omega_{4K^*}^\parallel\,,
\nonumber\\
 \phi_{2}^\parallel&=& -\frac{1}{12}\,a_2^\parallel 
                   +\frac{3}{4}\,\zeta_{3K^*}^\parallel
                   +\frac{3}{16}\,\widetilde\omega_{3K^*}^\parallel
                   -\frac{1}{8}\,\omega_{3K^*}^\parallel
                   + 7 \widetilde\omega_{4K^*}^\parallel\,,
\nonumber\\
 \psi_{1}^\parallel&=& -\frac{1}{4}\,a_2^\parallel 
                   -\frac{7}{12}\,\zeta_{3K^*}^\parallel
                   +\frac{3}{8}\,\omega_{3K^*}^\parallel
                   -\frac{21}{4}\, \widetilde\omega_{4K^*}^\parallel\,,
\nonumber\\
 \widetilde\psi_{2}^\parallel&=& 
                              \phantom{-} \frac{2}{3}\,\zeta_{3K^*}^\parallel
                             -\frac{9}{16}\,\widetilde\omega_{3K^*}^\parallel
                           + \frac{21}{4}\,\widetilde\omega_{4K^*}^\parallel\,,
\label{eq:NLO}
\end{eqnarray}
where the terms in $a_2^\parallel$, $\zeta_{3K^*}^\parallel$,
$\omega_{3K^*}^\parallel$ and $\widetilde\omega_{3K^*}^\parallel$ are mass
corrections.  $\omega_{3K^*}^\parallel$ and
$\widetilde\omega_{3K^*}^\parallel$ are twist-3 parameters and have
been discussed in Ref.~\cite{BJ07}.

The G-parity-breaking contributions, on the other hand, involve a 
different set of local operators and in particular 
$$
\bar q \gamma_z D_{\xi}gG^{\xi z} s 
   = -g^2 \sum_{\psi=u,d,s} (\bar q \gamma_z t^a s)  
(\bar \psi \gamma_z t^a \psi)\,,
$$
which determines the normalization and the leading conformal spin contribution 
to the DA $\Xi_{4;K^*}^\parallel(\underline{\alpha})$ defined in
   Eq.~(\ref{Xi}). Hence, a complete treatment of 
G-parity-breaking corrections to twist-4 DAs requires also the inclusion of
  $\Xi_{4;K^*}^\parallel$.
It is beyond the scope of this paper to work out the corresponding
relations between the matrix elements of local operators and expansion 
coefficients. Instead, we
adopt a different approach and estimate G-parity-breaking corrections
of conformal spin $j=4$ using the renormalon model of Ref.~\cite{BGG04}. 
The general idea of this technique is to estimate matrix elements of 
``genuine'' twist-4 operators by the quadratically divergent
contributions that appear when the matrix elements are defined using a
hard UV cut-off, see Ref.~\cite{BGG04} for details and further
references. In this way, three-particle twist-4 DAs
can be expressed in terms of the leading-twist DA $\phi_{2;K^*}^\parallel$:
\begin{eqnarray}
\label{RMthree}
\Phi_{4;K^*}^{\parallel,\rm R}(\underline{\alpha})&=&
-\frac{1}{6}\,\zeta_{4K^*}^\parallel\left[
\frac{\phi_{2;K^*}^\parallel(\alpha_1)}{1-\alpha_1}-
\frac{\phi_{2;K^*}^\parallel(\bar\alpha_2)}{1-\alpha_2}
\right],
\nonumber \\
\widetilde\Phi_{4;K^*}^{\parallel,\rm R}(\underline{\alpha})&=&
\phantom{-}\frac{1}{6}\,\zeta_{4K^*}^\parallel\left[
\frac{\phi_{2;K^*}^\parallel(\alpha_1)}{1-\alpha_1}+
\frac{\phi_{2;K^*}^\parallel(\bar\alpha_2)}{1-\alpha_2}
\right],
\nonumber \\
\Psi_{4;K^*}^{\parallel,\rm R}(\underline{\alpha})&=
&\phantom{-}\frac{1}{3}\,\zeta_{4K^*}^\parallel\left[
\frac{\alpha_2\phi_{2;K^*}^\parallel(\alpha_1)}{(1-\alpha_1)^2}
-\frac{\alpha_1\phi_{2;K^*}^\parallel(\bar\alpha_2)}{(1-\alpha_2)^2}
\right],
\nonumber\\
\widetilde\Psi_{4;K^*}^{\parallel,\rm R}(\underline{\alpha})&=
&\phantom{-}\frac{1}{3}\,\zeta_{4K^*}^\parallel\left[
\frac{\alpha_2\phi_{2;K^*}^\parallel(\alpha_1)}{(1-\alpha_1)^2}
+\frac{\alpha_1\phi_{2;K^*}^\parallel(\bar\alpha_2)}{(1-\alpha_2)^2}
\right],
\nonumber\\
\Xi_{4;K^*}^{\parallel,\rm R}(\underline{\alpha})&=&
-\frac{2}{3}\,\zeta_{4K^*}^\parallel
\left[\frac{\alpha_2\,\phi_{2;K^*}^\parallel(\alpha_1)}{1-\alpha_1}
-\frac{\alpha_1\,\phi_{2;K^*}^\parallel(\bar\alpha_2)}{1-\alpha_2}\right],
\label{3.21}
\end{eqnarray}
where, in contrast to Ref.~\cite{BGG04}, we do not assume that
$\phi_{2;K^*}^\parallel(u)$ 
is symmetric under the exchange $u\leftrightarrow \ub$.

The expressions in (\ref{RMthree}) do not rely on conformal expansion
and contain the contributions of all conformal partial waves.
Projecting onto the contributions with the lowest spin $j=3,4$, we obtain
\begin{equation}
\renewcommand{\arraystretch}{2.2}
\begin{array}[b]{l@{\quad}l@{\quad}l}
\ds\psi_0^{\parallel,\rm R} = 0\,, & \ds\psi_1^{\parallel,\rm R} = 
                      \phantom{-}\frac{7}{12}\zeta^\parallel_{4K^*}\,, &
\ds\psi_2^{\parallel,\rm R} =  
-\frac{7}{20}a_1^{\parallel}\zeta^\parallel_{4K^*}, \\
\ds\widetilde\psi_0^{\parallel,\rm R} = \frac{1}{3}\zeta^\parallel_{4K^*}, &
\ds\widetilde\psi_1^{\parallel,\rm R} = 
\phantom{-}\frac{7}{4}a_1^{\parallel}
                            \zeta^\parallel_{4K^*}, &
\ds\widetilde\psi_2^{\parallel,\rm R} =  -\frac{7}{12}\zeta^\parallel_{4K^*},\\
\ds\phi_0^{\parallel,\rm R} = \frac{1}{3}\zeta^\parallel_{4K^*}, & 
\ds\phi_1^{\parallel,\rm R} =-\frac{7}{18}\zeta^\parallel_{4K^*}, & 
\ds\phi_2^{\parallel,\rm R} =  -\frac{7}{9}\zeta^\parallel_{4K^*},\\
\ds\theta_0^{\parallel,\rm R} = 0, & 
\ds\theta_1^{\parallel,\rm R} = 
-\frac{7}{10}a_1^{\parallel}\zeta^\parallel_{4K^*},&
\ds\theta_2^{\parallel,\rm R} =  \phantom{-}\frac{7}{5}a_1^{\parallel}
                     \zeta^\parallel_{4K^*}\,,\\
\ds\xi^{\parallel,\rm R}_0 = 
\frac{1}{5}a_1^{\parallel} \zeta^\parallel_{4K^*}\,.
\end{array}
\label{thetas}
\renewcommand{\arraystretch}{1}
\end{equation}
These results have to be compared with those in Eqs.~(\ref{4.14}), 
(\ref{eq:NLO}) in
the limit that all contributions from twist-2 and twist-3 parameters
are set to 0.
It follows that in the renormalon model
\begin{equation}\label{3.23}
   \widetilde\omega_{4K^*}^\parallel = -\frac{1}{9}\,,
\end{equation}
which has the same sign as, but is larger than the result from a QCD
sum rule calculation, see Sec.~\ref{sec:5}.
Also note that in the renormalon model $\psi_0^\parallel = 
\theta_0^\parallel = 0$, in
contrast to (\ref{4.14}). This is due to the
fact that the contribution in
$\kappa_{4K^*}^\parallel$ in (\ref{4.14}) is obtained as the matrix
element of the operator (\ref{4.13})
which vanishes by the EOM (up to a total derivative), see 
(\ref{eq:BL04}). Therefore, against appearances, this contribution 
has to be
interpreted as ``kinematic'' mass correction induced by the
non-vanishing $K^*$-meson mass rather than a genuine twist-4 effect.
The complete expressions for the G-odd parameters
$\widetilde\psi_1^\parallel$, $\psi_2^\parallel$, $\theta_{1,2}^\parallel$ and
$\xi^\parallel_0$ will contain mass corrections in terms of
lower-twist parameters, whose determination is, as said before,  
beyond the scope of this paper.

We are now in a position to derive expressions for the chiral-even two-particle
DAs of twist 4, $\phi_{4;K^*}^\parallel$ and
$\psi_{4;K^*}^\parallel$,
which are defined in Eq.~(\ref{2.9}). From the
operator relations collected in App.~\ref{app:A} we obtain:
\begin{eqnarray}
\psi_{4;K^*}^\parallel(u) & = & \phi_{2;K^*}^\parallel(u) - 2
\frac{d}{du} \int_0^u d\alpha_1 \int_0^{\bar u} d\alpha_2
\,\frac{1}{\alpha_3} \left[2 \Phi_{4;K^*}^\parallel(\underline{\alpha}) +
\Psi_{4;K^*}^\parallel(\underline{\alpha}) \right]\nonumber\\
&&{}-\frac{m_s-m_q}{m_{K^*}}\,\frac{f_{K^*}^\perp}{f_{K^*}^\parallel}\, 
\frac{d}{du}\, \psi_{3;K^*}^\parallel\,,\\
\frac{d}{du}\,\phi_{4;K^*}^\parallel & = & 2\xi
\left(\psi_{4;K^*}^\parallel(u) -
\phi_{2;K^*}^\parallel(u)\right)
-4 \int_0^u dv\left[5\phi_{2;K^*}^\parallel(v) - 8
\phi_{3;K^*}^\perp(v) + 3 \psi_{4;K^*}^\parallel(v)\right] \nonumber\\
&&{}- 4 \frac{d}{du}\int_0^u d\alpha_1 \int_0^{\bar u} d\alpha_2
\,\frac{1}{\alpha_3^2}\,(\alpha_1 - \alpha_2 - \xi) \left[2
\Phi_{4;K^*}^\parallel(\underline{\alpha}) +
\Psi_{4;K^*}^\parallel(\underline{\alpha}) \right]\nonumber\\
&&{}+
\frac{m_s+m_q}{m_{K^*}}\,\frac{f_{K^*}^\perp}{f_{K^*}^\parallel}\,
\frac{d}{du}\, 2 \psi_{3;K^*}^\parallel\,.\label{4.26}
\end{eqnarray}
The latter equation has to be integrated with the boundary conditions
$\phi_{4;K^*}^\parallel(0) = 0 = \phi_{4;K^*}^\parallel(1)$ which
implies the relation (\ref{eq:BL04}) between $a_1^{\parallel}$ and
$\kappa_{4K^*}^\parallel$. The boundary condition arises from the
conversion of the matrix element of Eq.~(\ref{eq:oprel1}) for the
$K^*$ into a relation for $\phi_{4;K^*}^\parallel$. This derivation of
(\ref{eq:BL04}) is equivalent to that given in Ref.~\cite{BL04}.

$\psi_{4;K^*}^\parallel$ corresponds to the projection
$s=-\frac{1}{2}$ for both quark and antiquark and hence, in the
absence of quark-mass corrections in $m_s\pm m_q$, has an
expansion in terms of $C^{1/2}_n(\xi)$.
It can be split into contributions from
genuine twist-4 parameters which are defined in terms of local
twist-4 operators and kinematic Wandzura-Wilczek-type and 
mass corrections:
\begin{equation}
\psi^\parallel_{4;K^*}(u) = \psi^{\parallel,\rm T4}_{4;K^*}(u)
+ \psi^{\parallel,\rm WW}_{4;K^*}(u)\,.\label{4.27}
\end{equation}
$\psi^{\parallel,\rm WW}_{4;K^*}$ contains  corrections
explicitly proportional to $m_s\pm m_q$, of which we only keep
the leading terms in $(m_s\pm m_q)^1$, but neglect higher
powers.\footnote{The terms in $(m_s\pm m_q)^2$ actually diverge for
  $u\to 0,1$, which is however largely irrelevant for phenomenological
  applications.} We then find
\begin{eqnarray}
\psi_{4;K^*}^{\parallel,\rm T4}(u) & = &-
\frac{20}{3}\,\zeta_{4K^*}^\parallel C_2^{1/2}(\xi) + 
\left(10\theta_1^\parallel - 5 \theta_2^\parallel \right) C_3^{1/2}(\xi)\,,
\nonumber\\
\psi_{4;K^*}^{\parallel,\rm WW}(u) & = & 1 
+ \left( 12\kappa_{4K^*}^\parallel + \frac{9}{5}\,
   a_1^{\parallel}\right) C_1^{1/2}(\xi) 
+ \left(-1-\frac{2}{7}\,a_2^{\parallel}+ \frac{40}{3}\,
   \zeta_{3K^*}^\parallel\right) 
   C_2^{1/2}(\xi)\nonumber\\
&&{}
+ \left(-\frac{9}{5}\,a_1^{\parallel} -
   \frac{20}{3}\,\kappa_{3K^*}^\parallel -
   \frac{16}{3}\,\kappa_{4K^*}^\parallel\right) C_3^{1/2}(\xi)\nonumber\\
&&{}
+ \left( -\frac{27}{28}\,a_2^{\parallel} +
   \frac{5}{4}\,\zeta_{3K^*}^\parallel -
   \frac{15}{8}\,\omega_{3K^*}^\parallel -
   \frac{15}{16}\,\widetilde\omega_{3K^*}^\parallel \right)
   C_4^{1/2}(\xi)\nonumber\\
&&{} +
   6\,\frac{m_s-m_q}{m_{K^*}}\,\frac{f_{K^*}^\perp}{f_{K^*}^\parallel}
   \left\{ \xi + a_1^{\perp}\,\frac{1}{2}\,(3\xi^2-1) 
    + a_2^\perp\,\frac{1}{2}\,\xi (5\xi^2-3) + 
    \frac{5}{2}\, \kappa_{3K^*}^\perp (3\xi^2-1)\right. \nonumber\\
&&\left. \hspace*{3.5cm}  +\frac{5}{6}\, \omega_{3K^*}^\perp \xi (5\xi^2-3)
    - \frac{1}{16}\,\lambda_{3K^*}^\perp (35\xi^4-30\xi^2+3)\right\}.
\label{3.27}
\end{eqnarray}
$\phi_{4;K^*}^\parallel$, on the other hand, has no regular conformal
expansion and contains logarithms even in the chiral limit. We solve
the integral relation (\ref{4.26}) by substituting all DAs on the
right-hand side by their conformal expansion to NLO, implementing the
boundary condition
$\phi_{4;K^*}^\parallel(0)=0=\phi_{4;K^*}^\parallel(1)$ by eliminating
$\kappa_{4K^*}^\parallel$ in favour of $a_1^\parallel$ according to
(\ref{eq:BL04}), and dropping terms in $(m_s\pm m_q)^n$ with $n>1$. 
Like with $\psi_{4;K^*}^\parallel$, we distinguish between genuine
twist-4 and mass corrections and write
\begin{equation}
\phi^\parallel_{4;K^*}(u) = \phi^{\parallel,\rm T4}_{4;K^*}(u)
+ \phi^{\parallel,\rm WW}_{4;K^*}(u)\,.\label{4.29}
\end{equation}
We then find:
\begin{eqnarray}
\lefteqn{\phi_{4;K^*}^{\parallel,\rm T4}(u) = 30 u^2 \ub^2 \left\{
\frac{20}{9}\,\zeta_{4K^*}^\parallel + \left( - \frac{8}{15}\,
	   \theta_1^\parallel + \frac{2}{3}\,\theta_2^\parallel\right)
	     C_1^{5/2}(\xi)\right\}}\hspace*{1.3cm}\nonumber\\
&&{}
 -84 \widetilde\omega_{4K^*}^\parallel
\left\{\frac{1}{8}\,u \ub\,(21-13\xi^2)
+ u^3 (10-15 u + 6 u^2)\ln u
+ \ub^3 (10-15 \ub + 6 \ub^2)\ln \ub\right\} \nonumber\\
&&{}+ 80\psi_2^\parallel  \left\{ u^3 (2-u) \ln u - \ub^3
(2-\ub) \ln \ub - \frac{1}{8}\,(3\xi^2-11)\right\},\nonumber\\
\lefteqn{\phi_{4;K^*}^{\parallel,\rm WW}(u) =  30 u^2 \ub^2 \left\{ 
\frac{4}{5}\left( 1 + \frac{1}{21}\,a_2^\parallel +
           \frac{10}{9}\,\zeta_{3K^*}^\parallel\right)
            +\left( \frac{17}{50}\,a_1^\parallel 
	   + \frac{2}{5}\,\widetilde\lambda_{3K^*}^\parallel -
	   \frac{1}{5}\,\lambda_{3K^*}^\parallel\right)
	     C_1^{5/2}(\xi)\right.}\hspace*{1.4cm}\nonumber\\
&&{}\left.\hspace*{1.8cm}\ + \frac{1}{10}\,\left( \frac{9}{7}\, a_2^\parallel +
	     \frac{1}{9}\, \zeta_{3K^*}^\parallel +
	     \frac{7}{6}\,\omega_{3K^*}^\parallel -
	     \frac{3}{4}\,\widetilde\omega_{3K^*}^\parallel \right)
	     C_2^{5/2}(\xi) \right\}\nonumber\\
&&{}+ 2 \left\{ - 2 a_2^\parallel - \frac{14}{3}\,\zeta_{3K^*}^\parallel + 3
\omega_{3K^*}^\parallel\right\}
\left\{\frac{1}{8}\,u \ub\,(21-13\xi^2)\right.\nonumber\\
&&{}\left.\hspace*{4.5cm} + u^3 (10-15 u + 6 u^2)\ln u
+ \ub^3 (10-15 \ub + 6 \ub^2)\ln \ub\right\} \nonumber\\
&&{}+ 4 \left\{ a_1^\parallel - \frac{40}{3}\,\kappa_{3K^*}^\parallel
 \right\} \left\{ u^3 (2-u) \ln u - \ub^3
(2-\ub) \ln \ub- \frac{1}{8}\,(3\xi^2-11)\right\}\nonumber\\
&&{} + \frac{m_s+m_q}{m_{K^*}}\,\frac{f_{K^*}^\perp}{f_{K^*}^\parallel}\,
6 u \ub \left\{ 2 (3 + 16 a_2^\perp) + \frac{10}{3}( \kappa_{3K^*}^\perp
- a_1^\perp) C_1^{3/2}(\xi)\right.\nonumber\\
&&\left.\hspace*{4cm} + \left( \frac{5}{9}\,\omega_{3K^*}^\perp -
a_2^\perp \right) C_2^{3/2}(\xi) - \frac{1}{10}\,\lambda_{3K^*}^\perp
C_3^{3/2}(\xi) \right\}\nonumber\\
&&{} + 24\,\frac{m_s+m_q}{m_{K^*}}\,\frac{f_{K^*}^\perp}{f_{K^*}^\parallel}\,
\left\{ (1-3 a_1^\perp + 6 a_2^\perp) u^2 \ln u+
(1+3 a_1^\perp + 6 a_2^\perp) \ub^2 \ln \ub \right\}
\nonumber\\
&&
{}+\frac{m_s-m_q}{m_{K^*}}\,\frac{f_{K^*}^\perp}{f_{K^*}^\parallel}\,
6 u \ub \left\{ -\left( 10 \kappa_{3K^*}^\perp +
\frac{82}{5}\,a_1^\perp \right) C_1^{3/2}(\xi)\right.
\nonumber\\
&& \hspace*{2cm}{}+ 20 \left( \frac{10}{189} + \frac{1}{3}\, a_2^\perp -
\frac{1}{21}\, \omega_{3K^*}^\perp \right) C_2^{3/2}(\xi) + 
\left( \frac{7}{54}\,\lambda_{3K^*}^\perp + \frac{2}{5}\,a_1^\perp
\right) C_3^{3/2}(\xi)
\nonumber\\
&& \hspace*{2cm}\left. + \left( \frac{1}{5}\,a_2^\perp - \frac{2}{315}-
\frac{1}{21}\, \omega_{3K^*}^\perp\right) C_4^{3/2}(\xi) 
+ \frac{2}{135}\,\lambda_{3K^*}^\perp C_5^{3/2}(\xi)\right\}
\nonumber\\
&&{}+ \frac{m_s-m_q}{m_{K^*}}\,\frac{f_{K^*}^\perp}{f_{K^*}^\parallel}\,
\left\{  (5u^2-23 - 54 a_1^\perp - 108 a_2^\perp) \ln\ub\right.
\nonumber\\
&&\left.\hspace*{3cm}
- (5\ub^2-23 + 54 a_1^\perp - 108 a_2^\perp) \ln u\right\}.
\label{3.29}
\end{eqnarray}
Recall that $\ub = 1-u$ and $\xi=2u-1$. 
The DAs for $\bar K^*=(q\bar s)$ mesons are obtained by replacing $u$ by $1-u$.
Both (\ref{4.27}) and (\ref{4.29}) agree, for the $\rho$ meson, with
the results obtained in Ref.~\cite{BB98}.

The above expressions provide a self-consistent model of the twist-4 DAs
which includes the first three terms of the conformal expansion.
As mentioned above, one shortcoming of the model is
that G-parity-breaking terms in $\psi_2^\parallel$ and 
$\theta_{1,2}^\parallel$ are not known exactly, but only in the
renormalon model, Eq.~(\ref{thetas}), which misses meson mass
corrections.
Numerically, the neglected parameters may be of the same size as the
included ones.

An estimate of the
size of higher orders in the conformal expansion
can be obtained using the full renormalon model for
$\psi^{\parallel,\rm T4}_{4;K^*}$ and $\phi^{\parallel,\rm T4}_{4;K^*}$.
In this case, one has \cite{BGG04}
\begin{eqnarray}
\psi^{\parallel,
\rm T4,R}_{4;K^*}(u) & = & -\frac{2}{3}\,\zeta_{4K^*}^\parallel \,
\frac{d}{du} \left\{ u \int_u^1
dv\,\frac{\phi_{2;K^*}^\parallel(v)}{v^2} - \bar u \int_0^u
dv\,\frac{\phi_{2;K^*}^\parallel(v)}{\bar v^2} \right\},\nonumber\\
\phi^{\parallel,\rm T4,R}_{4;K^*}(u) & = & \frac{8}{3}\,\zeta_{4K^*}^\parallel
\left\{
\int_0^u dv\,\frac{\phi_{2;K^*}^\parallel(v)}{\bar v^2} \left[ \bar u
  + \bar u^2 + (u-v)\ln\,\frac{u-v}{\bar v}\right]\right.\nonumber\\
&&{}
\left. \hspace*{1.2cm} + 
\int_u^1 dv\,\frac{\phi_{2;K^*}^\parallel(v)}{v^2} \left[ u
  + u^2 + (v-u)\ln\,\frac{v-u}{v}\right]\right\}.
\end{eqnarray}
As explained
in Ref.~\cite{BGG04}, the renormalon model does not take into account 
the damping of higher conformal-spin contributions by the increasing
anomalous dimensions and, therefore, only provides an upper bound for 
their contribution. The most important effect of these corrections is
to significantly enhance the end-point behaviour of 
higher-twist DAs in some cases, which can
be important in phenomenological applications, for instance
$D\to(\pi,K)$ form factors \cite{Ddecays}.

\section{Chiral-Odd Distributions}\label{sec:4}
\setcounter{equation}{0}

The analysis of chiral-odd DAs proceeds along similar lines and,
except for the inclusion of G-odd contributions, replicates that
performed in Ref.~\cite{BB98}. We first
define the relevant three-particle DAs:
\begin{eqnarray}
\lefteqn{\langle 0|\bar q(z) \sigma_{\alpha\beta}
         gG_{\mu\nu}(vz)
         s(-z)|K^*(P,\lambda)\rangle}
\nonumber\\
&=& f_{K^*}^\perp m_{K^*}^2 \frac{e^{(\lambda)} z }{2 (p  z)}
    [ p_\alpha p_\mu g^\perp_{\beta\nu}
     -p_\beta p_\mu g^\perp_{\alpha\nu}
     -p_\alpha p_\nu g^\perp_{\beta\mu}
     +p_\beta p_\nu g^\perp_{\alpha\mu} ]
      \Phi^\perp_{3;K^*}(v,pz)
\nonumber\\
&&{}+ f_{K^*}^\perp m_{K^*}^2
    [ p_\alpha e^{(\lambda)}_{\perp\mu}g^\perp_{\beta\nu}
     -p_\beta e^{(\lambda)}_{\perp\mu}g^\perp_{\alpha\nu}
     -p_\alpha e^{(\lambda)}_{\perp\nu}g^\perp_{\beta\mu}
     +p_\beta e^{(\lambda)}_{\perp\nu}g^\perp_{\alpha\mu} ]
      \Phi^{\perp(1)}_{4;K^*}(v,pz)
\nonumber\\
&&{}+ f_{K^*}^\perp m_{K^*}^2
    [ p_\mu e^{(\lambda)}_{\perp\alpha}g^\perp_{\beta\nu}
     -p_\mu e^{(\lambda)}_{\perp\beta}g^\perp_{\alpha\nu}
     -p_\nu e^{(\lambda)}_{\perp\alpha}g^\perp_{\beta\mu}
     +p_\nu e^{(\lambda)}_{\perp\beta}g^\perp_{\alpha\mu} ]
      \Phi^{\perp(2)}_{4;K^*}(v,pz)
\nonumber\\
&&{}+ \frac{f_{K^*}^\perp m_{K^*}^2}{pz}
    [ p_\alpha p_\mu e^{(\lambda)}_{\perp\beta}z_\nu
     -p_\beta p_\mu e^{(\lambda)}_{\perp\alpha}z_\nu
     -p_\alpha p_\nu e^{(\lambda)}_{\perp\beta}z_\mu
     +p_\beta p_\nu e^{(\lambda)}_{\perp\alpha}z_\mu ]
     \Phi^{\perp(3)}_{4;K^*}(v,pz)
\nonumber\\
&&{}+ \frac{f_{K^*}^\perp m_{K^*}^2}{pz}
    [ p_\alpha p_\mu e^{(\lambda)}_{\perp\nu}z_\beta
     -p_\beta p_\mu e^{(\lambda)}_{\perp\nu}z_\alpha
     -p_\alpha p_\nu e^{(\lambda)}_{\perp\mu}z_\beta
     +p_\beta p_\nu e^{(\lambda)}_{\perp\mu}z_\alpha ]
      \Phi^{\perp(4)}_{4;K^*}(v,pz)+\dots\,,
\hspace*{0.6cm}
\label{eq:T3}
\end{eqnarray}
\begin{eqnarray}
\langle 0|\bar q(z)
         gG_{\mu\nu}(vz)
         s(-z)|K^*(P,\lambda)\rangle
&=&i f_{K^*}^\perp m_{K^*}^2
 [e^{(\lambda)}_{\perp\mu}p_\nu-e^{(\lambda)}_{\perp\nu}p_\mu] 
\Psi^\perp_{4;K^*}(v,pz)+\dots\,,
\nonumber\\
\langle 0|\bar q(z)
         ig\widetilde G_{\mu\nu}(vz)\gamma_5
         s(-z)|K^*(P,\lambda)\rangle
& =& i f_{K^*}^\perp m_{K^*}^2
 [e^{(\lambda)}_{\perp\mu}p_\nu-e^{(\lambda)}_{\perp\nu}p_\mu]
  \widetilde\Psi^\perp_{4;K^*}(v,pz)+\dots\,,
\nonumber\\
\langle 0|\bar q(z)\sigma_{\mu\nu}
         gD_\alpha G_{\alpha\beta}(vz)s(-z)|K^*(P,\lambda)\rangle
& =& i f_{K^*}^\perp m_{K^*}^2 \left[ e^{(\lambda)}_{\perp\mu} p_\nu -
         e^{(\lambda)}_{\perp\nu} p_\mu\right] p_\beta\,
         \Xi_{4;K^*}^\perp(v,pz)+\dots
\nonumber\\[-15pt]
\label{eq:2.21}
\end{eqnarray}
The twist-3 DA $\Phi^\perp_{3;K^*}$ was already mentioned in
Sec.~\ref{sec:2}; the twist-4 DAs are related to those defined in
Ref.~\cite{BB98} by $\Phi^{\perp(i)}_{4;K^*}= T_i^{(4)}$,
$\Psi^\perp_{4;K^*} = S$ and $\widetilde\Psi^\perp_{4;K^*} =
\widetilde S$; $\Xi_{4;K^*}^\perp$ was first introduced in
Ref.~\cite{BGG04}. As usual, the dots denote terms of higher twist.

As the matrix element in (\ref{eq:T3}) is G-odd, the
DAs $\Phi^{\perp(i)}_{4;K^*}$ are, in the limit of equal mass quarks, 
antisymmetric under the
exchange of $\alpha_1$ and $\alpha_2$, whereas $\Psi^\perp_{4;K^*}$
and $\widetilde\Psi^\perp_{4;K^*}$ are symmetric. 
In order to resolve the conformal structure of $\Phi^{\perp(i)}_{4;K^*}$,
it is useful to exploit the fact that
$\sigma_{\mu\nu}\gamma_5$ is not independent of $\sigma_{\mu\nu}$, and
to define the ``dual'' matrix element 
\begin{equation}\label{4.34}
\langle 0 | \bar q(z) i \sigma_{\alpha\beta} \gamma_5
g\wt{G}_{\mu\nu}(vz) s(-z) | K^*\rangle = \mbox{r.h.s.\ of
  (\ref{eq:T3}) with
  $\Phi^{\perp(i)}_{4;K^*}\to\widetilde\Phi^{\perp(i)}_{4;K^*}$}.
\end{equation}
One easily finds
\begin{eqnarray}
\Phi^\perp_{3;K^*} & = & \phantom{-}\wt\Phi^\perp_{3;K^*}\,,\quad
\wt\Phi^{\perp(1)}_{4;K^*}\ =\ -\Phi^{\perp(3)}_{4;K^*}\,,\quad
\wt\Phi^{\perp(2)}_{4;K^*}\ =\ -\Phi^{\perp(4)}_{4;K^*}\,,
\nonumber\\
\wt\Phi^{\perp(3)}_{4;K^*} & = & -\Phi^{\perp(1)}_{4;K^*}\,,\quad
\wt\Phi^{\perp(4)}_{4;K^*}\ =\ -\Phi^{\perp(2)}_{4;K^*}\,.
\end{eqnarray}
$\Phi^{\perp(1)}_{4;K^*}$ and $\wt\Phi^{\perp(1)}_{4;K^*}$
correspond to the Lorentz spin projection $s=+1/2$ for both quark
fields and  $s=0$ for the gluon. Hence the conformal
expansion reads, to NLO:
\begin{eqnarray}
\Phi_{4;K^*}^{\perp(1)}(\underline{\alpha})
 & = & 
120\alpha_1\alpha_2\alpha_3 [ \phi_0^\perp + 
\phi_1^\perp(\alpha_1-\alpha_2) + 
\phi_2^\perp (3\alpha_3-1)]\,,
\nonumber\\
-\Phi^{\perp(3)}_{4;K^*}(\underline{\alpha})\ =\ 
\widetilde\Phi_{4;K^*}^{\perp(1)}(\underline{\alpha}) 
& = & 
120 \alpha_1\alpha_2\alpha_3 [ \widetilde\phi_0^\perp +
  \widetilde\phi_1^\perp(\alpha_1-\alpha_2) + 
\widetilde\phi_2^\perp (3\alpha_3-1)]\,.
\end{eqnarray}
Here $\phi_{0,2}^\perp$ are G-violating and $\phi_1^\perp$ is
G-conserving; the same applies to $\widetilde\phi_i^\perp$.

The conformal expansion of $\Xi^\perp_{4;K^*}$ starts at $j=4$ and
reads, to NLO accuracy:
\begin{equation}
\Xi^\perp_{4;K^*}(\underline{\alpha}) = 840 \alpha_1\alpha_2\alpha_3^2
\xi^\perp_0\,.
\end{equation}

The remaining DAs do not correspond to fixed values of the
Lorentz-spin projection $s$. Like in the chiral-even case, we introduce
auxiliary amplitudes with a regular conformal expansion:
\begin{eqnarray}
\langle 0 | \bar q(z) \gamma_z \gamma_p gG_{\mu\nu}(vz)
s(-z)|K^*(P,\lambda)\rangle & = & if_{K^*}^\perp m_{K^*}^2 
(pz)[e^{(\lambda)}_{\perp\mu} p_\nu -
e^{(\lambda)}_{\perp\nu} p_\mu] \Psi^{\ud}(v,pz),\nonumber\\
\langle 0 | \bar q(z) \gamma_z \gamma_p i\gamma_5 g
\widetilde{G}_{\mu\nu}(vz)
s(-z)|{K^*}(P,\lambda)\rangle & = & if_{K^*}^\perp m_{K^*}^2 (pz) 
[e^{(\lambda)}_{\perp\mu} p_\nu -e^{(\lambda)}_{\perp\nu} p_\mu] 
\wt{\Psi}^{\ud}(v,pz),\hspace*{0.5cm}
\end{eqnarray}
and, similarly, two more distributions $\Psi^{\du}$ and $\wt{\Psi}^{\du}$ 
by replacing $\gamma_z\gamma_p\rightarrow \gamma_p\gamma_z$. 
The DAs in (\ref{eq:T3}) and
  (\ref{eq:2.21}) are then given by:
\begin{eqnarray}
{\Psi_{4;K^*}^\perp}(\underline{\alpha}) & = & \frac{1}{2}\, ({
  \Psi}^{\ud}(\underline{\alpha})  +
{ \Psi}^{\du}(\underline{\alpha}))\,,\quad
\phantom{-}\wt \Psi_{4;K^*}^\perp(\underline{\alpha}) 
\ = \ \frac{1}{2}\, (\wt{
  \Psi}^{\ud}(\underline{\alpha})  +
\wt{ \Psi}^{\du}(\underline{\alpha}))\,,\nonumber\\
\Phi_{4;K^*}^{\perp(4)}(\underline{\alpha}) & = & \frac{1}{2}\, ({
  \Psi}^{\ud}(\underline{\alpha})  -
{ \Psi}^{\du}(\underline{\alpha}))\,,\quad
-\Phi_{4;K^*}^{\perp(2)}(\underline{\alpha})\ =\ 
\wt\Phi_{4;K^*}^{\perp(4)}(\underline{\alpha}) 
\ = \ \frac{1}{2}\, (\wt{ \Psi}^{\ud}(\underline{\alpha})  -
\wt{ \Psi}^{\du}(\underline{\alpha}))\,.\nonumber\\[-12pt]\label{eq:obvious}
\end{eqnarray}
${\Psi}^{\uparrow\downarrow}$ and ${\Psi}^{\downarrow\uparrow}$
have a regular expansion in terms of conformal  polynomials, to wit:
\begin{eqnarray}
{\Psi}^{\uparrow\downarrow}(\underline{\alpha}) & = &
60\alpha_2\alpha_3^2 \left[ \psi_{0}^{\uparrow\downarrow} + 
\psi_{1}^{\uparrow\downarrow}(\alpha_3-3\alpha_1) +
\psi_{2}^{\uparrow\downarrow}\left(\alpha_3-\frac{3}{2}\alpha_2\right)\right],
\nonumber\\
{\Psi}^{\downarrow\uparrow}(\underline{\alpha}) & = &
60\alpha_1\alpha_3^2 \left[ \psi_{0}^{\downarrow\uparrow} +
\psi_{1}^{\downarrow\uparrow}(\alpha_3-3\alpha_2) +
\psi_{2}^{\downarrow\uparrow}\left(\alpha_3-\frac{3}{2}\alpha_1\right)\right].
\end{eqnarray}
Again G-parity ensures that for the $\rho$ and $\phi$ mesons 
$\psi_i^{\uparrow\downarrow}\equiv \psi_i^{\downarrow\uparrow}$.
For $K^*$, we write
\begin{equation}
\psi_i^{\uparrow\downarrow} = \psi^\perp_i + \theta^\perp_i,\qquad 
\psi_i^{\downarrow\uparrow} = \psi^\perp_i - \theta^\perp_i,
\end{equation}
where the $\theta_i^\perp$ correspond to SU(3)-breaking corrections that 
also violate G-parity. Introducing an analogous decomposition of
$\wt\Psi^{\ud}$ and $\wt\Psi^{\du}$ in terms of
$\widetilde{\psi}^\perp_i$ and $\widetilde{\theta}^\perp_i$, we then find
\begin{eqnarray}
\lefteqn{{\Psi}_{4;K^*}^\perp(\underline{\alpha}) = 
 30 \alpha_3^2\left\{ \psi^\perp_{0}(1-\alpha_3)
     +\psi^\perp_{1}\left[\alpha_3(1-\alpha_3)-
6\alpha_1\alpha_2\right]\right.}
\nonumber\\
& & \left.
            {} +\psi^\perp_{2}\left[\alpha_3(1-\alpha_3)-
\frac{3}{2}(\alpha_1^2
                               +\alpha_2^2)\right]-
		    (\alpha_1-\alpha_2)\left[ \theta^\perp_0 + \alpha_3
		      \theta^\perp_1 + \frac{1}{2}\,(5 \alpha_3-3)
		      \theta^\perp_2\right]\right\},
\nonumber\\
\lefteqn{\wt{\Psi}_{4;K^*}^\perp(\underline{\alpha}) = 
 30 \alpha_3^2\left\{ \wt\psi^\perp_{0}(1-\alpha_3)
     +\wt\psi^\perp_{1}\left[\alpha_3(1-\alpha_3)-
6\alpha_1\alpha_2\right]\right.}
\nonumber\\
& & \left.
            {} +\wt\psi^\perp_{2}\left[\alpha_3(1-\alpha_3)-
\frac{3}{2}(\alpha_1^2
                               +\alpha_2^2)\right]-
		    (\alpha_1-\alpha_2)\left[ \wt\theta^\perp_0 + \alpha_3
		      \wt\theta^\perp_1 + \frac{1}{2}\,(5 \alpha_3-3)
		      \wt\theta^\perp_2\right]\right\},
\nonumber\\
\lefteqn{{\Phi}^{\perp(2)}_{4;K^*}(\underline{\alpha})  = 
 -30 \alpha_3^2\left\{ \wt\theta^\perp_{0}(1-\alpha_3)
          +\wt\theta^\perp_{1}\left[\alpha_3(1-\alpha_3)-
6\alpha_1\alpha_2\right]
\right.}\nonumber\\
& &
   \left. {}  +\wt\theta^\perp_{2}\left[\alpha_3(1-\alpha_3)-
\frac{3}{2}(\alpha_1^2
                               +\alpha_2^2)\right]-
		    (\alpha_1-\alpha_2)\left[ \wt\psi^\perp_0 + \alpha_3
		      \wt\psi^\perp_1 + \frac{1}{2}\,(5 \alpha_3-3)
		   \wt \psi^\perp_2\right]\right\}\,,
\nonumber\\
\lefteqn{{\Phi}^{\perp(4)}_{4;K^*}(\underline{\alpha})  = 
 30 \alpha_3^2\left\{ \theta^\perp_{0}(1-\alpha_3)
          +\theta^\perp_{1}\left[\alpha_3(1-\alpha_3)-
6\alpha_1\alpha_2\right]
\right.}\nonumber\\
& &
   \left. {}  +\theta^\perp_{2}\left[\alpha_3(1-\alpha_3)-
\frac{3}{2}(\alpha_1^2
                               +\alpha_2^2)\right]-
		    (\alpha_1-\alpha_2)\left[ \psi^\perp_0 + \alpha_3
		      \psi^\perp_1 + \frac{1}{2}\,(5 \alpha_3-3)
		    \psi^\perp_2\right]\right\}.
\hspace*{20pt}
\end{eqnarray}

Our next task is to relate $\psi^\perp_i$ and $\theta^\perp_i$ to
local matrix elements. To LO in the conformal expansion, the relevant
matrix elements are \cite{BB98}
\begin{eqnarray}
\langle 0 | \bar q g G_{\alpha\beta} s | K^*(P,\lambda)\rangle & = & i
f_K^\perp m_{K^*}^2 \zeta_{4K^*}^\perp \left(\epsilon_\alpha^{(\lambda)}
P_\beta - \epsilon_\beta^{(\lambda)} P_\alpha\right),
\nonumber\\
\langle 0 | \bar q g \widetilde{G}_{\alpha\beta} i\gamma_5 s | 
K^*(P,\lambda)\rangle & = & i
f_K^\perp m_{K^*}^2 \widetilde{\zeta}_{4K^*}^\perp 
\left(\epsilon_\alpha^{(\lambda)}
P_\beta - \epsilon_\beta^{(\lambda)} P_\alpha\right),
\nonumber\\
\langle 0 | \bar q g G_{\alpha\mu} \sigma_{\beta\mu} s | 
K^*(P,\lambda)\rangle & = & f_K^\perp m_{K^*}^2 \left\{\frac{1}{2} 
\kappa_{3K^*}^\perp 
\left(\epsilon_\alpha^{(\lambda)} P_\beta + \epsilon_\beta^{(\lambda)} 
P_\alpha\right) + \kappa_{4K^*}^\perp 
\left(\epsilon_\alpha^{(\lambda)} P_\beta - \epsilon_\beta^{(\lambda)} 
P_\alpha\right)\right\},
\nonumber\\[-10pt]
\end{eqnarray}
where again $\zeta$ denotes G-parity-conserving parameters and
$\kappa$ G-parity-breaking ones. The twist-3 parameter
$\kappa_{3K^*}^\perp$ was investigated in Ref.~\cite{BJ07}.
Taking the local limit of
(\ref{eq:T3}) and (\ref{eq:2.21}) and comparing with the above, one
finds
\begin{equation}
\begin{array}[b]{r@{\ =\ }l@{\quad}r@{\ =\ }l}
\psi_0^\perp & \phantom{-}\zeta_{4K^*}^\perp \,,
& \wt\psi_0^\perp & \phantom{-}\wt\zeta_{4K^*}^\perp \,,\\[12pt]
\phi_0^\perp & \ds\phantom{-}\frac{1}{6}\kappa_{3K^*}^\perp +
               \frac{1}{3}\kappa_{4K^*}^\perp \,,
& \wt\phi_0^\perp & \ds\phantom{-}\frac{1}{6}\kappa_{3K^*}^\perp -
               \frac{1}{3}\kappa_{4K^*}^\perp\,, \\[12pt]
\theta_0^\perp & \ds-\frac{1}{6}\kappa_{3K^*}^\perp -
               \frac{1}{3}\kappa_{4K^*}^\perp \,,
& \wt\theta_0^\perp & \ds-\frac{1}{6}\kappa_{3K^*}^\perp +
               \frac{1}{3}\kappa_{4K^*}^\perp\,.
\end{array}
\label{4.43}
\end{equation}
Whereas $\zeta_{4K^*}^\perp$, $\wt\zeta_{4K^*}^\perp$ and
$\kappa_{3K^*}^\perp$ are independent parameters, $\kappa_{4K^*}^\perp$
depends on quark masses and $a_1^\perp$ by a relation that is 
analogous to Eq.~(\ref{eq:BL04}) and was obtained in Ref.~\cite{BZ06a}:
\begin{equation}\label{4.45}
\kappa^\perp_{4K^*} = \frac{1}{10}\,a_1^{\perp} +
\frac{f_{K^*}^\parallel}{f_{K^*}^\perp}\,\frac{m_s-m_q}{12 m_{K^*}}
 - \frac{m_s^2-m_q^2}{4 m_{K^*}^2}\,.
\end{equation}
Like with $\kappa_{4K^*}^\parallel$, 
the scale-dependence of $\kappa_{4K^*}^\perp$ follows from that of the
parameters on the right-hand side and is given by
\begin{eqnarray}
\kappa^\perp_{4K^*}(\mu^2) 
&=& 
\kappa^\perp_{4K^*}(\mu_0^2)
+ \left( L^{8/(3\beta_0)} - 1\right) \frac{1}{10}\,a_1^{\perp}(\mu_0^2)
+ \left( L^{8/(3\beta_0)} - 1\right) 
\frac{f_{K^*}^\parallel}{f_{K^*}^\perp(\mu^2_0)}\,
\frac{[m_s-m_q](\mu_0^2)}{12 m_{K^*}} 
\nonumber\\
&&{}
- \left( L^{8/\beta_0} - 1\right)\frac{[m_s^2-m_q^2](\mu^2_0)}{4 m_{K^*}^2}\,
\end{eqnarray}
with $L=\alpha_s(\mu^2)/\alpha_s(\mu_0^2)$.
In the limit of zero quark mass, the parameters $\zeta_4^T$, $\wt\zeta_4^T$
renormalise multiplicatively \cite{BBK89}:
\begin{eqnarray}
\left(\zeta_4^T + \wt{\zeta}_4^T\right)(\mu^2) = L^{\gamma^+/{\beta_0}}
\left(\zeta_4^T + \wt{\zeta}_4^T\right)(\mu_0^2),& \qquad & \gamma_+ = 3 C_A -
\frac{8}{3}\, C_F,\nonumber\\
\left(\zeta_4^T - \wt{\zeta}_4^T\right)(\mu^2) = L^{\gamma^-/{\beta_0}}
\left(\zeta_4^T - \wt{\zeta}_4^T\right)(\mu_0^2),& \qquad & \gamma_- = 4 C_A -
4 C_F.
\end{eqnarray}
Again, these simple scaling relations will receive corrections from
terms proportional to the quark masses. These corrections are unknown,
but based on the experience with pseudoscalar twist-4 matrix elements
\cite{BBL06} we do not expect them to be large.

The calculation of the NLO G-even corrections is pretty
involved and presented in detail in App.~B of Ref.~\cite{BB98}. 
The upshot is
that the six coefficients $\psi_{1,2}^\perp$, $\wt\psi_{1,2}^\perp$,
$\phi_1^\perp$ and $\wt\phi_1^\perp$ involve three additional genuine twist-4
parameters $\langle\!\langle Q^{(1,3,5)}\rangle\!\rangle$:
\begin{eqnarray}
\phi_1^\perp
& = & \phantom{-}\frac{9}{44}\, a_2^\perp + \frac{1}{8}\,
\omega_{3K^*}^\perp +\frac{63}{220}\,\langle\!\langle Q^{(1)}\rangle\!\rangle -
\frac{119}{44}\, \langle\!\langle Q^{(3)}\rangle\!\rangle\,,
\nonumber\\
\wt\phi_1^\perp
& = & -\frac{9}{44}\, a_2^\perp + \frac{1}{8}\,
\omega_{3K^*}^\perp - 
\frac{63}{220}\, \langle\!\langle Q^{(1)}\rangle\!\rangle -
\frac{35}{44}\, \langle\!\langle Q^{(3)}\rangle\!\rangle\,,
\nonumber\\
\psi_1^\perp
& = & \phantom{-}\frac{3}{44}\, a_2^\perp + \frac{1}{12}\,
\omega_{3K^*}^\perp + \frac{49}{110}\, 
\langle\!\langle Q^{(1)}\rangle\!\rangle -
\frac{7}{22}\, \langle\!\langle Q^{(3)}\rangle\!\rangle 
+ \frac{7}{3}\, \langle\!\langle Q^{(5)}\rangle\!\rangle\,,
\nonumber\\
\wt\psi_1^\perp
& = & -\frac{3}{44}\, a_2^\perp + \frac{1}{12}\,
\omega_{3K^*}^\perp - \frac{49}{110}\, 
\langle\!\langle Q^{(1)}\rangle\!\rangle +
\frac{7}{22}\, \langle\!\langle Q^{(3)}\rangle\!\rangle
 + \frac{7}{3}\, \langle\!\langle Q^{(5)}\rangle\!\rangle\,,
\nonumber\\
\psi_2^\perp 
& = & -\frac{3}{22}\, a_2^\perp - \frac{1}{12}\, 
\omega_{3K^*}^\perp + \frac{28}{55}\, 
\langle\!\langle Q^{(1)}\rangle\!\rangle +
\frac{7}{11}\, \langle\!\langle Q^{(3)}\rangle\!\rangle + 
\frac{14}{3}\, \langle\!\langle Q^{(5)}\rangle\!\rangle\,,
\nonumber\\
\wt\psi_2^\perp 
& = & \phantom{-}\frac{3}{22}\, a_2^\perp - \frac{1}{12}\,
\omega_{3K^*}^\perp - \frac{28}{55}\, 
\langle\!\langle Q^{(1)}\rangle\!\rangle -
\frac{7}{11}\, \langle\!\langle Q^{(3)}\rangle\!\rangle 
+ \frac{14}{3}\, \langle\!\langle Q^{(5)}\rangle\!\rangle\,.
\label{4.46}
\end{eqnarray}
The precise definition of $\langle\!\langle Q^{(i)}\rangle\!\rangle$ is
given in Ref.~\cite{BB98}. $\omega_{3K^*}^\perp$ is a twist-3
parameter and defined in Ref.~\cite{BJ07}.
Existing numerical determinations of these parameters from QCD sum
rules are far from
being precise, so we decide to estimate them using the renormalon
model of Ref.~\cite{BGG04} instead. The model entails the following 
expressions of
the three-particle twist-4 DAs in terms of $\phi_{2;K^*}^\perp$:
\begin{eqnarray}
\Phi_{4;K^*}^{\perp(1),\rm R}(\underline{\alpha}) 
& = & -\Phi_{4;K^*}^{\perp(3),\rm R}(\underline{\alpha})\ =\ 
\zeta_{4K^*}^\perp\left[
\frac{\alpha_2\phi^\perp_{2;K^*}(\alpha_1)}{(1-\alpha_1)^2}
-\frac{\alpha_1\phi^\perp_{2;K^*}
(\bar\alpha_2)}{(1-\alpha_2)^2}
\right],
\nonumber\\
 \Phi_{4;K^*}^{\perp(2),\rm R}(\underline{\alpha})& = &
   \phantom{-}\Phi_{4;K^*}^{\perp(4),\rm R}(\underline{\alpha})\ =\ 
-\frac12 \zeta_{4K^*}^\perp\left[
\frac{\phi^\perp_{2;K^*}(\alpha_1)}{(1-\alpha_1)}
-\frac{\phi^\perp_{2;K^*}(\bar\alpha_2)}{(1-\alpha_2)}
\right],
\nonumber\\
\Psi_{4;K^*}^{\perp,\rm R}(\underline{\alpha}) & = &   
-\wt\Psi_{4;K^*}^{\perp,\rm R}(\underline{\alpha})\ =\ 
\frac12 \zeta_{4K^*}^\perp\left[
\frac{\phi^\perp_{2;K^*}(\alpha_1)}{1-\alpha_1}+
\frac{\phi^\perp_{2;K^*}(\bar\alpha_2)}{1-\alpha_2}
\right],
\nonumber\\
\Xi^{\perp,\rm R}_{4;K^*}(\underline{\alpha}) &=& -
 2\zeta_{4K^*}^\perp
\left[\frac{\alpha_2}{1-\alpha_1}\,\phi^\perp_{2;K^*}(\alpha_1)
-\frac{\alpha_1}{1-\alpha_2}\phi^\perp_{2;K^*}(\bar\alpha_2)
\right],
\end{eqnarray}
which is similar to the results for the chiral-even DAs, Eq.~(\ref{3.21}).
The model implies the following estimates for the G-conserving twist-4
parameters in (\ref{4.43}) and (\ref{4.46}):
\begin{equation}
\begin{array}[b]{
r@{\ =\ }l@{\quad}r@{\ =\ }l@{\quad}r@{\ =\ }l@{\quad}r@{\ =\ }l}
\wt\zeta_{4K^*}^{\perp,\rm R} & -\zeta_{4K^*}^\perp\,, &
\ds\langle\!\langle Q^{(1)}\rangle\!\rangle^{\rm R}
& \ds-\frac{10}{3}\zeta_{4K^*}^\perp\,, & 
\langle\!\langle Q^{(3)}\rangle\!\rangle^{\rm R}
& -\zeta_{4K^*}^\perp\,, &
\langle\!\langle Q^{(5)}\rangle\!\rangle^{\rm R}
& 0\,.
\end{array}
\end{equation}
The G-parity-violating parameters are given by:
\begin{equation}
\begin{array}[b]{r@{\ =\ }l@{\quad}r@{\ =\ }l@{\quad}r@{\ =\ }l@{\quad}r@{\ =\ }l}
\phi_0^{\perp,\rm R} & 0\,, & \wt\phi_0^{\perp,\rm R} & 0\,, &
\phi_2^{\perp,\rm R} & 
\ds-\frac{21}{20}\,\zeta_{4K^*}^{\perp} a_1^{\perp}\,, & 
\wt\phi_2^{\perp,\rm R} & 
\ds-\frac{21}{20}\,\zeta_{4K^*}^{\perp} a_1^{\perp}\,,\\[15pt]
\theta_0^{\perp,\rm R} & 0\,, & \wt\theta_0^{\perp,\rm R} & 0\,, & 
\theta_1^{\perp,\rm R} & 
\ds-\frac{21}{10}\,\zeta_{4K^*}^{\perp} a_1^{\perp}\,,& 
\wt\theta_1^{\perp,\rm R} & 
\ds\phantom{-}\frac{21}{10}\,\zeta_{4K^*}^{\perp} a_1^{\perp}\,,\\[15pt]
\theta_2^{\perp,\rm R} & 
\ds\frac{21}{5}\,\zeta_{4K^*}^{\perp} a_1^{\perp}\,,& 
\wt\theta_2^{\perp,\rm R} & \ds-\frac{21}{5}\,\zeta_{4K^*}^{\perp}
a_1^{\perp}\,, & \xi_0^{\perp,\rm R} & 
\ds\phantom{-}\frac{3}{5}\zeta_{4K^*}^{\perp} a_1^{\perp}\,.\\[-0.5cm]
\end{array}
\end{equation}
As with chiral-even distributions, the above results provide an
estimate only for the genuine twist-4 contributions, 
but miss any mass corrections in terms of lower-twist parameters. This
is also the reason why $\phi_0^\perp$, $\theta_0^\perp$, 
$\widetilde\phi_0^\perp$ and $\widetilde\theta_0^\perp$ vanish in the
renormalon model, in contrast to Eq.~(\ref{4.43}).

Let us now turn to the two-particle twist-4 DAs $\phi_{4;K^*}^\perp$
and $\psi_{4;K^*}^\perp$ defined in Eq.~(\ref{2.10}). From the
operator relations given in App.~\ref{app:A} we obtain:
\begin{eqnarray}
\psi^\perp_{4;K^*}(u) & = & -\phi^\perp_{2;K^*}(u) + 2
\psi^\parallel_{3;K^*}(u) 
\nonumber\\
&&{}-2\, \frac{d}{du} \int_0^ud\alpha_1
\int_0^{\bar u} d\alpha_2 \left\{
\frac{1}{\alpha_3^2}\,(\alpha_1-\alpha_2-\xi)
  \Psi^\perp_{4;K^*}(\underline{\alpha})\right.
\nonumber\\
&&{}\left.\hspace*{2cm} - \frac{1}{\alpha_3} \left(
  \Phi^{\perp(2)}_{4;K^*}(\underline{\alpha}) - 
\Phi^{\perp(3)}_{4;K^*}(\underline{\alpha})\right)\right\}
 +
2\,\frac{f_{K^*}^\parallel}{f_{K^*}^\perp}\,\frac{m_s+m_q}{m_{K^*}}\, 
\phi^\perp_{3;K^*}(u)\,,\label{4.20}\\
\frac{d}{du}\,\phi^\perp_{4;K^*}(u) & = & - 2 \xi \left\{
\psi^\perp_{4;K^*}(u) + \phi^\perp_{2;K^*}(u)\right\} + 8 \int_0^u dv
\{\psi^\perp_{4;K^*}(v) - \phi^\perp_{2;K^*}(v)\}
\nonumber\\
&&{}-\frac{d}{du}\int_0^u d\alpha_1 \int_0^{\bar u} d\alpha_2
\,\frac{4}{\alpha_3} \left\{ \frac{\alpha_1-\alpha_2-\xi}{\alpha_3}
\left[ \Phi^{\perp(2)}_{4;K^*}(\underline{\alpha}) -
  \Phi^{\perp(3)}_{4;K^*}(\underline{\alpha}) \right] -
\Psi^\perp_{4;K^*}(\underline{\alpha}) \right\}
\nonumber\\
&&{}+4\,\frac{f_{K^*}^\parallel}{f_{K^*}^\perp}\,
\frac{m_s-m_q}{m_{K^*}}\,\phi^\perp_{3;K^*}(u)\,.\label{4.21}  
\end{eqnarray}
The boundary condition for $\phi^\perp_{4;K^*}$ is
$\phi^\perp_{4;K^*}(0) = 0 = \phi^\perp_{4;K^*}(1)$, which implies
the relation (\ref{4.45}). In the renormalon model, one obtains exact
expressions for these DAs \cite{BGG04}:
\begin{eqnarray}
\psi^{\perp,\rm T4, R}_{4;K^*}(u) & = & 0\,,\nonumber\\
\phi^{\perp.\rm T4, R}_{4;K^*}(u) & = & 8 \zeta_{4K^*}^\perp \left[ \int_0^u dv
  \left( \bar u + (u-v) \ln\,\frac{u-v}{\bar v}\right)
  \frac{\phi^\perp_{2;K^*}(v)}{\bar v^2} \right.
\nonumber\\
&&\left.\hspace*{2cm} + \int_u^1 dv \left( u + (v-u) 
\ln\,\frac{v-u}{v}\right)\frac{\phi^\perp_{2;K^*}(v)}{v^2}\right].
\end{eqnarray}

Like the chiral-even DA $\psi_{4;K^*}^\parallel$, 
$\psi_{4;K^*}^\perp$ corresponds to the projection
$s=-\frac{1}{2}$ for both quark and antiquark and hence, in the
absence of quark-mass corrections in $m_s\pm m_q$, has an
expansion in terms of $C^{1/2}_n(\xi)$.
The full $\psi_{4;K^*}^{\perp}$ contains corrections explicitly
proportional to $m_s\pm m_q$, of which we only keep the leading term
in $(m_s\pm m_q)^1$. To NLO in the conformal expansion, (\ref{4.20})
yields:
\begin{eqnarray}
\psi^{\perp}_{4;K^*}(u)
& = & 
1 + \left(12 \kappa_{4K^*}^\perp - \frac{3}{5}\,
a_1^\perp \right) C_1^{1/2}(\xi)
+\left( -1 + \frac{3}{7}\, a_2^\perp-10 \left\{ \zeta^\perp_{4K^*} +
\widetilde{\zeta}^\perp_{4K^*}\right\}\right) C_2^{1/2}(\xi)
\nonumber\\
&&{}+ \left\{ -5\kappa^\perp_{3K^*} - 12 \kappa_{4K^*}^\perp -
\frac{1}{3}\,\lambda_{3K^*}^\perp + \frac{3}{5}\, a_1^\perp
+ 5 \left[ \theta_1^\perp + \widetilde{\theta}_1^\perp -
\frac{1}{2}\left( \theta_2^\perp +
\widetilde{\theta}_2^\perp\right) \right] \right\}
C_3^{1/2}(\xi)
\nonumber\\
&&+ \left(-\frac{5}{4}\,\omega^\perp_{3K^*} -
\frac{3}{7}\,a_2^\perp\right) C_4^{1/2}(\xi)
+ \frac{1}{3}\,\lambda^\perp_{3K^*} C_5^{1/2}(\xi)
\nonumber\\
&& + \frac{m_s+m_s}{m_{K^*}}\,
\frac{f^\parallel_{K^*}}{f^\perp_{K^*}}\, 
\left\{ 3 \left(1+6 a_2^\parallel\right) 
        + 3 a_1^\parallel C_1^{1/2}(\xi)
        + 5 \left( 4 \zeta_{3K^*}^\parallel - 3 a_2^\parallel\right)
	C_2^{1/2}(\xi) \right.
\nonumber\\
&& \hspace*{1cm}\left. + 5 \left( 4 \kappa_{3K^*}^\parallel - \frac{3}{4}\,
\lambda_{3K^*}^\parallel +
\frac{3}{2}\,\widetilde{\lambda}_{3K^*}^\parallel \right)
C_3^{1/2}(\xi)
+ \frac{15}{4}\left( 2 \omega_{3K^*}^\parallel -
\widetilde{\omega}_{3K^*}^\parallel \right) C_4^{1/2}(\xi)\right\}
\nonumber\\ 
&&{} + 6\,\frac{m_s+m_s}{m_{K^*}}\,
\frac{f^\parallel_{K^*}}{f^\perp_{K^*}}\, 
\left\{ \left( 1 - 3 a_1^\parallel + 6 a_2^\parallel\right) u \ln u
+ \left( 1 +3 a_1^\parallel + 6 a_2^\parallel\right) \bar u \ln \bar
u \right\}
\nonumber\\
&&{}- 6\,\frac{m_s-m_s}{m_{K^*}}\,
\frac{f^\parallel_{K^*}}{f^\perp_{K^*}}\, u \bar u \left( 9
a_1^\parallel + 10 \xi a_2^\parallel \right)
\nonumber\\
&&{}+  6\,\frac{m_s-m_s}{m_{K^*}}\,
\frac{f^\parallel_{K^*}}{f^\perp_{K^*}}\, 
\left\{ \left( 1 - 3 a_1^\parallel + 6 a_2^\parallel\right) u \ln u
- \left( 1 +3 a_1^\parallel + 6 a_2^\parallel\right) \bar u \ln \bar
u \right\}.\label{4.23}
\end{eqnarray}
Recall that $\bar u = 1-u$ and $\xi = 2u-1$. The above expression
refers to a $K^*=(s\bar q)$ meson; for $\bar K^* = (q\bar s)$, one has
to replace $u$ by $1-u$.

The explicit formula for $\phi^\perp_{4;K^*}$ from (\ref{4.21}) is very
long and complicated, so we only give the result for $m_s\pm m_q\to
0$:
\begin{eqnarray}
\phi^\perp_{4;K^*}(u) 
& = & 
30 u^2\bar u^2 \left\{ \left(\frac{4}{3}\, \zeta^\perp_{4K^*} -
\frac{8}{3}\, \widetilde{\zeta}^\perp_{4K^*} + \frac{2}{5} +
\frac{4}{35}\, a_2^\perp\right) \right.
\nonumber\\
&&{}\hspace*{1cm} + \left( \frac{3}{25}\,a_1^\perp +
\frac{1}{3}\,\kappa_{3K^*}^\perp - \frac{1}{45}\, \lambda_{3K^*}^\perp
- \frac{1}{15}\,\theta_1^\perp  + \frac{7}{30}\,\theta_2^\perp
+\frac{1}{5}\, \widetilde{\theta}_1^\perp - 
\frac{3}{10}\,\widetilde{\theta}_2^\perp\right) C_1^{5/2}(\xi)
\nonumber\\
&&\hspace*{1cm} +\left.\left(\frac{3}{35}\,a_2^\perp +
\frac{1}{60}\,{\omega}_{3K^*}^\perp \right) C_2^{5/2}(\xi) 
-\frac{4}{1575}\,\lambda_{3K^*}^\perp C_3^{5/2}(\xi)\right\}
\nonumber\\
&+& \left( 5 \kappa_{3K^*}^\perp - a_1^\perp - 
20 \widetilde{\phi}_2^\perp\right)
\left\{-4 u^3 (2-u) \ln u + 4 \bar u^3 (2-\bar u) \ln \bar u +
\frac{1}{2}\,u \bar u \xi (3\xi^2-11)\right\}
\nonumber\\
&+&  \left( 2 \omega_{3K^*}^\perp - \frac{36}{11}\, a_2^\perp 
-\frac{252}{55}\, \langle\!\langle Q^{(1)}\rangle\!\rangle -
\frac{140}{11}\, \langle\!\langle Q^{(3)}\rangle\!\rangle\right)
\nonumber\\
&&\hspace*{0.4cm}\times\left\{ u^3
(6 u^2-15u+10)\ln u + \bar u^3 (6 \bar u^2-15\bar u+10)\ln\bar u -
\frac{1}{8}\,u\bar u \left( 13\xi^2-21\right)\right\}.
\nonumber\\[-10pt]\label{4.24}
\end{eqnarray}

Both (\ref{4.23}) and (\ref{4.24}) agree, for the $\rho$ meson, with
the results obtained in Ref.~\cite{BB98}.
The numerics of the above DAs will be discussed in the next section.

\section{Models for Distribution Amplitudes}\label{sec:5}
\setcounter{equation}{0}

In this section we compile the numerical estimates of all necessary parameters 
and present explicit models of the twist-4 two-particle
DAs introduced in Secs.~\ref{sec:3} and \ref{sec:4}.
The important point is that these DAs are related to three-particle 
ones by exact QCD EOM and have to be used together: this
guarantees the consistency of the approximation.
Our model thus introduces a minimum number of non-perturbative 
parameters, which are defined as  matrix elements of certain local operators 
between the vacuum and the meson state, and which we estimate using 
QCD sum rules. More
sophisticated models can be constructed in a systematic way by adding 
contributions of higher conformal partial waves when estimates of the relevant 
non-perturbative matrix elements will become available. 

Our  approach involves the implicit assumption that the conformal partial 
wave expansion is well convergent. This can be justified rigorously 
at large scales, since the anomalous dimensions of all involved 
operators increase logarithmically with the conformal spin $j$, but 
is non-trivial at relatively low scales of order $\mu \sim (1$--$2)\,$GeV
which we choose as reference scale. An upper bound for the contribution
of higher partial waves can be obtained from the renormalon model.

Since orthogonal polynomials of high orders are rapidly oscillating 
functions, a truncated expansion in conformal partial waves is, 
almost necessarily, oscillatory as well. Such a behaviour is 
clearly unphysical,
but this does not constitute a real problem since  physical observables 
are given by convolution integrals of DAs with
smooth coefficient functions. 
A classical example for this
feature is the $\gamma\gamma^*$-meson form factor, which is governed
by the quantity
$$
\int du\,\frac{1}{u}\, \phi(u) \sim \sum a_i,
$$
where the coefficients $a_i$ are exactly the ``reduced matrix elements''
in the conformal expansion.
The oscillating terms are averaged over and strongly suppressed. 
Stated otherwise: 
models of DAs should generally be understood as
distributions (in the mathematical sense).    

\begin{table}[tb]
\renewcommand{\arraystretch}{1.3}
\addtolength{\arraycolsep}{3pt}
$$
\begin{array}{l || c | c ||  l | l || c | c }
\hline
& \multicolumn{2}{c||}{\rho} & \multicolumn{2}{c||}{K^*}  &  
\multicolumn{2}{c}{\phi}\\
\cline{2-7}
& \mu = 1\,{\rm GeV} & \mu = 2\,{\rm GeV} & \mu = 1\,{\rm GeV} & \mu =
2\,{\rm GeV} & \mu = 1\,{\rm GeV} & \mu = 2\,{\rm GeV}\\
\hline
f_V^\parallel\:[{\rm MeV}] & 216(3) & 216(3) &
                \multicolumn{1}{c|}{220(5)} & \multicolumn{1}{c||}{220(5)} &
                             215(5) & 215(5)\\
f_V^\perp\:[{\rm MeV}] & 165(9) & 145(4) & 
                \multicolumn{1}{c|}{185(9)} & \multicolumn{1}{c||}{163(8)} &
                             186(9) & 164(8)\\
\hline
a_1^\parallel & 0 & 0 & \phantom{-}0.03(2) & \phantom{-}0.02(2) & 0 & 0 
\\
a_1^\perp & 0 & 0 & \phantom{-}0.04(3) & \phantom{-}0.03(3) & 0 & 0
\\
a_2^\parallel & 0.15(7) & 0.10(5) & \phantom{-}0.11(9) & 
\phantom{-}0.08(6) & 0.18(8) & 0.13(6)
\\
a_2^\perp & 0.14(6) & 0.11(5) & \phantom{-}0.10(8) &
\phantom{-}0.08(6) & 0.14(7) & 0.11(5)
\\\hline
\zeta_{3V}^\parallel 
& 0.030(10) & 0.020(9) & \phantom{-}0.023(8) & \phantom{-}0.015(6) & 
0.024(8) & 0.017(6)
\\
\widetilde\lambda_{3V}^\parallel 
& 0 & 0 & \phantom{-}0.035(15)& \phantom{-}0.017(8) & 0 & 0
\\
\widetilde\omega_{3V}^\parallel 
& -0.09(3) & -0.04(2) & -0.07(3) &  -0.03(2) & -0.045(15) & -0.022(8)
\\
\kappa_{3V}^\parallel 
& 0 & 0 & \phantom{-}0.000(1) & -0.001(2) & 0 & 0
\\
\omega_{3V}^\parallel 
& 0.15(5) & 0.09(3) & \phantom{-}0.10(4) & \phantom{-}0.06(3) &
0.09(3) & 0.06(2)
\\
\lambda_{3V}^\parallel 
& 0 & 0 & -0.008(4) & -0.004(2) & 0 & 0
\\
\kappa_{3V}^\perp 
& 0 & 0 & \phantom{-}0.003(3) & -0.001(2) & 0 & 0 
\\
\omega_{3V}^\perp 
& 0.55(25) & 0.37(19) & \phantom{-}0.3(1) & \phantom{-}0.2(1) &  
0.20(8) & 0.15(7)
\\
\lambda_{3V}^\perp 
& 0 & 0 & -0.025(20) & -0.015(10) & 0 & 0\\\hline 
\end{array}
$$
\renewcommand{\arraystretch}{1}
\addtolength{\arraycolsep}{-3pt}
\vspace*{-10pt}
\caption[]{\small Decay constants and twist-2 and -3 hadronic parameters 
at the scale  $\mu= 1\,{\rm GeV}$ and scaled up to $\mu= 2\,{\rm
  GeV}$. The sign of the twist-3 parameters
corresponds to the sign convention for the strong coupling defined by
the covariant derivative $D_\mu = \partial_\mu - i g A^a_\mu t^a$; they
change sign if $g$ is fixed by $D_\mu = \partial_\mu + i g
A^a_\mu t^a$. Numbers taken from Ref.~\cite{BJ07}, see also
Ref.~\cite{BJZ} for a detailed discussion of the decay
constants.}
\label{tab1}
\end{table}
\begin{table}[tb]
\renewcommand{\arraystretch}{1.5}
\addtolength{\arraycolsep}{3pt}
$$
\begin{array}{l || l | l ||  l | l || l | l }
\hline
& \multicolumn{2}{c||}{\rho} & \multicolumn{2}{c||}{K^*}  &  
\multicolumn{2}{c}{\phi}\\
\cline{2-7}
& \mu = 1\,{\rm GeV} & \mu = 2\,{\rm GeV} & \mu = 1\,{\rm GeV} & \mu =
2\,{\rm GeV} & \mu = 1\,{\rm GeV} & \mu = 2\,{\rm GeV}\\
\hline
\zeta_4^\parallel & \phantom{-}0.07(3) & \phantom{-}0.06(2) & 
                    \phantom{-}0.02(2) & \phantom{-}0.02(2) & 
                    \phantom{-}0.00(2) & \phantom{-}0.00(2)\\
\widetilde{\omega}^\parallel_4 & -0.03(1) & -0.02(1) &
                               -0.02(1) & -0.01(1) & -0.02(1) &
			       -0.01(1)\\
\zeta^\perp_4 & -0.03(5) & -0.02(3) & -0.01(3) & -0.01(2) 
                                    & -0.01(3) & -0.01(2)\\
\widetilde{\zeta}^\perp_4 & -0.08(5) & -0.05(3) & -0.05(4) 
                          & -0.04(2) & -0.03(4) & -0.02(2)\\
\kappa_{4K^*}^\parallel & \phantom{-}0 & \phantom{-}0 
                 & -0.025(5) & -0.020(4) & \phantom{-}0 & \phantom{-}0\\
\kappa_{4K^*}^\perp & \phantom{-}0 & \phantom{-}0 & \phantom{-}0.013(5) &
\phantom{-}0.011(5) & \phantom{-}0 & \phantom{-}0\\
\hline
\end{array}
$$
\renewcommand{\arraystretch}{1}
\addtolength{\arraycolsep}{-3pt}
\vspace*{-10pt}
\caption[]{\small Twist-4 parameters at the scale $\mu=1\,$GeV and
  scaled up to 2~GeV. Sign convention for the strong coupling $g$ as for
  twist-3 parameters in Tab.~\ref{tab1}.}\label{tab2}
\end{table}

All relevant numerical input parameters for our model DAs are given in
Tabs.~\ref{tab1} and \ref{tab2}, at the scale $\mu=1\,$GeV, 
which is appropriate for
QCD sum-rule results, and at the scale $\mu=2\,$GeV, 
using the scaling relations given in Secs.~\ref{sec:3} and \ref{sec:4}, 
to facilitate the comparison with future lattice
determinations of these quantities. 

The parameters related to twist-2 matrix
elements have been determined using various methods; see the
discussion in Ref.~\cite{BJ07}. Matrix elements of twist-3 operators
were also discussed in Ref.~\cite{BJ07}. Twist-4 matrix elements for
the $\rho$ were estimated
a long time ago from QCD sum rules \cite{BB98,BBK89,BK86}. 
In this paper, we perform a complete reanalysis of these parameters
and also include G-parity-breaking effects relevant for the $K^*$ and
SU(3) breaking for the $\phi$ meson. The corresponding sum rules and
plots are given in the appendices. 

{}For the chiral-even parameter $\zeta^\parallel_{4\rho}$ we find 
$\zeta^\parallel_{4\rho} = 0.07\pm 0.03$, which agrees with our older
result $\zeta^\parallel_{4\rho} = 0.15\pm 0.10$ \cite{BB98} within  
errors. The change is due to updated input parameters.
Another parameter, $\widetilde{\omega}^\parallel_{4\rho}$, was
estimated, in Ref.~\cite{BB98}, from a correlation function of
currents with different chirality, by dividing the leading
contribution (a dimension-5 power correction) by the typical hadronic
scale. The result $\widetilde{\omega}^\parallel_{4\rho} = 0.1\pm 0.1$
is a crude estimate. In this paper we obtain
$\widetilde{\omega}^\parallel_{4\rho} = -0.03\pm 0.01$, from a careful
analysis of various sum rules.
This result is smaller than the previous estimate and negative, 
in agreement with the prediction based on the renormalon model
(\ref{3.23}). Th absolute size is smaller than the renormalon-model
prediction, which is not
significant, however, as the intrinsic renormalisation scale at which
the model is valid is not fixed. 

Another important result is that we find
$\zeta^\perp_4+\widetilde{\zeta}^\perp_4\neq 0$. This parameter is usually 
set to zero, based on the observation that the
leading contribution to the correlation function of the corresponding
quark-quark-gluon operator
with the electromagnetic current vanishes \cite{BBK89}. 
Similarly, as discussed in
Ref.~\cite{BGG04}, there is no leading renormalon 
contribution to this operator, which implies 
$\zeta^\perp_4+\widetilde{\zeta}^\perp_4=0$ in the renormalon model.
In App.~\ref{app:C} we carefully 
investigate a number of different sum rules for $\zeta^\perp_4\pm 
\widetilde{\zeta}^\perp_4$, which are mutually consistent and yield
the results given in Tab.~\ref{tab2}, with
$\zeta^\perp_4+\widetilde{\zeta}^\perp_4\neq 0$. 
On the other hand, our result for 
$\zeta^\perp_4-\widetilde{\zeta}^\perp_4$ is consistent with
older estimates based on the analysis of the leading contribution to
the chiral-odd sum rules \cite{BBK89}, albeit a factor two smaller.  

\begin{figure}
$$\epsfxsize=0.46\textwidth\epsffile{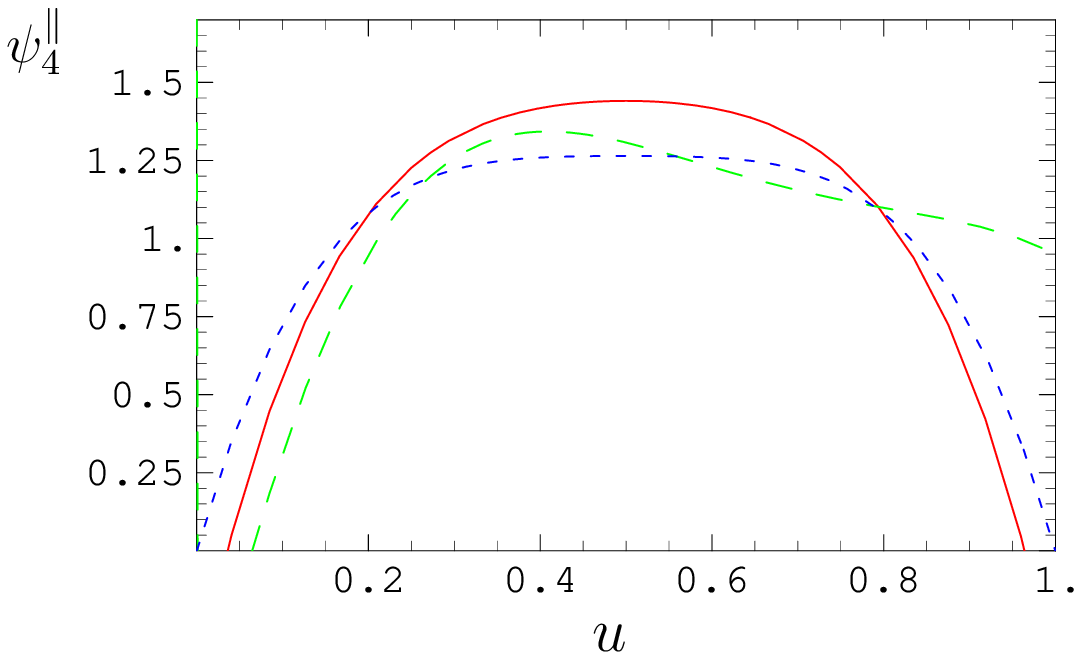}\quad
\epsfxsize=0.46\textwidth\epsffile{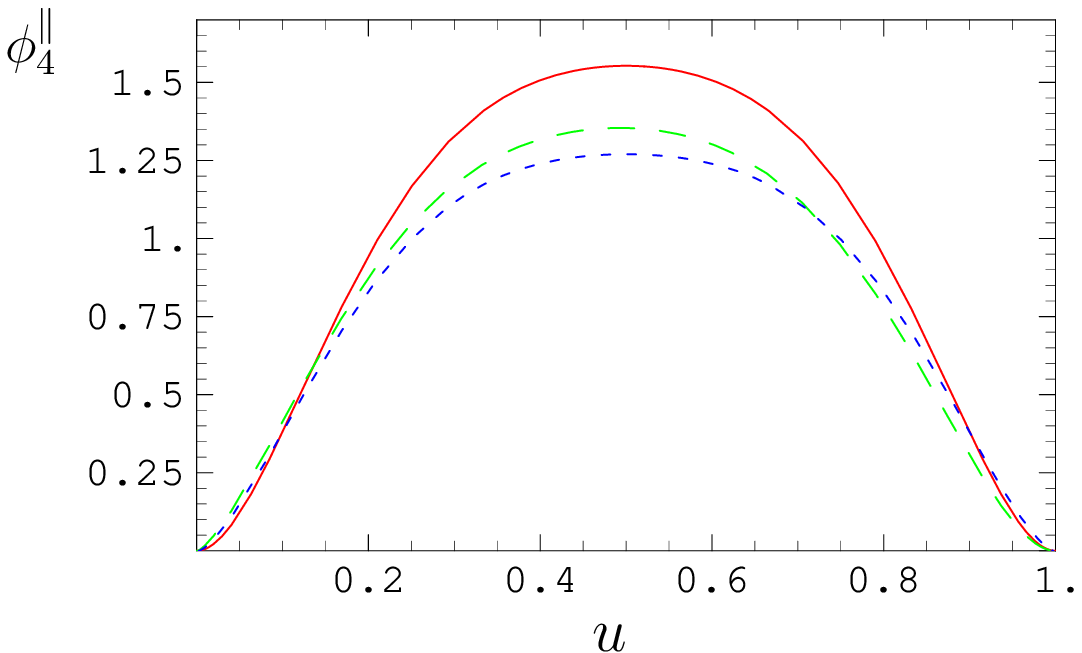}$$
\caption{\small [Colour online] 
Left panel: $\psi_{4}^\parallel$, (\ref{4.27}), as a function of $u$
  for the central
  value of the hadronic parameters, for $\mu=1\,$GeV. Solid [red]
  line: $\psi_{4;\rho}^\parallel$, dashed [green]:
  $\psi_{4;K^*}^\parallel$, short-dashed [blue[: $\psi_{4;\phi}^\parallel$. 
Right panel: same for $\phi_4^\parallel$, (\ref{4.29}).}\label{fig1}
$$\epsfxsize=0.46\textwidth\epsffile{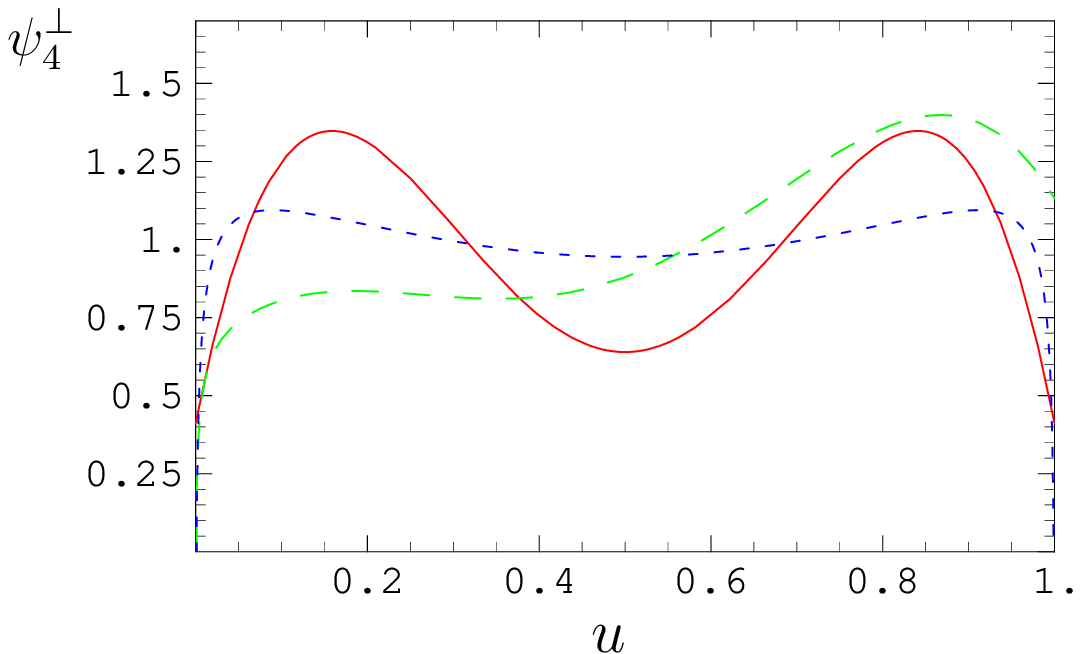}\quad
\epsfxsize=0.46\textwidth\epsffile{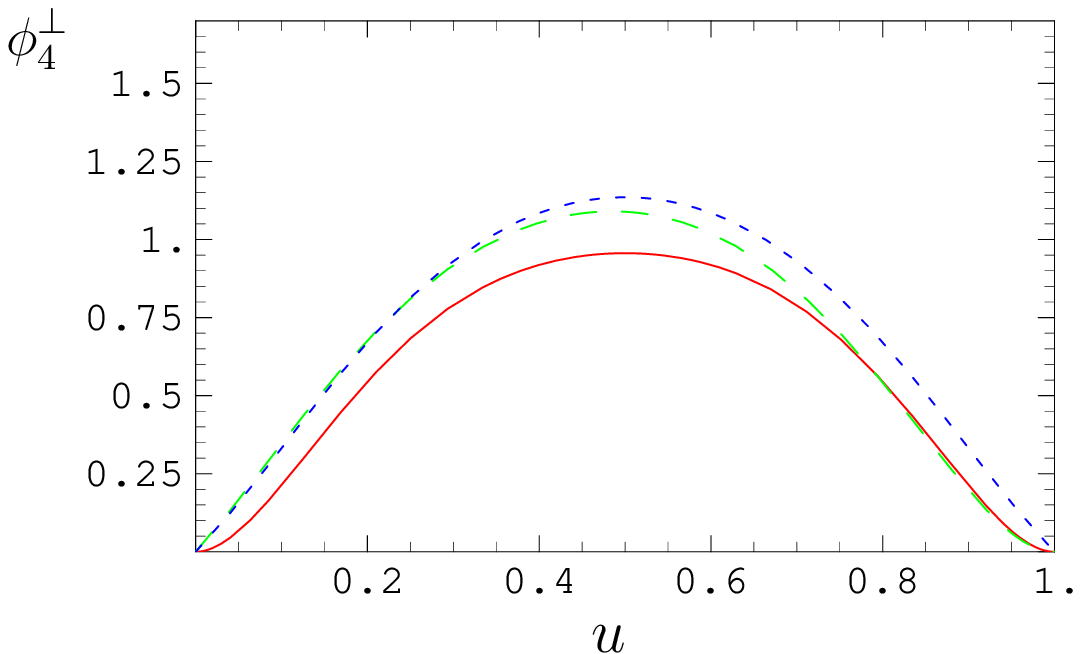}$$
\caption{\small  [Colour online] Same as Fig.~\ref{fig1} for
  $\psi_4^\perp$, (\ref{4.23}), and
  $\phi_4^\perp$, (\ref{4.24}).}\label{fig2}
$$\epsfxsize=0.46\textwidth\epsffile{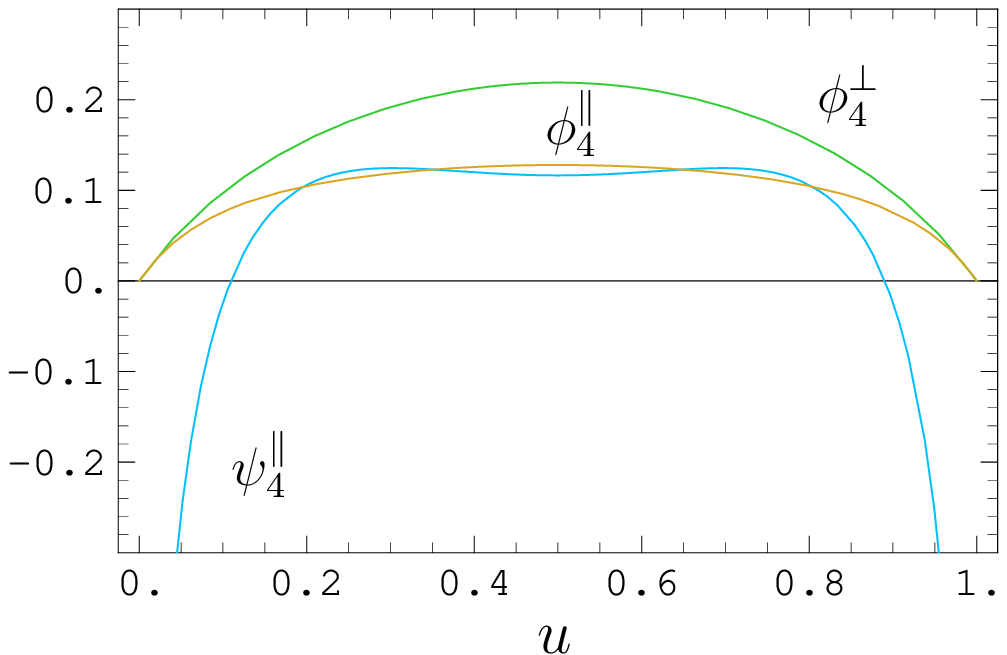}$$
\caption{\small [Colour online] Renormalon-model predictions for
  $\phi^{\parallel,\perp}_{4;\rho}$ and $\psi^\parallel_{4;\rho}$;
  $\psi^\perp_{4;\rho}=0$ in the renormalon model.}\label{fig3} 
\end{figure}

The resulting four two-particle twist-4 DAs, as given by 
(\ref{4.27}), (\ref{4.29}), (\ref{4.23}) and (\ref{4.24}),
 are
 shown in Figs.~\ref{fig1} and \ref{fig2}. We use the
renormalon-model predictions for all matrix elements which are not
known from a direct calculation. For $\zeta_4^{\perp,{\rm R}}$, we use
  $\zeta_4^{\perp,{\rm R}}=(\zeta^\perp_4-\widetilde\zeta^\perp_4)/2$. 
The SU(3) breaking is moderate in  $\phi^{\parallel,\perp}_4$, but 
obviously 
more pronounced for $\psi^{\parallel,\perp}_4$. This
feature is mainly due to the different shape of the asymptotic DAs which 
vanish at the end-points for  $\phi_4$, but are non-zero for $\psi_4$.
As is seen from the behaviour of
$\psi^\parallel_{4;K^*}$ in particular, Fig.~\ref{fig1}, the finite
mass corrections in $m_s$ change the shape of the DA noticeably for
$u\to 1$; this result is dominated by the terms linear in
$m_s$. Keeping all quark masses, the behaviour very close to the
end-points is given by $m_q(m_s-m_q) \ln \ub$ for $u\to 1$ and
$m_s(m_s-m_q) \ln u$ for $u\to 0$.  
For $u\to 0$ the logarithmic term is dominant and
causes the marked asymmetry in the dashed (green) curve in the left
panel of Fig.~\ref{fig1}. For the $\phi$ meson, the logarithms vanish
as $m_q\to m_s$. A similar effect is at play for
$\psi^\perp_{4K^*}$, Fig.~\ref{fig2}, but is slightly less marked
numerically. Due to the dominance of these finite-mass corrections,
the renormalon model alone gives only a poor description of the full DAs, see
Fig.~\ref{fig3}. In particular the size of the apex at $u=1/2$ is
considerably underestimated. The apex is actually
dominated by the contribution of $a_0^{\parallel,\perp}=1$ to the DAs,
see Eqs.~(\ref{3.27}) and (\ref{3.29}). Technically speaking, this
term is a mass correction and hence not included in the renormalon
model. Despite this shortcoming, the renormalon model is very useful 
for estimating otherwise only
poorly constrained higher-conformal waves of twist-4 DAs, and in
particular G-parity-breaking parameters.

\section{Summary and Conclusions}\label{sec:6}
\setcounter{equation}{0}

In this paper we have studied the twist-4 two- and three-particle
distribution amplitudes of $\rho$, $K^*$ and $\phi$ mesons in QCD and 
expressed them in a model-independent way by a minimal number of 
non-perturbative parameters. The work presented here is an extension of
Refs.~\cite{BBKT,BB98,BJ07} and completes the analysis of SU(3)-breaking
corrections by also including G-parity-breaking corrections in
$m_s-m_q$ to twist-4 distribution amplitudes. Our main results are the
expressions for twist-4 two-particle distribution amplitudes,
Eqs.~(\ref{4.27}), (\ref{4.29}), (\ref{4.23}), (\ref{4.24}), and the
complete set of twist-4 input parameters, Tab.~\ref{tab2}. With these
results, a complete set of light-meson DAs of twist 2, 3 and 4 is
available for both pseudoscalar and vector mesons.

Our approach consists of two components.
One is the use of the QCD equations of motion, which allow 
dynamically dependent DAs to be expressed in terms of independent 
ones. The other ingredient is conformal
expansion, which makes it possible to separate transverse and
longitudinal variables in the wave functions, the former ones being
governed by renormalisation-group equations, the latter ones being
described in terms of irreducible representations of the corresponding
symmetry group.
We have derived expressions for all twist-4 two- and
three-particle distribution amplitudes to NLO in the
conformal expansion, including both chiral corrections ${\mathcal
  O}(m_s+m_q)$ and G-parity-breaking corrections ${\mathcal
  O}(m_s-m_q)$; the corresponding formulas are given in
Secs.~\ref{sec:3} and \ref{sec:4}.
We have also generalized the renormalon model of Ref.~\cite{BGG04} to describe
SU(3)-breaking contributions to high-order conformal partial waves.

We have done a complete reanalysis of the numerical values of the
relevant higher-twist hadronic parameters from QCD sum rules. 
Our sum rules can be compared, in the chiral limit, with
existing calculations for the $\rho$ \cite{BB98,BK86}. 
Whenever possible, we have aimed at
determining these matrix elements from more than one sum rule; we
find mutually consistent results, which provides a consistency check
of the approach. 
Our final numerical results, at the scales 1 and 2~GeV, are collected
in Tab.~\ref{tab2}. Any substantial improvement of these results will require
input from alternative non-perturbative methods, in particular lattice
calculations. 

We hope that our results will contribute to a better understanding of
SU(3)-breaking effects in hard exclusive processes and in particular
in the decays of $B_{u,d}$ and $B_s$ mesons into final states containing
light vector mesons.

\section*{Acknowledgments}

The work of P.B.\ is supported in part by the EU networks
contract Nos.\ MRTN-CT-2006-035482, {\sc Flavianet}, and
MRTN-CT-2006-035505, {\sc Heptools}.

\appendix

\section*{Appendices}
\renewcommand{\theequation}{\Alph{section}.\arabic{equation}}
\renewcommand{\thetable}{\Alph{table}}
\renewcommand{\thefigure}{\Alph{figure}}
\setcounter{section}{0}
\setcounter{table}{0}
\setcounter{figure}{0}

\section{Non-Local Operator Identities}\label{app:A}
\setcounter{equation}{0}

For completeness, we quote the following non-local operator identities
from Refs.~\cite{BBKT,BZ06a}:
\begin{eqnarray}
\frac{\partial}{\partial x_\mu}\, \bar q(x)\gamma_\mu s(-x)
& = &{} - i \int_{-1}^1 dv\, v \bar q (x) x_\alpha
gG_{\alpha\mu}(vx) \gamma_\mu s(-x)\nonumber\\
& &  + i(m_s+m_q) \bar q(x) s(-x),\label{eq:oprel1}\\
\partial_\mu \{\bar q(x)\gamma_\mu s(-x)\}
& = & {} - i\int_{-1}^1 dv\, \bar q(x) x_\alpha
gG_{\alpha\mu}(vx) \gamma_\mu s(-x)\nonumber\\
& & {} - i(m_s-m_q)\bar q(x) s(-x),\label{eq:oprel2}\\
\partial_\mu \bar q(x) \sigma_{\mu\nu} s(-x) & = &
-i\,\frac{\partial\phantom{x_\nu}}{\partial x_\nu} \,\bar q(x)
s(-x) + \int_{-1}^1 dv\, v \bar q(x)
x_\rho gG_{\rho\nu}(vx)s(-x)\nonumber\\
& & {} -i\int_{-1}^1 dv\, \bar q(x) x_\rho gG_{\rho\mu}(vx)
\sigma_{\mu\nu} s(-x)\nonumber\\
& &  - (m_s+m_q) \bar q(x) \gamma_\nu s(-x),\label{eq:oprel3}\\
\frac{\partial\phantom{x_\nu}}{\partial x_\mu} \,\bar q(x)
\sigma_{\mu\nu} s(-x) & = & -i\partial_\nu \bar q(x)
s(-x) +
\int_{-1}^1 dv\, \bar q(x) x_\rho gG_{\rho\nu}(vx)s(-x)\nonumber\\
& & {} -
i\int_{-1}^1 dv\, v \bar q(x) x_\rho gG_{\rho\mu}(vx)
\sigma_{\mu\nu} s(-x)\nonumber\\
& & {}+ (m_s-m_q) \bar q(x) \gamma_\nu  s(-x).
\label{eq:oprel4}
\end{eqnarray}
Here $\partial_\mu$ is the total derivative defined as
$$
\partial_\mu \left\{ \bar q(x)\Gamma s(-x)\right\} \equiv
\left.\frac{\partial}{\partial y_\mu}\,\left\{ \bar q(x+y) [x+y,-x+y]
    \Gamma s(-x+y)\right\}\right|_{y\to 0}.
$$
By taking matrix elements of the above relations between the vacuum
and the meson state, one obtains exact integral
representations for those DAs that are not dynamically independent.

\section{Chiral-Even Twist-4 Parameters}\label{app:B}
\setcounter{equation}{0}

In this appendix we calculate the parameters $\zeta^\parallel_{4V}$
and $\widetilde\omega^\parallel_{4V}$ defined in Eqs.~(\ref{3.12}) and
(\ref{eq:w4A}). We shall obtain them from QCD sum rules,
using various correlation functions of either identical currents,
so-called {\em diagonal} sum rules, or different currents, so-called
{\em non-diagonal} sum rules. We shall further distinguish between
{\em pure-parity} (PP) and {\em mixed-parity} (MP) sum rules,
depending on the parity of hadronic states that contribute to these
correlation functions.

Let us first consider the non-diagonal correlation function
\begin{equation}\label{B1}
z^\mu z^\nu i\int d^4 y e^{-ipy} \langle 0 | T \bar q(z) g \widetilde
G_{\mu\alpha}(vz) \gamma^\alpha \gamma_5 s(0) \bar s(y) \gamma_\nu
q(y) |0\rangle = (pz)^2 \int {\mathcal D}\underline{\alpha}\, 
e^{-ipz(\alpha_2 + v \alpha_3)} \pi(\underline{\alpha}),
\end{equation}
where both currents have the same chirality. The integration measure
${\cal D}\underline{\alpha}$ is defined in (\ref{eq:measure}). This is a MP
correlation function, with both $J^P=1^-$ and $0^+$ states contributing.
We have calculated the OPE including condensates up to dimension 6:
\begin{eqnarray}
\pi(\underline{\alpha})  
&=& 
-\frac{\alpha_s}{2\pi^3}p^2 \ln \frac{\mu^2}{-p^2}
\alpha_1\alpha_2\alpha_3 
\Bigg\{\frac{1}{\alpha_1(1-\alpha_1)}+\frac{1}{\alpha_2(1-\alpha_2)}\Bigg\}
  -\frac{1}{6p^2} \left\langle \frac{\alpha_s}{\pi}G^2\right\rangle \, 
   \delta(\alpha_3)
 \nonumber\\&&
 +\frac{1}{3p^2}\frac{\alpha_s}{\pi}m_q\langle \bar q q\rangle
   \Big[\bar\alpha_3(\alpha_3-3)\delta(\alpha_2)-
   \alpha_3\bar\alpha_3\delta'(\alpha_2)\Big]\nonumber\\
&&
 +\frac{1}{3p^2}\frac{\alpha_s}{\pi}m_s\langle \bar s s\rangle
   \Big[\bar\alpha_3(\alpha_3-3)\delta(\alpha_1)-
   \alpha_3\bar\alpha_3\delta'(\alpha_1)\Big] 
\nonumber\\&&{}
   -\frac{2}{3p^2}\frac{\alpha_s}{\pi} \left[ \alpha_3^2 \ln 
\frac{\mu^2}{-p^2} -\alpha_3^2\ln (\alpha_3\bar \alpha_3) + 
\bar \alpha_3 \alpha_3\right]
\Big[m_s\langle \bar q q\rangle \delta(\alpha_2)+ 
m_q\langle \bar s s\rangle \delta(\alpha_1)\Big]
  \nonumber\\&&{} 
 +\frac{1}{p^4}\Big[\frac{8}{27}\pi\alpha_s \langle \bar s s\rangle^2
   + \frac{1}{3}
    m_s\langle \bar s\sigma g G s\rangle\Big]\delta(\alpha_1)\delta(\alpha_3)
\nonumber\\&&{}
 +\frac{1}{p^4}\Big[\frac{8}{27}\pi\alpha_s \langle \bar q q\rangle^2
   + \frac{1}{3}
    m_q\langle \bar q\sigma g G q\rangle\Big]\delta(\alpha_2)\delta(\alpha_3)
\nonumber\\&&{}
  -\frac{16}{9p^4}\pi\alpha_s \langle \bar q q\rangle \langle \bar s s
   \rangle \delta(\alpha_1)\delta(\alpha_2)\,.
\label{B2}
\end{eqnarray}
{}In the local limit and zero quark masses, the result agrees with the 
calculation in Ref.~\cite{BK86}.
In the product of $\delta$-functions
$\delta'(\alpha_{1,2})\delta(\alpha_1+\alpha_2+\alpha_3-1)$,
$\delta$ has to be integrated over before $\delta'$.

{}From (\ref{B2}) we obtain the following sum rules:
\begin{eqnarray}
(f_{K^*}^\parallel)^2 m_{K^*}^2 \zeta_{4K^*}^\parallel e^{-m_{K^*}^2/M^2}
& = & 
-\frac{\alpha_s}{18\pi^3}\, M^4
  \left\{1-\Gamma(2,s_0/M^2)\right\}
 + \frac{4}{9}\,\frac{\alpha_s}{\pi}\,\left( m_q \langle \bar q
  q\rangle + m_s \langle \bar s s\rangle\right) 
\nonumber\\
& & + \frac{2}{9}\,\frac{\alpha_s}{\pi}\,\left( m_s \langle \bar q
  q\rangle + m_q \langle \bar s s\rangle\right) \left\{ \frac{8}{3} +
\gamma_E - \ln
  \,\frac{M^2}{\mu^2} + \Gamma(0,s_0/M^2)\right\}
\nonumber\\
& & + \frac{1}{6}\,\left\langle \frac{\alpha_s}{\pi}\,G^2\right\rangle +
\frac{1}{3M^2}\, \left( m_q \langle\bar q gG\sigma q\rangle +  m_s
\langle\bar s gG\sigma s\rangle\right)
\nonumber\\  
&& + \frac{8\pi\alpha_s}{27M^2}\, \left\{  \langle\bar q q\rangle^2 +
\langle\bar s s\rangle^2 \right\} - \frac{16\pi\alpha_s}{9M^2 }\,
  \langle\bar q q\rangle\langle\bar s s\rangle\,,
\label{B3}\\
(f_{K^*}^\parallel)^2 m_{K^*}^2 
\widetilde\omega_{4K^*}^\parallel e^{-m_{K^*}^2/M^2}
& = & 
\frac{5\alpha_s}{2592\pi^3}\, M^4
  \left\{1-\Gamma(2,s_0/M^2)\right\}
 - \frac{19}{648}\,\frac{\alpha_s}{\pi}\,\left( m_q \langle \bar q
  q\rangle + m_s \langle \bar s s\rangle\right) 
\nonumber\\
& & + \frac{11}{324}\,\frac{\alpha_s}{\pi}\,\left( m_s \langle \bar q
  q\rangle + m_q \langle \bar s s\rangle\right) \left\{ \frac{8}{3} +
\gamma_E - \ln
  \,\frac{M^2}{\mu^2} + \Gamma(0,s_0/M^2)\right\}
\nonumber\\
& & - \frac{1}{27}\,\left\langle \frac{\alpha_s}{\pi}\,G^2\right\rangle -
\frac{2}{27M^2}\, \left( m_q \langle\bar q gG\sigma q\rangle +  m_s
\langle\bar s gG\sigma s\rangle\right)
\nonumber\\  
&& - \frac{16\pi\alpha_s}{243M^2}\, \left\{  \langle\bar q q\rangle^2 +
\langle\bar s s\rangle^2 \right\} - \frac{40\pi\alpha_s}{81M^2 }\,
  \langle\bar q q\rangle\langle\bar s s\rangle\,,
\label{B4}
\end{eqnarray}
where
$$\Gamma(a,s_0/M^2) = \frac{1}{(M^2)^a} \,\int_{s_0}^\infty ds s^{a-1}
e^{-s/M^2}.$$
(\ref{B3}) follows from (\ref{B2}) by integration over ${\cal
  D}\underline{\alpha}$ with the weight factor 1, and (\ref{B4}) by integration
with  weight factor $(\alpha_3-4/9)/2$, see Eq.~(\ref{eq:w4A}).
The sum rules for $\rho$ are obtained by letting $s\to q$ and those
for $\phi$ by letting $q\to s$. 

\begin{figure}
$$\epsfxsize=0.48\textwidth\epsffile{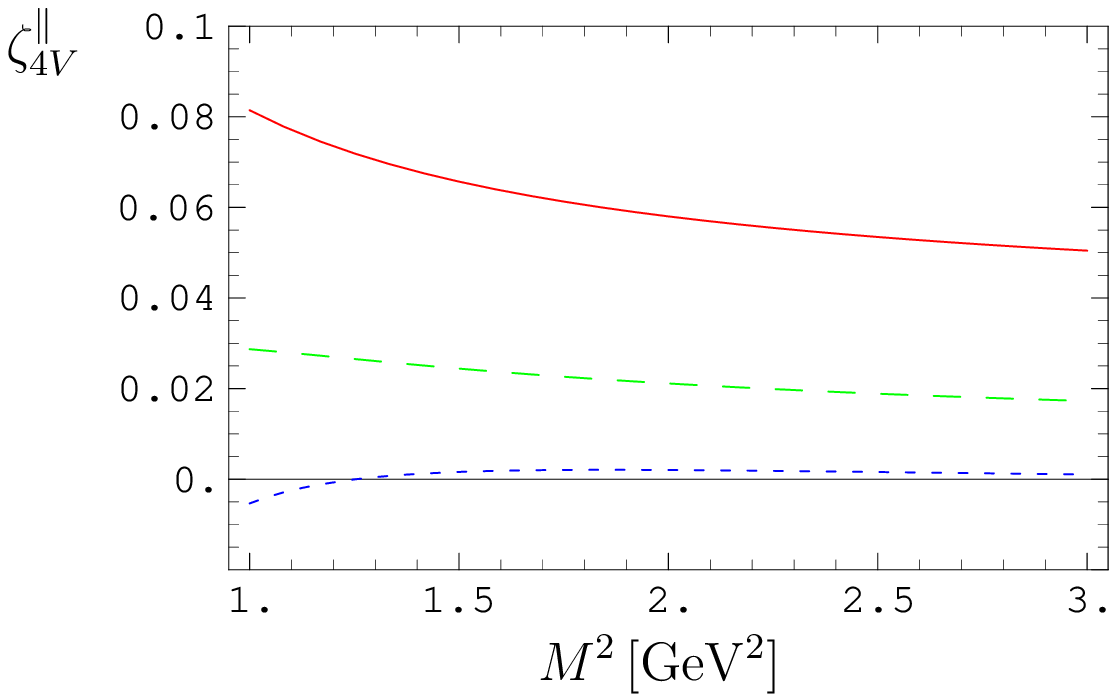}\quad
\epsfxsize=0.48\textwidth\epsffile{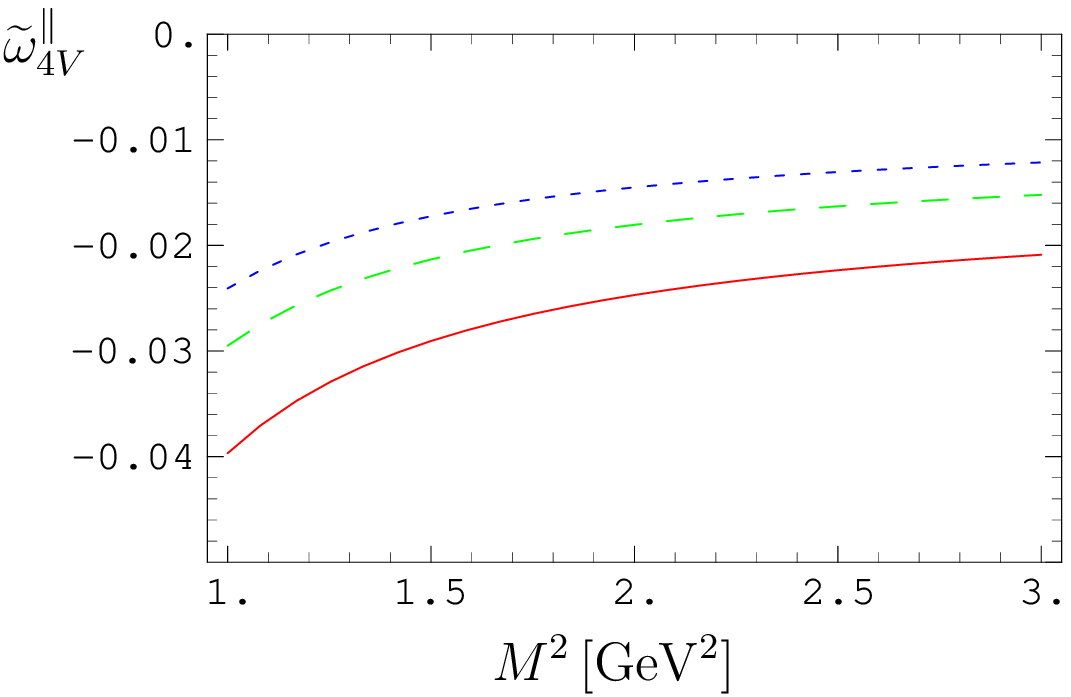}$$
\vskip-15pt
\caption[]{\small [Colour online] $\zeta_{4V}^\parallel$ (left) and
  $\widetilde\omega_{4V}^\parallel$ (right) from the non-diagonal MP sum
  rules (\ref{B3}) and (\ref{B4}) as functions of
  $M^2$, for central values of the input parameters. 
Solid [red] lines: $\rho$ ($s_0 = 1.5\,{\rm GeV}^2$), 
long dashes [green]: $K^*$ ($s_0 = 1.8\,{\rm GeV}^2$), short dashes [blue]:
$\phi$ ($s_0 = 2\,{\rm GeV}^2$). All parameters are evaluated at the
  scale $\mu=1\,$GeV.}\label{figA}
\end{figure}

We evaluate the above sum rules using the input parameters collected
in Tabs.~\ref{tab1} and \ref{tab:cond}; the results, for central
values of the input parameters, are shown in Fig.~\ref{figA}.
\begin{table}[tb]
\renewcommand{\arraystretch}{1.3}
\addtolength{\arraycolsep}{3pt}
$$
\begin{array}{r@{\:=\:}l||r@{\:=\:}l}
\hline
\quark & (-0.24\pm0.01)^3\,\mbox{GeV}^3 & \squark & (1-\delta_3)\,\quark\\
\mixed & m_0^2\,\quark &  \smixed & (1-\delta_5)\mixed\\[6pt]
\displaystyle \gluon & (0.012\pm 0.006)\,
{\rm GeV}^4 & \langle g^3 f G^3 \rangle & (0.08\pm 0.02)\,{\rm
  GeV}^6\quad \protect{\cite{G3}}\\[6pt]\hline
\multicolumn{4}{c}{m_0^2 = (0.8\pm 0.1)\,{\rm GeV}^2,\quad \delta_3
  = 0.2\pm 0.2, \quad \delta_5 = 0.2\pm 0.2}\\\hline
\multicolumn{4}{c}{\overline{m}_s(2\,\mbox{GeV}) = (100\pm
20)\,\mbox{MeV}~~~\longleftrightarrow~~~ \overline{m}_s(1\,\mbox{GeV})
= (133\pm 27)\,\mbox{MeV}}\\\hline
\multicolumn{4}{c}{\alpha_s(m_Z) = 0.1176\pm 0.002  ~\longleftrightarrow~
\alpha_s(1\,\mbox{GeV}) = 0.497\pm0.005}\\\hline
\end{array}
$$
\renewcommand{\arraystretch}{1}
\addtolength{\arraycolsep}{-3pt}
\vskip-10pt
\caption[]{\small Input parameters for sum rules at the
  renormalization scale $\mu=1\,$GeV. The value of $m_s$ is obtained
  from unquenched lattice calculations with $N_f=2$ light quark flavours
as summarized in Ref.~\cite{mslatt}, which agrees with the 
results from QCD sum-rule calculations \cite{jamin}. 
$\alpha_s(m_Z)$ is the PDG
  average \cite{PDG}.}\label{tab:cond}
\end{table}
The sum rules are dominated by the contribution of the gluon
condensate; the impact of the specific value of the continuum
threshold $s_0$ is only moderate. The figure also shows that the
impact of SU(3)-breaking is very relevant: the contributions of the
quark and mixed condensate reduce the values of $\zeta_{4V}^\parallel$
and $\widetilde\omega_{4V}^\parallel$. In the Borel-window $1\,{\rm
  GeV}^2 < M^2 < 2\,{\rm GeV}^2$, and including the input-parameter
uncertainties given in Tab.~\ref{tab:cond}, we find, at the scale
$\mu=1\,$GeV: 
\begin{eqnarray}
\zeta_{4\rho}^\parallel & = & 0.07\pm 0.03\,,\qquad 
\widetilde\omega_{4\rho}^\parallel~~ =  -0.03\pm 0.01\,,\nonumber\\
\zeta_{4K^*}^\parallel & = & 0.02\pm 0.02\,,\qquad 
\widetilde\omega_{4K^*}^\parallel  =  -0.02\pm 0.01\,,\nonumber\\
\zeta_{4\phi}^\parallel & = & 0.00\pm 0.02\,,\qquad 
\widetilde\omega_{4\phi}^\parallel~~  =  -0.02\pm 0.01\,.\label{resdiagonal}
\end{eqnarray}
We have added in quadrature all individual sources of uncertainty. The
total error is dominated by that of the gluon condensate.

For $\zeta_{4K^*}^\parallel$, we also consider diagonal sum rules which
can be obtained from the correlation function
\begin{eqnarray}
\Pi_{\mu\nu}^{V} &=& i\int\! d^4 x\, e^{ipx}\,
  \bra 0| T J_\mu^{V}(x)  (J_\nu^{V})^\dagger(0)|0\ket\, 
    = \, p_\mu p_\nu \,\Pi_{0}^{V}(p^2) -g_{\mu\nu} \, \Pi_{1}^{V}(p^2)\,, 
\label{4-2}
\end{eqnarray}
with the current $J_\mu^V = 
\bar q \, g\widetilde G_{\mu\alpha}\gamma_\alpha\gamma_5 s$. For $\rho$,
$\Pi_{0,1}^V$ was calculated in Ref.~\cite{BK86}, while the
SU(3)-corrections were calculated in  Ref.~\cite{BBL06}, including
contributions from condensates up to dimension 8.
The suitability of this correlation function for extracting
$\zeta_{4K^*}^\parallel$ is not immediately obvious: 
$\Pi_0^V$ contains contributions not only of vector mesons, but also
of hybrid $0^+$ mesons, whose coupling to $J_\mu^V$ is much larger
than that of the $K^*$, ruling out the possibility to construct a
MP sum rule for $\zeta_{4K^*}^\parallel$. This situation is
qualitatively different from that of the non-diagonal correlation
function (\ref{B1}), where the presence of $\bar s \gamma_\nu q$
removes all contributions from hybrid mesons.
We hence focus on the PP function $\Pi_{1}^{V}$. 
From Ref.~\cite{BBL06}, we quote
\begin{eqnarray}
\Pi_{1}^{V} 
&=& 
\frac{\alpha_s}{240 \pi^3}\, p^6 \ln \frac{\mu^2}{-p^2} -
                 \frac{1}{36} \gluon p^2\ln \frac{\mu^2}{-p^2}
\nonumber\\
&&{}+  \frac{\alpha_s}{6\pi}\big[m_q \quark + 
m_s \squark\big] p^2\ln \frac{\mu^2}{-p^2}
    +  \frac{\alpha_s}{18\pi}\big[m_s \quark + 
m_q \squark\big] p^2\ln \frac{\mu^2}{-p^2}
\nonumber\\
&&{}+ \frac{8\pi\alpha_s}{9} \quark\squark - 
\frac{1}{192\pi^2}\, \bra g^3 f G^3\ket
\nonumber\\
&&{} -\frac{19}{144} \frac{\alpha_s}{\pi} 
\big[m_q \mixed + m_s \smixed\big]\ln \frac{\mu^2}{-p^2} 
\nonumber\\
&&{} - \frac{19}{144}
\frac{\alpha_s}{\pi}\big[m_s \mixed + m_q \smixed\big]\ln \frac{\mu^2}{-p^2}
\nonumber\\
&&{}   +\frac{25\pi\alpha_s}{162p^2} m_0^2\big[\quark^2 + \squark^2\big]
    - \frac{181\pi\alpha_s}{162p^2} m_0^2 \quark\squark
\nonumber\\
&&{} + \frac{\pi}{18p^2}\gluon \big[m_q \quark + m_s \squark\big]
     + \frac{\pi}{6p^2}\gluon \big[m_s \quark + m_q \squark\big]\,.         
\label{4-Pi0}                 
\end{eqnarray} 
The PP sum rule for $\zeta_{4K^*}^\parallel$ is
\begin{equation}\label{d1}
\left(f_K^\parallel\right)^2 m_{K^*}^6 (\zeta_{4K^*}^\parallel)^2\,
e^{-m_{K^*}^2/M^2} = {\cal B}_{\rm sub} \Pi_1^V\,,
\end{equation}
where ${\cal B}_{\rm sub}\Pi_1^V$ is the continuum-subtracted 
Borel transform of $\Pi_1^V$, which we define as
$$
{\cal B}_{\rm sub} \int_0^\infty ds \,\frac{\rho(s)}{s-p^2} =
\int_0^{s_0} ds \,e^{-s/M2} \rho(s)\,,
$$ 
in terms of the dispersive representation of $\Pi_1^V$.
For $\rho$, the above sum rule was derived
and analysed in Ref.~\cite{BK86}. It features a large negative
contribution from the gluon condensate which, for Borel parameters
$M^2\sim 2\,{\rm GeV}^2$ and continuum thresholds $s_0$ between 1.3
and 3$\,{\rm GeV}^2$, drives the right-hand side of (\ref{d1})
negative. In Ref.~\cite{BK86} it was argued that this large negative
contribution signals the presence of a larger mass scale $\sim 2\,$GeV
in the spectral density and is indicative for a breakdown of
quark-hadron duality, at least if the usual simple continuum model
with only one resonance, the $\rho$, is used. A remedy is the use of a
more appropriate continuum model including higher mass
states like $\rho(1450)$. This automatically increases $s_0$, but it
also turns out that the coupling of $\rho(1450)$ to the gluonic
current $J_\mu^V$ is larger than that of $\rho(770)$, which does not
really help the
determination of $\zeta_{4\rho}^\parallel$. An alternative is to
analyse (\ref{d1}) for small $M^2\approx 1\,{\rm GeV}^2$, where
duality still works and the suppression of higher-mass resonances is
effective. Numerically, the sum rule is then dominated by the gluon
and the dimension-8 condensate $\langle\bar q q\rangle\langle \bar q
\sigma gG q\rangle$. Clearly such a sum rule cannot give an accurate
estimate of $\zeta_{4V}^\parallel$, so we only use it as a consistency check
for the results obtained from (\ref{B3}) and, in particular, the large
SU(3) breaking.
\begin{figure}
$$\epsfxsize=0.48\textwidth\epsffile{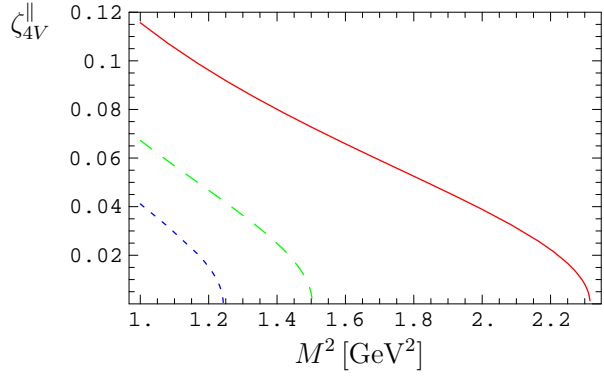}$$
\vskip-15pt
\caption[]{\small  [Colour online] $\zeta_{4V}^\parallel$ from
  (\ref{d1}). Solid [red] line:
  $\rho$, long dashes [green]: $K^*$, short dashes [blue]: $\phi$. Same input
  parameters as in Fig.~\ref{figA}.}\label{figB}
\end{figure}
The results from (\ref{d1}) are shown in Fig.~\ref{figB}. Note that
the breakdown of duality sets in the earlier, the heavier the
meson. Although it is not possible to extract precise values for
$\zeta_{4V}^\parallel$, we see that the values are not inconsistent
with the results from the non-diagonal sum rule, (\ref{resdiagonal}),
and that in particular the relative hierarchy, $\zeta_{4\rho}^\parallel >
\zeta_{4K^*}^\parallel > \zeta_{4\phi}^\parallel$, is reproduced.

Our final results are given in (\ref{resdiagonal}).
A comparison with  earlier determinations is
given in Sec.~\ref{sec:5}.

\section{Chiral-odd Twist-4 Parameters}\label{app:C}
\setcounter{equation}{0}

In this appendix we calculate 
\begin{equation}\label{C.0}
\zeta_{\pm}^\perp \equiv \zeta_{4K^*}^\perp \pm
\widetilde\zeta_{4K^*}^\perp\,.
\end{equation}
Like for $\zeta_{4K^*}^\parallel$, we
consider both non-diagonal and diagonal sum rules -- the former for
all mesons, the latter only for $\rho$. We also consider PP
and MP sum rules. To distinguish between the
results of these sum rules, the following notation proves convenient:
\begin{equation}\label{notation}
\left. \zeta_\pm^\perp\right|_{{\rm D(ND)},{\rm PP(MP)}}\,.
\end{equation}

\begin{figure}
$$\epsfxsize=0.48\textwidth\epsffile{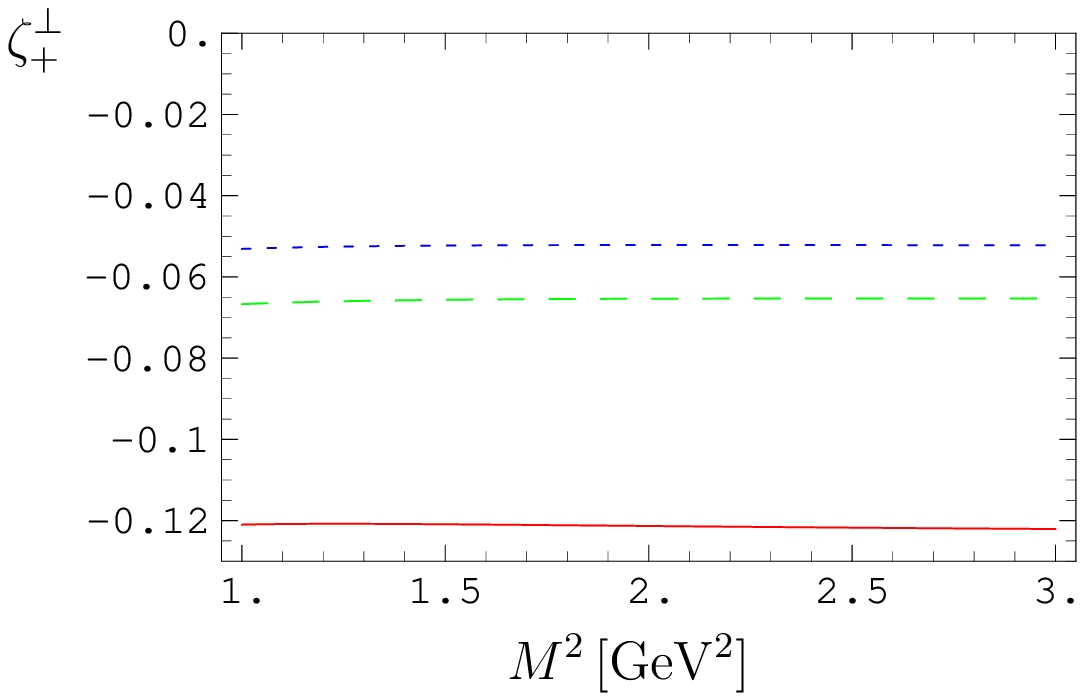}\quad
\epsfxsize=0.48\textwidth\epsffile{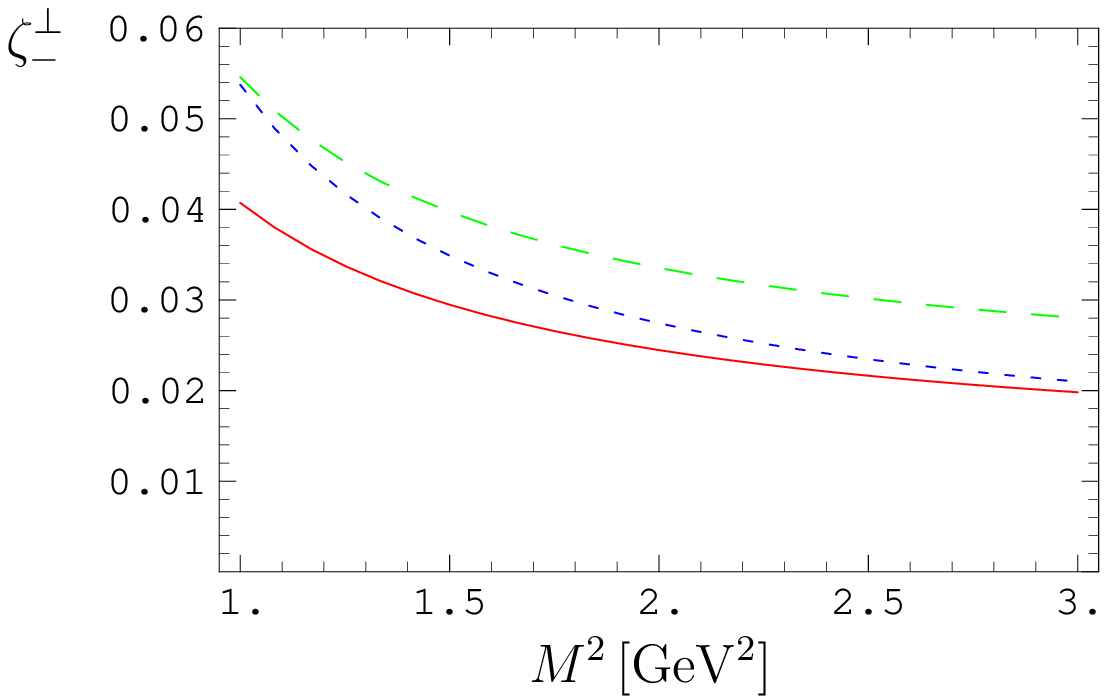}$$
\vskip-15pt
\caption[]{\small [Colour online] 
$\left.\zeta_+^\perp\right|_{\rm ND,PP}$ (left) 
and $\left.\zeta_-^\perp\right|_{\rm ND,PP}$ (right) from the
PP sum rule (\ref{pp1}), as functions of $M^2$, for
$\mu=1\,$GeV and  central
values of input parameters. Solid [red] lines: $\rho$ ($s_0=1.2\,{\rm
  GeV}^2$), long dashes [green]: $K^*$ ($s_0=1.4\,{\rm GeV}^2$), short
dashes [blue]:
$\phi$ ($s_0=1.8\,{\rm GeV}^2$).}\label{figC}
\end{figure}

Let us start with the non-diagonal sum rules for $\zeta_\pm^\perp$, yielding
$\left. \zeta_\pm^\perp\right|_{{\rm ND},{\rm PP(MP)}}$. 
The relevant correlation function is 
\begin{eqnarray}
\Pi^\pm_{\alpha\beta\mu\nu} &=& i \int d^4y e^{-ipy} \langle 0 | 
T [\bar q g(G_{\mu\nu} \pm i
\widetilde G_{\mu\nu}\gamma_5) s](0) [\bar s \sigma_{\alpha\beta}
q](y) | 0 \rangle
\nonumber\\
& = & \frac{1}{p^2}\Big\{\left[(p_\mu g_{\nu\alpha}
  p_\beta) 
  - ( \mu\leftrightarrow \nu)\right] - \left[ \alpha\leftrightarrow
  \beta \right] \Big\}\Pi^{\pm}_{V}\nonumber\\
&&{}+\frac{1}{p^2}\Big\{\left[(p_\mu g_{\nu\alpha}
  p_\beta) 
  - ( \mu\leftrightarrow \nu)\right] - \left[ \alpha\leftrightarrow
  \beta \right] + p^2 (g_{\alpha\mu} g_{\beta\nu} - g_{\alpha\nu}
g_{\beta\mu}) \Big\}\Pi^{\pm}_{A}.\label{C.2}
\end{eqnarray}
The invariant functions $\Pi^{\pm}_{V}$ and $\Pi^{\pm}_{A}$ contain 
contributions of  $1^-$ and $1^+$ mesons,
respectively, and can be separated by considering the two
projections
\begin{eqnarray}
(pz)^2 \Pi^{\pm}_1 
&\equiv
& z^\mu z^\alpha g^{\nu\beta}\Pi^\pm_{\alpha\beta\mu\nu} = 
          -2 \frac{(pz)^2}{p^2}\Big[\Pi^{\pm}_{V} + \Pi^{\pm}_{A}\Big],
\nonumber\\
    \phantom{(pz)^2} \Pi^{\pm}_2 
&\equiv
& g^{\mu\alpha}g^{\nu\beta}\Pi^\pm_{\alpha\beta\mu\nu} =
       -6 \Big[\Pi^{\pm}_{V} - \Pi^{\pm}_{A}\Big].        
\label{eq:project22}
\end{eqnarray}
In calculating $\Pi^\pm_{1,2}$, we use dimensional regularization and
the identity
\begin{equation}
   G_{\mu\nu} - i \widetilde G_{\mu\nu}\gamma_5 
= \frac{1}{4}\{\sigma_{\mu\nu},\sigma_{\rho\sigma}\} G^{\rho\sigma} 
\end{equation}
in order to avoid ambiguities with the definition of the
epsilon-tensor and the $\gamma_5$ matrix in $d$ dimensions. 
Here $\{\ldots,\ldots\}$ denotes the anti-commutator.
We find
\begin{eqnarray}
\Pi^-_V 
&=& 
 - \frac{\alpha_s}{48\pi^3} p^4 \ln \frac{\mu^2}{-p^2}
 -\frac{1}{3}\frac{\alpha_s}{\pi}\Big[m_s\langle \bar q q\rangle + 
                    m_q\langle \bar s s\rangle\Big] \ln \frac{\mu^2}{-p^2}
 - \frac{8\pi}{9p^2}\alpha_s \langle \bar s s \rangle \langle \bar q
 q\rangle
 - \frac{1}{24}\,\left\langle\frac{\alpha_s}{\pi}\,G^2\right\rangle
\nonumber\\
&&{}
    +\frac{1}{12p^2 }\Big[m_q\langle \bar q\sigma g G q\rangle + 
      m_s\langle \bar s\sigma g G s\rangle\Big]
    + \frac{1}{6p^2}\Big[m_q\langle \bar s\sigma g G s\rangle + 
     m_s\langle \bar q\sigma g G q\rangle   \Big]
\nonumber\\
&&{}-\frac{\alpha_s}{9\pi}\,\left[m_s \langle\bar q q\rangle + m_q
  \langle\bar s s\rangle \right]
    - \frac{\alpha_s}{6\pi}\,\left[m_s \langle\bar s s\rangle + m_q
  \langle\bar q q\rangle \right],
\nonumber\\
\Pi^-_A 
&=& - 
   \frac{\alpha_s}{48\pi^3} p^4 \ln \frac{\mu^2}{-p^2}
   +\frac{1}{3}\frac{\alpha_s}{\pi}\Big[m_s\langle \bar q q\rangle + 
    m_q\langle \bar s s\rangle\Big] \ln \frac{\mu^2}{-p^2}
   +\frac{8\pi}{9p^2}\alpha_s \langle \bar s s \rangle \langle 
   \bar q q\rangle
   - \frac{1}{24}\,\left\langle\frac{\alpha_s}{\pi}\,G^2\right\rangle
\nonumber\\
&&{}
    +\frac{1}{12p^2 }\Big[m_q\langle \bar q\sigma g G q\rangle + 
    m_s\langle \bar s\sigma g G s\rangle\Big]
    - \frac{1}{6p^2}\Big[m_q\langle \bar s\sigma g G s\rangle + 
    m_s\langle \bar q\sigma g G q\rangle   \Big]
\nonumber\\
&&{}-\frac{\alpha_s}{9\pi}\,\left[m_s \langle\bar q q\rangle + m_q
  \langle\bar s s\rangle \right]
    - \frac{\alpha_s}{6\pi}\,\left[m_s \langle\bar s s\rangle + m_q
  \langle\bar q q\rangle \right],
\nonumber\\
\Pi^+_V 
&=& 
   - \frac{\alpha_s}{72\pi^3} p^4 \ln \frac{\mu^2}{-p^2}
   - \frac{1}{12} \left\langle \frac{\alpha_s}{\pi}G^2\right\rangle 
     \ln \frac{\mu^2}{-p^2}
   - \frac{1}{9}\frac{\alpha_s}{\pi}\Big[m_s\langle \bar q q\rangle + 
     m_q\langle \bar s s\rangle\Big] \ln \frac{\mu^2}{-p^2}
\nonumber\\
&&{}
 +\frac{1}{9}\frac{\alpha_s}{\pi}\Big[m_q\langle \bar q q\rangle + 
 m_s\langle \bar s s\rangle\Big] \ln \frac{\mu^2}{-p^2}
 - \frac{8\pi}{9p^2}\alpha_s \langle \bar s s \rangle \langle \bar q q\rangle
 + \frac{8\pi}{27p^2}\alpha_s \Big[\langle \bar s s\rangle^2 + 
 \langle \bar q q \rangle^2\Big]
\nonumber\\
&&{} 
    +\frac{1}{12p^2 }\Big[m_q\langle \bar q\sigma g G q\rangle + 
    m_s\langle \bar s\sigma g G s\rangle\Big]
    - \frac{5\alpha_s}{27\pi}\,\left[ m_s\langle\bar q q\rangle + m_q
      \langle \bar s s\rangle \right],
\nonumber\\
\Pi^+_A 
&=& + \frac{\alpha_s}{72\pi^3} p^4 \ln \frac{\mu^2}{-p^2}
    + \frac{1}{12} \left\langle \frac{\alpha_s}{\pi}G^2\right\rangle 
      \ln \frac{\mu^2}{-p^2}
    -\frac{1}{9}\frac{\alpha_s}{\pi}\Big[m_s\langle \bar q q\rangle + 
     m_q\langle \bar s s\rangle\Big] \ln \frac{\mu^2}{-p^2}
\nonumber\\
&&{}
    -\frac{1}{9}\frac{\alpha_s}{\pi}\Big[m_q\langle \bar q q\rangle + 
    m_s\langle \bar s s\rangle\Big] \ln \frac{\mu^2}{-p^2}
    - \frac{8\pi}{9p^2}\alpha_s \langle \bar s s \rangle \langle 
      \bar q q\rangle
    - \frac{8\pi}{27p^2}\alpha_s \Big[\langle \bar s s\rangle^2 + 
      \langle \bar q q \rangle^2\Big]
\nonumber\\
&&{} 
    -\frac{1}{12p^2 }\Big[m_q\langle \bar q\sigma g G q\rangle + 
    m_s\langle \bar s\sigma g G s\rangle\Big]
    - \frac{5\alpha_s}{27\pi}\,\left[ m_s\langle\bar q q\rangle + m_q
      \langle \bar s s\rangle \right].
\end{eqnarray}
Again, the correlation functions for the $\rho$ meson are obtained by
$s\to q$, and those for $\phi$ by $q\to s$.
The above functions allow one to derive the PP sum rules
\begin{eqnarray}\label{pp1}
\left( f_{K^*}^\perp\right)^2 m_{K^*}^4
\left.\zeta_{\pm}^\perp\right|_{\rm ND,PP}
e^{-m_{K^*}^2/M^2} & = & {\cal B}_{\rm sub}\, \Pi_V^{\pm}\,,
\end{eqnarray}
and correspondingly for axial-vector mesons. 
As discussed in Ref.~\cite{BZ05}, there are actually two strange $1^+$  
mesons, $K_1(1270)$ and $K_1(1400)$, which are usually
interpreted as mixture of a ${}^3 P_1$ state, the $K_a$, and a
${}^1 P_1$ state, the $K_b$ \cite{Suzuki,Goldman}:
\begin{eqnarray*}
K_1(1270) & = & K_a \cos \theta_K - K_b \sin \theta_K,\\
K_1(1400) & = & K_a \sin \theta_K + K_b \cos \theta_K.
\end{eqnarray*}
The results of Refs.~\cite{Suzuki,Goldman} indicate that the system is
close to ideal mixing, i.e.\ $\theta_K \approx 45^\circ$. To the
accuracy needed here it is then sufficient to replace the
two resonances by one effective one with the mass $m_{K_1} = 1.34\,$GeV
\cite{Goldman}. We find that the sum rules
for $\zeta_+^\perp$ and its axial-vector equivalent are dominated by
the gluon condensate contribution, which implies
\begin{equation}
(f_{K_1}^\perp)^2 m_{K_1}^4 \left.\zeta^\perp_+(K_1)\right|_{\rm ND,PP}\approx 
-(f_{K^*}^\perp)^2 m_{K^*}^4 \left.\zeta^\perp_+\right|_{\rm ND,PP}
\end{equation}
with $\sim 30\%$ accuracy. For $\zeta_-^\perp$, no single contribution
is dominant, but one still finds that $\zeta_-^\perp$ and
$\zeta_-^\perp(K_1)$ have opposite sign. This is similar to the
situation with PP sum rules for the G-odd twist-4 parameters
$\kappa_{4K^*}^\perp$ and $\kappa_{4K_1}^\perp$ discussed in
Ref.~\cite{BZ06a}, and has some impact on the relative size of
continuum contributions in PP vs.\ MP sum rules. We
will come back to this point below. In Fig.~\ref{figC}, we show the
results for $\left.\zeta_{\pm}^\perp\right|_{\rm ND,PP}$.
The values of $s_0$ are chosen in such a way as to ensure maximum
stability of $\left.\zeta_+^\perp\right|_{\rm ND,PP}$ in the Borel
parameter. Note that the sum
rules (\ref{pp1}) are quite sensitive to the value of the continuum
threshold, which we vary by $\pm 0.3\,{\rm GeV}^2$. 
Including this uncertainty and the variation in $M^2$, in the interval
$1\,{\rm GeV}^2<M^2<2\,{\rm GeV}^2$, and the error
induced by the hadronic input
parameters, Tabs.~\ref{tab2} and \ref{tab:cond}, we find, at the scale
$\mu=1\,$GeV:
\begin{eqnarray}
\left.\zeta_+^\perp(\rho)\right|_{\rm ND,PP} & = & -0.12\pm 0.04,\quad 
\left.\zeta_-^\perp(\rho)\right|_{\rm ND,PP}~~ = 0.03\pm 0.02\,,
\nonumber\\
\left.\zeta_+^\perp(K^*)\right|_{\rm ND,PP} & = & -0.07\pm 0.02,\quad 
\left.\zeta_-^\perp(K^*)\right|_{\rm ND,PP} = 0.04\pm 0.02\,,
\nonumber\\
\left.\zeta_+^\perp(\phi)\right|_{\rm ND,PP} & = & -0.05\pm 0.02,\quad 
\left.\zeta_-^\perp(\phi)\right|_{\rm ND,PP}~~ = 0.04 \pm 0.02\,.
\label{C.9}
\end{eqnarray}
We have added all individual uncertainties in quadrature. 
The bulk of SU(3) breaking in these quantities is due to the factor
$m_V^4 (f_V^\perp)^2$ in (\ref{pp1}), with $m_{K^*}^4
(f_{K^*}^\perp)^2/(m_\rho^4 (f_\rho^\perp)^2) = 2.2$ and $m_{\phi}^4
(f_{\phi}^\perp)^2/(m_\rho^4 (f_\rho^\perp)^2) = 3.8$, which explains
the relative hierarchy
$|\zeta_+^\perp(\rho)|>|\zeta_+^\perp(K^*)|>|\zeta_+^\perp(\phi)|$.
For $\zeta_-^\perp$, one has a cancellation of several terms which renders the
interpretation of the hierarchy of the 
curves in Fig.~\ref{figC} less clear-cut. 

Let us now turn to the extraction of $\zeta_\pm^\perp$ from
MP sum rules, which contain contributions from both vector
and axial-vector mesons. These sum rules are derived from the
combinations $(\Pi_V^\pm + \Pi_A^\pm)/p^2$ and, thanks to the factor
$1/p^2$, benefit from a smaller mass dimension than the PP sum
rules:
\begin{equation}\label{mp}
(f_{K^*}^\perp)^2 m_{K^*}^2 \left.\zeta^\perp_\pm\right|_{\rm ND,MP} 
e^{-m_{K^*}^2/M^2} = {\cal
    B}_{\rm sub}\, \frac{1}{p^2}\, \left(\Pi_V^\pm + \Pi_A^\pm\right).
\end{equation}
\begin{figure}
$$\epsfxsize=0.48\textwidth\epsffile{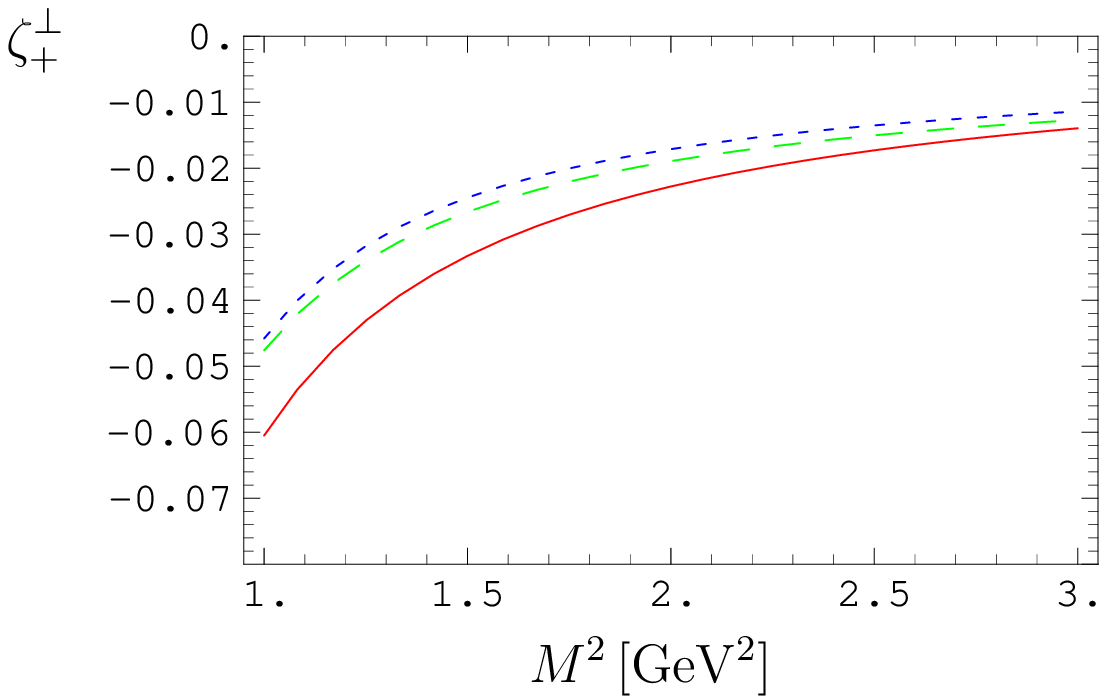}\quad
\epsfxsize=0.48\textwidth\epsffile{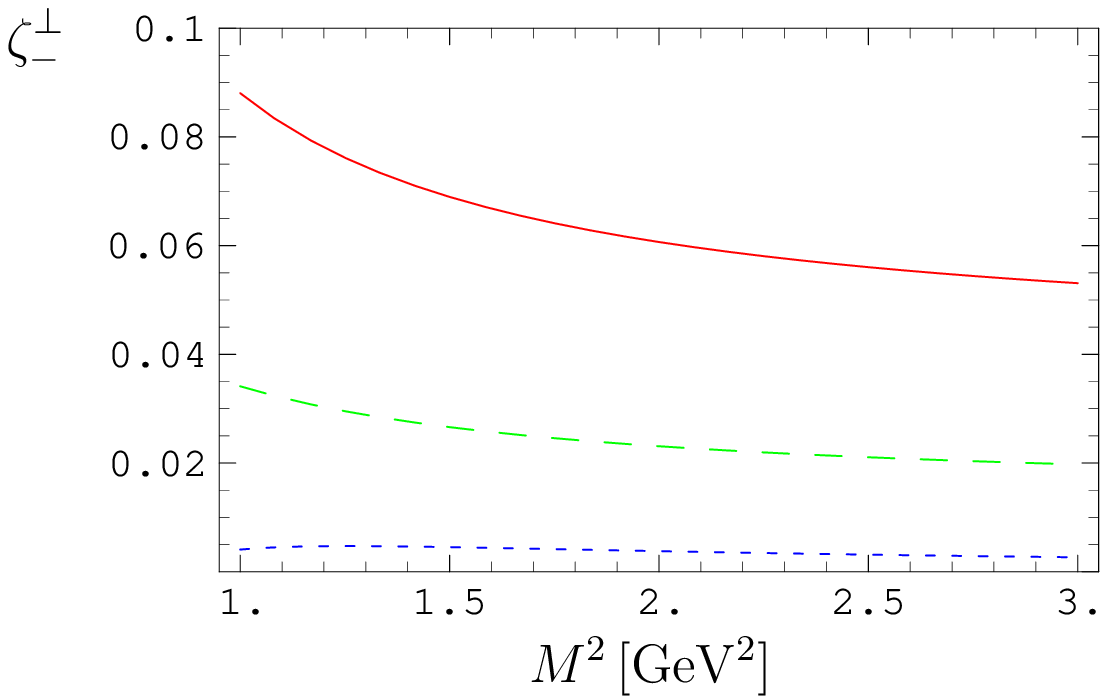}$$
\vskip-15pt
\caption[]{\small [Colour online] 
$\left.\zeta_+^\perp\right|_{\rm ND,MP}$ (left) and 
$\left.\zeta_-^\perp\right|_{\rm ND,MP} $ (right) from the
  MP sum rule (\ref{mp}), as functions of $M^2$, for
$\mu=1\,$GeV and  central
values of input parameters. Solid [red] lines: $\rho$ ($s_0=1.0\,{\rm
  GeV}^2$), long dashes [green]: $K^*$ ($s_0=1.2\,{\rm GeV}^2$), short
dashes [blue]:
$\phi$ ($s_0=1.5\,{\rm GeV}^2$).}\label{figD}
\end{figure}
The corresponding results are shown in Fig.~\ref{figD}. The sum rule for
$\left.\zeta_+^\perp\right|_{\rm ND,MP}$ consists of only two terms:
the quark-condensate and the four-quark-condensate contributions, with
the latter dominant. Such a sum rule, sensitive to higher-dimensional
condensates, is not reliable and we do
not include its results into our final value for $\zeta_+^\perp$.  
$\left.\zeta_-^\perp\right|_{\rm ND,MP}$, on the other hand, does not
receive any contribution from the four-quark condensate, and is
dominated by the gluon condensate. 
In contrast to
$\left.\zeta_\pm^\perp\right|_{\rm ND,PP}$,
$\left.\zeta_\pm^\perp\right|_{\rm ND,MP}$ is rather insensitive
to the precise value of the continuum threshold $s_0$. The reason for
this is the different sign of vector and axial-vector contributions to
the hadronic side of the sum rule: as mentioned above, the $1^+$ and
$1^-$ matrix elements tend to have different sign, and hence the resonance
contributions to the sum rule tend to cancel, reducing the size of the
continuum contribution. We choose $s_0$ for the MP sum rules slightly
below that for PP sum rules, to account for the lower mass of the
$1^+$ ground state as compared to the first $1^-$ excitation. 
Using again the hadronic input parameters from Tab.~\ref{tab2} and
Tab.~\ref{tab:cond}, and varying $s_0$ by $\pm 0.3\,{\rm GeV}^2$, and $M^2$
in the window $1\,{\rm GeV}^2 < M^2 < 2\,{\rm GeV}^2$, 
we obtain the following results for 
$\left.\zeta_-^\perp\right|_{\rm ND,MP}$ (at the scale $\mu=1\,$GeV):
\begin{equation}\label{C.10}
\left.\zeta_-^\perp(\rho)\right|_{\rm ND,MP} =  0.07\pm 0.03\,,\quad
\left.\zeta_-^\perp(K^*)\right|_{\rm ND,MP} =  0.03\pm 0.02\,,\quad
\left.\zeta_-^\perp(\phi)\right|_{\rm ND,MP} =  0\pm 0.02\,.
\end{equation}
We have added all individual uncertainties in quadrature. 
We do not give any results for $\left.\zeta_+^\perp\right|_{\rm
  ND,MP}$ because this sum rule is dominated by higher-dimension
  condensates and hence not reliable. Again SU(3) breaking in the
  above numbers is dominated by the overall factors $m_V^2 (f_V^\perp)^2$ in
  (\ref{mp}). Both PP and MP non-diagonal sum rules agree
  about the signs of $\zeta_\pm^\perp$: $\zeta_-^\perp>0$
  and $\zeta_+^\perp<0$.

\begin{figure}[tb]
$$\epsfxsize=0.48\textwidth\epsffile{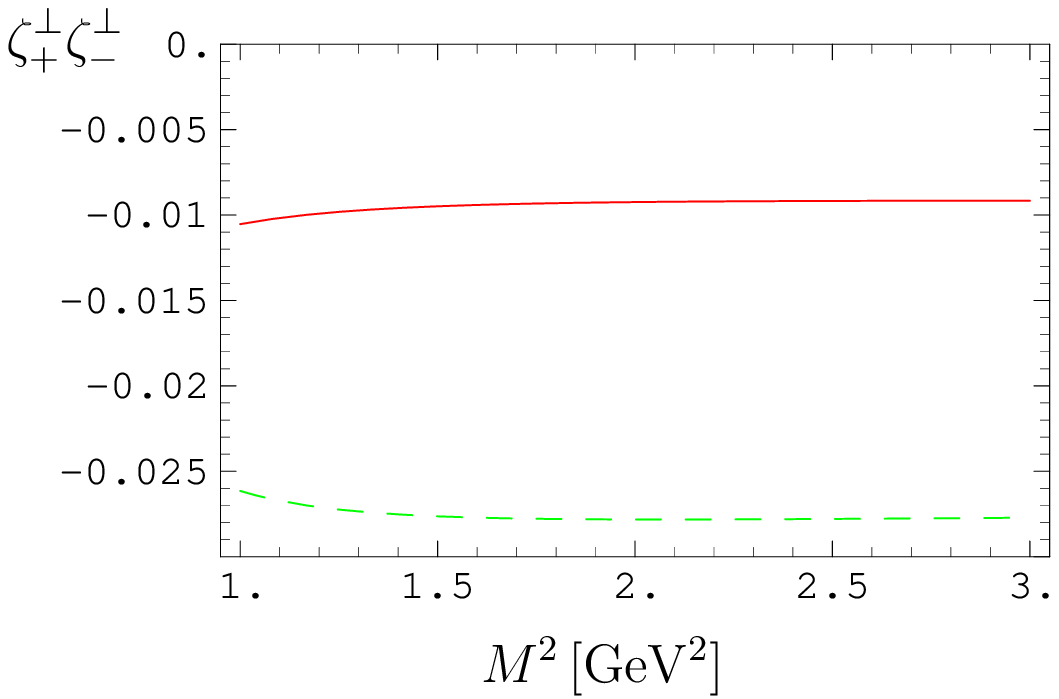}$$
\vskip-15pt
\caption[]{\small [Colour online] 
$(\zeta_+^\perp \zeta_-^\perp)_D$ for $\rho$ from the PP sum rule
  (\ref{x2}) (solid [red] curve) and the MP sum rule (\ref{x4})
  (dashed [green] curve).}\label{figE}
$$\epsfxsize=0.48\textwidth\epsffile{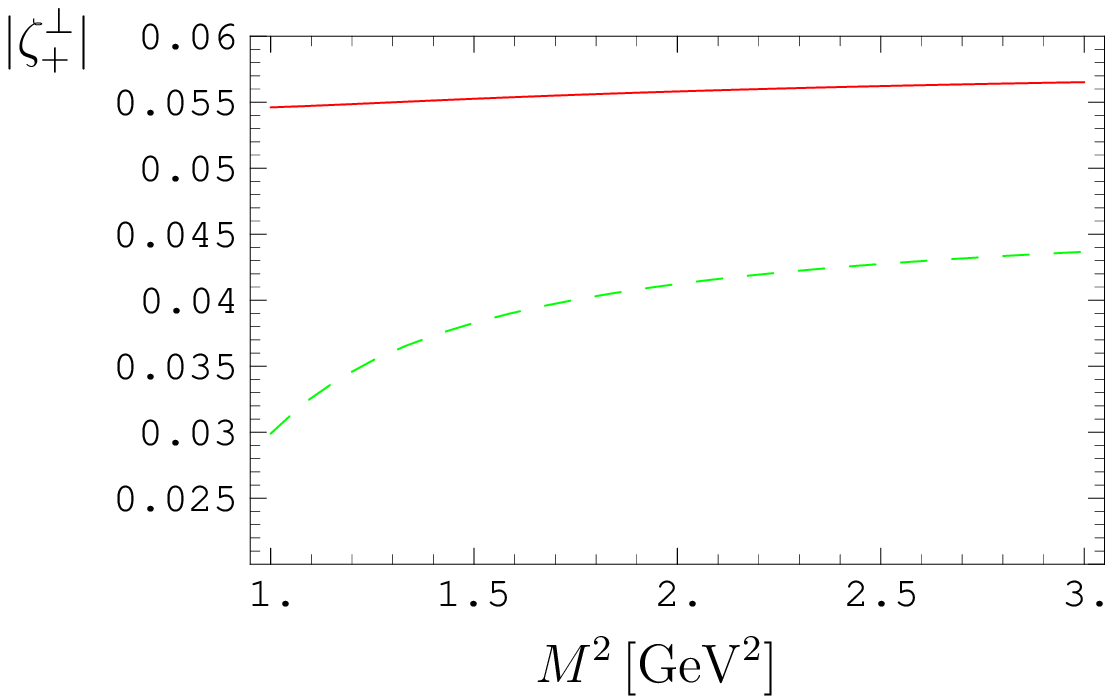}\quad
\epsfxsize=0.48\textwidth\epsffile{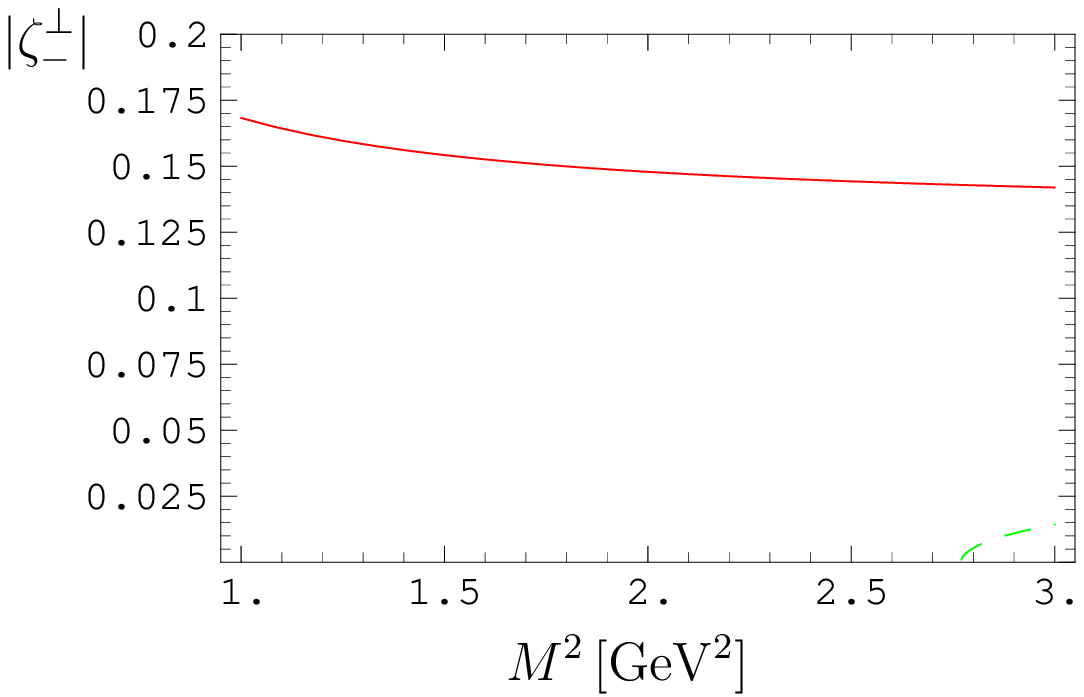}$$
\vskip-15pt
\caption[]{\small [Colour online] 
$|\zeta_+^\perp|_{\rm D}$ (left) and $|\zeta_-^\perp|_{\rm
    D}$ (right), for the $\rho$, 
   from the sum rules (\ref{x1}) and (\ref{x3}). Solid [red]
    curves: PP ($s_0=1.2\,{\rm GeV}^2$), dashed [green] curves: MP sum rules
    ($s_0=1.0\,{\rm GeV}^2$).}\label{figF}
\end{figure}

Let us now discuss the diagonal sum rules for $\zeta^\perp_\pm$ which
can be derived from the correlation function of two 
quark-antiquark-gluon currents:
\begin{eqnarray}
\Pi^{\pm\pm}_{\alpha\beta\mu\nu} &=& i \int d^4y e^{-ipy} \langle 0 | T\{ 
[\bar q g(G_{\mu\nu} \pm i \widetilde G_{\mu\nu}\gamma_5) s](0) 
[\bar s g(G_{\alpha\beta} \pm i \widetilde G_{\alpha\beta}\gamma_5)
  q](y) \} | 0 \rangle
\nonumber\\
& = & \frac{1}{p^2}\Big\{\left[(p_\mu g_{\nu\alpha}
  p_\beta) 
  - ( \mu\leftrightarrow \nu)\right] - \left[ \alpha\leftrightarrow
  \beta \right] \Big\}\Pi^{\pm\pm}_{V}\nonumber\\
&&{}+\frac{1}{p^2}\Big\{\left[(p_\mu g_{\nu\alpha}
  p_\beta) 
  - ( \mu\leftrightarrow \nu)\right] - \left[ \alpha\leftrightarrow
  \beta \right] + p^2 (g_{\alpha\mu} g_{\beta\nu} - g_{\alpha\nu}
g_{\beta\mu}) \Big\}\Pi^{\pm\pm}_{A}.\label{C.12}
\end{eqnarray}
Like for the non-diagonal correlation function (\ref{C.2}), 
the invariant functions $\Pi^{\pm\pm}_{V}$ and $\Pi^{\pm\pm}_{A}$ 
contain contributions of  $J^P=1^-$ and $1^+$ states,
respectively, and can be separated by the  projections
\begin{eqnarray}
(pz)^2 \Pi^{\pm\pm}_1 
&\equiv& 
z^\mu z^\alpha g^{\nu\beta}\Pi^{\pm\pm}_{\alpha\beta\mu\nu} = 
          -2 \frac{(pz)^2}{p^2}\Big[\Pi^{\pm\pm}_{V} + \Pi^{\pm\pm}_{A}\Big],
\nonumber
\\
\phantom{(pz)^2} \Pi^{\pm\pm}_2 
&\equiv& 
g^{\mu\alpha}g^{\nu\beta}\Pi^{\pm\pm}_{\alpha\beta\mu\nu} =
       -6 \Big[\Pi^{\pm\pm}_{V} - \Pi^{\pm\pm}_{A}\Big].        
\label{eq:project27}
\end{eqnarray}
Obviously $\Pi^{+-}_{V(A)} = \Pi^{-+}_{V(A)}$. We calculate these
invariant functions neglecting mass corrections and find
\begin{eqnarray}
 \Pi^{++}_{V} &=&  
  -\frac{\alpha_s}{480\pi^3}p^6 \ln\frac{-p^2}{\mu^2}  
 +\frac{\langle g^3 f G^3\rangle}{48\pi^2}
 - \frac{g^2}{324}  \langle \bar q q\rangle \langle \bar q\sigma g G q\rangle 
  \frac{1}{p^2}\,,
\nonumber\\
 \Pi^{++}_{A} &=&  
  -\frac{\alpha_s}{480\pi^3}p^6 \ln\frac{-p^2}{\mu^2}
 -\frac{\langle g^3 f G^3\rangle}{48\pi^2}  
 - \frac{g^2}{324} 
 \langle \bar q q\rangle \langle \bar q\sigma g G q\rangle 
 \frac{1}{p^2}\,,
\nonumber\\
 \Pi^{+-}_{V} &=&
+\frac{1}{24}\left\langle\frac{\alpha_s}{\pi}G^2\right\rangle p^2 
\ln\frac{-p^2}{\mu^2} 
+\frac{4\pi}{9}\alpha_s \langle \bar q q\rangle^2
  + \frac{\langle g^3 f G^3\rangle}{24\pi^2} \left[\ln
 \frac{\mu^2}{-p^2}-
\frac{1}{2}\right]
\nonumber\\&&{}
 + \frac{23 g^2}{216} 
 \langle \bar q q\rangle \langle \bar q\sigma g G q\rangle 
 \frac{1}{p^2}\,,
 \nonumber\\
 \Pi^{+-}_{A} &=&
-\frac{1}{24}\left\langle\frac{\alpha_s}{\pi}G^2\right\rangle p^2 
\ln\frac{-p^2}{\mu^2}
 +\frac{4\pi}{9}\alpha_s \langle \bar q q\rangle^2
  - \frac{\langle g^3 f G^3\rangle}{24\pi^2} 
\left[\ln \frac{\mu^2}{-p^2}-\frac{1}{2}\right]
\nonumber\\&&{}
 + \frac{5g^2}{216}
 \langle \bar q q\rangle \langle \bar q\sigma g G q\rangle  
 \frac{1}{p^2}\,,
\nonumber\\
 \Pi^{--}_{V} &=&
 -\frac{\alpha_s}{480\pi^3}p^6 \ln\frac{-p^2}{\mu^2} 
+ \frac{\langle g^3 f G^3\rangle}{48\pi^2} 
 - \frac{29g^2}{324}
 \langle \bar q q\rangle \langle \bar q\sigma g G q\rangle  
 \frac{1}{p^2}\,,
\nonumber\\
 \Pi^{--}_{A} &=&
 -\frac{\alpha_s}{480\pi^3}p^6 \ln\frac{-p^2}{\mu^2} 
- \frac{\langle g^3 f G^3\rangle}{48\pi^2} 
+ \frac{13g^2}{324}
 \langle \bar q q\rangle \langle \bar q\sigma g G q\rangle  
 \frac{1}{p^2}\,.
\end{eqnarray}
Without quark-mass corrections, the above results only allow the
calculation of the $\rho$ couplings.
We construct PP and MP sum rules for
$(\zeta_+^\perp)^2$ and $(\zeta_-^\perp)^2$, and also for the product
$\zeta_+^\perp \zeta_-^\perp$:
\begin{eqnarray}
(f_\rho^\perp)^2 m_\rho^6 \left.(\zeta_\pm^\perp)^2\right|_{\rm D,PP}
  & = & {\cal B}_{\rm sub} \Pi^{\pm\pm}_V\,,\label{x1}\\
(f_\rho^\perp)^2 m_\rho^6 \left.(\zeta_+^\perp\zeta_-^\perp)\right|_{\rm D,PP}
  & = & {\cal B}_{\rm sub} \Pi^{+-}_V\,,\label{x2}\\
(f_\rho^\perp)^2 m_\rho^4 \left.(\zeta_\pm^\perp)^2\right|_{\rm D,MP}
  & = & {\cal B}_{\rm sub} \frac{1}{p^2}\,\left(\Pi^{\pm\pm}_V +
  \Pi^{\pm\pm}_A\right),\label{x3}\\
(f_\rho^\perp)^2 m_\rho^4 \left.(\zeta_+^\perp\zeta_-^\perp)\right|_{\rm D,MP}
  & = & {\cal B}_{\rm sub} \frac{1}{p^2}\,\left(\Pi^{+-}_V +
  \Pi^{+-}_A\right).\label{x4}
\end{eqnarray}
The results are shown in Figs.~\ref{figE} and \ref{figF}.
As Fig.~\ref{figE} indicates, both (\ref{x2}) and (\ref{x4}) predict
different sign for $\zeta^\perp_+$ and $\zeta^\perp_-$, which agrees
with the result from non-diagonal sum rules, Figs.~\ref{figC} and
\ref{figD}. In Fig.~\ref{figF} we plot the results from (\ref{x1}) and
(\ref{x3}). All four sum rules receive contributions only from
perturbation theory and the dimension 8 condensate $\langle\bar q
q\rangle \langle \bar q \sigma g G q\rangle$. For both sum rules
for $\zeta^\perp_+$, perturbation theory is dominant, whereas for
those for $\zeta^\perp_-$, the condensate is dominant. For
$\left. \zeta^\perp_-\right|_{\rm D,MP}$ it comes with a negative
sign, which explains the fact that the dashed curve
in Fig.~\ref{figF} (right) only starts at large $M^2$: below that, the result
is imaginary. A possible explanation is the contribution of
hybrid $1^\pm$ states to the sum rules, both for
$\left. \zeta^\perp_-\right|_{\rm D,MP}$ and
$\left. \zeta^\perp_-\right|_{\rm D,PP}$. Indeed, replacing the
parametrisation on the left-hand side of (\ref{x3}) by a contribution
of such a state, with a mass $\sim 1.5\,$ or $2\,$GeV, and
correspondingly larger $s_0\approx 4\,{\rm GeV}^2$, the coupling
becomes real. In any case, the dominance of the dimension-8 term in
$\langle\bar q
q\rangle \langle \bar q \sigma g G q\rangle$ in these sum rules renders
them unreliable and hence we discard their results. As for
$\left. \zeta^\perp_+\right|_{\rm D}$, the sum rule is very dependent
on $s_0$, which totally dominates the error budget. This 
is another manifestation of the fact that the currents
in (\ref{C.12}) have a strong coupling to hybrid states. We finally
find
\begin{equation}\label{C.21}
\left.\zeta_+^\perp\right|_{\rm D,PP} = -0.06\pm 0.02,\qquad
\left.\zeta_+^\perp\right|_{\rm D,MP} = -0.04\pm 0.02\,.
\end{equation}

Our last task is to give final results for $\zeta_\pm^\perp$ and,
equivalently, $\zeta_{4V}^\perp$ and $\widetilde{\zeta}_{4V}^\perp$,
according to (\ref{C.0}). 
Let us first discuss the $\rho$ parameters, as only for those we
  have information from both diagonal and non-diagonal sum rules. For
  $\zeta^\perp_+$, we have three results,
  $\left.\zeta^\perp_+\right|_{\rm D,PP}$  and 
$\left.\zeta^\perp_+\right|_{\rm D,MP}$ from (\ref{C.21}), dominated
  by perturbation theory, and $\left.\zeta^\perp_+\right|_{\rm
  ND,PP}=-0.12\pm 0.04$, dominated by the gluon condensate
  contribution. As none of these results is a priori ``better'' than
  the others, we average over all of them to obtain our final result
\begin{equation}
\zeta^\perp_+(\rho,1\,{\rm GeV}) = -0.10\pm 0.06\,.
\end{equation}
The average is smaller than the MP sum result alone, which we take
into account when arriving at our final results for $K^*$ and $\phi$
from (\ref{C.9}):
\begin{equation}
\zeta^\perp_+(K^*,1\,{\rm GeV}) = -0.06\pm 0.03\,,\qquad
\zeta^\perp_+(\phi,1\,{\rm GeV}) = -0.04\pm 0.03\,.
\end{equation}
As for $\zeta^\perp_-$, the diagonal sum rules have to be discarded,
whereas the non-diagonal ones yield
$\left.\zeta^\perp_-\right|_{\rm ND,PP}=0.03\pm 0.02$, with the most
relevant contributions from perturbation theory and the dimension 6
condensate $\langle\bar q q\rangle^2$, and 
$\left.\zeta^\perp_-\right|_{\rm ND,MP}=0.07\pm 0.03$, with the most
relevant contributions from perturbation theory and the gluon
condensate. Here we obtain our final result as a straight average of
PP and MP sum rules and find, from (\ref{C.9}) and (\ref{C.10}):
\begin{equation}
\zeta^\perp_-(\rho,1\,{\rm GeV}) =  0.05\pm 0.05\,,\quad
\zeta^\perp_-(K^*,1\,{\rm GeV})  =  0.04\pm 0.04\,,\qquad
\zeta^\perp_-(\phi,1\,{\rm GeV}) =  0.02\pm 0.04\,.
\end{equation}
{}From (\ref{C.0}), we also find the final results for $\zeta^\perp_{4V}$
and $\widetilde{\zeta}^\perp_{4V}$:
\begin{eqnarray}
\zeta^\perp_{4\rho}(1\,{\rm GeV}) & = & -0.03\pm 0.05\,,\quad
\zeta^\perp_{4K^*}(1\,{\rm GeV})    =   -0.01\pm 0.03\,,\quad
\zeta^\perp_{4\phi}(1\,{\rm GeV})   =   -0.01\pm 0.03\,,
\nonumber\\
\widetilde\zeta^\perp_{4\rho}(1\,{\rm GeV}) & = & -0.08\pm 0.05\,,\quad
\widetilde\zeta^\perp_{4K^*}(1\,{\rm GeV})    =   -0.05\pm 0.04\,,\quad
\widetilde\zeta^\perp_{4\phi}(1\,{\rm GeV})   =   -0.03\pm 0.04\,.
\nonumber\\[-10pt]
\end{eqnarray}
A comparison of our results with those from previous calculations is
given in Sec.~\ref{sec:5}.


\begin{thebibliography}{99}


\bibitem{exclusive}
V.~L.~ Chernyak and A.~R.~Zhitnitsky,
JETP Lett.\  {\bf 25} (1977) 510;
Sov.\ J.\ Nucl.\ Phys.\  {\bf 31} (1980) 544;\\
A.~V.~Efremov and A.~V.~Radyushkin,
Phys.\ Lett.\ B {\bf 94} (1980) 245;
Theor.\ Math.\ Phys.\  {\bf 42} (1980) 97;\\
G.~P.~Lepage and S.~J.~Brodsky,
Phys.\ Lett.\ B {\bf 87} (1979) 359;
Phys.\ Rev.\ D {\bf 22} (1980) 2157;\\
V.~L.~Chernyak, A.~R.~Zhitnitsky and V.~G.~Serbo,
JETP Lett.\  {\bf 26} (1977) 594;
Sov.\ J.\ Nucl.\ Phys.\  {\bf 31} (1980) 552.

\bibitem{BLreport} 
S.~J.~Brodsky and G.~P.~Lepage,
Adv.\ Ser.\ Direct.\ High Energy Phys.\  {\bf 5} (1989) 93.

\bibitem{LCQ} 
S.~J.~Brodsky, H.~C.~Pauli and S.~S.~Pinsky,
  Phys.\ Rept.\  {\bf 301} (1998) 299
  [arXiv:hep-ph/9705477].

\bibitem{LCSR}
For recent applications, see:\\
 P.~Ball and R.~Zwicky,
  JHEP {\bf 0110} (2001) 019
 [arXiv:hep-ph/0110115];
Phys.\ Rev.\ D {\bf 71} (2005) 014015
  [arXiv:hep-ph/0406232];
  Phys.\ Rev.\ D {\bf 71} (2005) 014029
 [arXiv:hep-ph/0412079];\\
P.~Ball and E.~Kou,
JHEP {\bf 0304} (2003) 029
[arXiv:hep-ph/0301135];\\
A.~Khodjamirian, T.~Mannel and N.~Offen,
Phys.\ Lett.\ B {\bf 620} (2005) 52
[arXiv:hep-ph/0504091];\\
 P.~Ball and R.~Zwicky,
   Phys.\ Lett.\ B {\bf 625} (2005) 225
   [arXiv:hep-ph/0507076];\\
A.~Khodjamirian, T.~Mannel and N.~Offen,
  Phys.\ Rev.\  D {\bf 75} (2007) 054013
  [arXiv:hep-ph/0611193];\\
P.~Ball and G.~W.~Jones,
  arXiv:0706.3628 [hep-ph].

\bibitem{BBNS}
M.~Beneke, G.~Buchalla, M.~Neubert and C.~T.~Sachrajda,
  Phys.\ Rev.\ Lett.\  {\bf 83} (1999) 1914
  [arXiv:hep-ph/9905312].

\bibitem{BB00}
S.~W.~Bosch and G.~Buchalla,
  Nucl.\ Phys.\  B {\bf 621} (2002) 459
  [arXiv:hep-ph/0106081].

\bibitem{BFS}
M.~Beneke, T.~Feldmann and D.~Seidel,
  Nucl.\ Phys.\  B {\bf 612} (2001) 25
  [arXiv:hep-ph/0106067].

\bibitem{BJZ}
  P.~Ball and R.~Zwicky,
  JHEP {\bf 0604} (2006) 046
  [arXiv:hep-ph/0603232];\\
  P.~Ball and R.~Zwicky,
  Phys.\ Lett.\ B {\bf 642} (2006) 478
  [arXiv:hep-ph/0609037];\\
P.~Ball, G.~W.~Jones and R.~Zwicky,
  Phys.\ Rev.\  D {\bf 75} (2007) 054004
  [arXiv:hep-ph/0612081].

\bibitem{BF90}
  V.~M.~Braun and I.~E.~Filyanov,
  Z.\ Phys.\ C {\bf 48} (1990) 239.

\bibitem{string}
  I.~I.~Balitsky and V.~M.~Braun,
  Nucl.\ Phys.\ B {\bf 311} (1989) 541.

\bibitem{BKM03}
  V.~M.~Braun, G.~P.~Korchemsky and D.~M\"uller,
  Prog.\ Part.\ Nucl.\ Phys.\  {\bf 51} (2003) 311
  [arXiv:hep-ph/0306057].

\bibitem{Andersen99}
  J.~R.~Andersen,
  Phys.\ Lett.\ B {\bf 475} (2000) 141
  [arXiv:hep-ph/9909396].

\bibitem{BGG04}
  V.~M.~Braun, E.~Gardi and S.~Gottwald,
  Nucl.\ Phys.\ B {\bf 685} (2004) 171
  [arXiv:hep-ph/0401158].

\bibitem{BBL06}
P.~Ball, V.~M.~Braun and A.~Lenz,
  JHEP {\bf 0605} (2006) 004
  [arXiv:hep-ph/0603063].

\bibitem{BBKT}
  P.~Ball, V.~M.~Braun, Y.~Koike and K.~Tanaka,
  Nucl.\ Phys.\ B {\bf 529} (1998) 323
 [arXiv:hep-ph/9802299].

\bibitem{BB98}
  P.~Ball and V.~M.~Braun,
  Nucl.\ Phys.\ B {\bf 543} (1999) 201
 [arXiv:hep-ph/9810475].
 
\bibitem{BJ07}
 P.~Ball and G.~W.~Jones,
  JHEP {\bf 0703} (2007) 069  
[arXiv:hep-ph/0702100].

\bibitem{BB03} 
P.~Ball and M.~Boglione,
  Phys.\ Rev.\ D {\bf 68} (2003) 094006
  [arXiv:hep-ph/0307337].

\bibitem{BL04}
  V.~M.~Braun and A.~Lenz,
  Phys.\ Rev.\ D {\bf 70} (2004) 074020
  [arXiv:hep-ph/0407282].

\bibitem{BZ05}
P.~Ball and R.~Zwicky,
  Phys.\ Lett.\  B {\bf 633} (2006) 289
  [arXiv:hep-ph/0510338].

\bibitem{BZ06a}
P.~Ball and R.~Zwicky,
JHEP {\bf 0206} (2006) 034
 [arXiv:hep-ph/0601086].

\bibitem{PB98}
  P.~Ball,
  JHEP {\bf 9901} (1999) 010
  [arXiv:hep-ph/9812375].


\bibitem{BBK89}
  I.~I.~Balitsky, V.~M.~Braun and A.~V.~Kolesnichenko,
  Nucl.\ Phys.\ B {\bf 312} (1989) 509.

\bibitem{Cbasis}
  V.~M.~Braun, S.~E.~Derkachov, G.~P.~Korchemsky and A.~N.~Manashov,
  Nucl.\ Phys.\  B {\bf 553} (1999) 355
  [arXiv:hep-ph/9902375];\\
  V.~M.~Braun, G.~P.~Korchemsky and A.~N.~Manashov,
  Nucl.\ Phys.\  B {\bf 603} (2001) 69
  [arXiv:hep-ph/0102313].


\bibitem{soper}
J.~B.~Kogut and D.~E.~Soper,
  Phys.\ Rev.\ D {\bf 1} (1970) 2901.

\bibitem{Ddecays}
P.~Ball,
  Phys.\ Lett.\  B {\bf 641} (2006) 50
  [arXiv:hep-ph/0608116].

\bibitem{BK86}
  V.~M.~Braun and A.~V.~Kolesnichenko,
  Phys.\ Lett.\ B {\bf 175} (1986) 485
  [Sov.\ J.\ Nucl.\ Phys.\  {\bf 44} (1986) 489].

\bibitem{G3}
A.~R.~Zhitnitsky,
  Sov.\ J.\ Nucl.\ Phys.\  {\bf 41} (1985) 846
  [Yad.\ Fiz.\  {\bf 41} (1985) 1331].

\bibitem{mslatt}
F.~Knechtli,
  Acta Phys.\ Polon.\ B {\bf 36} (2005) 3377
  [arXiv:hep-ph/0511033].

\bibitem{jamin}
E.~Gamiz {\it et al.},
Phys.\ Rev.\ Lett.\  {\bf 94} (2005) 011803
[arXiv:hep-ph/0408044];\\
S.~Narison,
  Phys.\ Rev.\  D {\bf 74} (2006) 034013
  [arXiv:hep-ph/0510108].

\bibitem{PDG}
 W.~M.~Yao {\it et al.}  [Particle Data Group],
  J.\ Phys.\ G {\bf 33} (2006) 1.

\bibitem{Suzuki}
M.~Suzuki,
Phys.\ Rev.\ D {\bf 47} (1993) 1252.

\bibitem{Goldman}
L.~Burakovsky and T.~Goldman,
Phys.\ Rev.\ D {\bf 57} (1998) 2879
[arXiv:hep-ph/9703271].

\end{thebibliography}
\end{document}